\documentclass[10pt,journal,compsoc]{IEEEtran} 

\def\myvspace{2mm}

\def\singleparamA{
\def\lnsp{1.5}
\def\taband{&}
\def\invtaband{}
\def\endtab{\end{tabular}}
\def\mylnbksp{}
\def\mylnbk{}
\def\myarrow{\longrightarrow}
\def\myemptycell{~\\}
}

\def\singleparam{
\singleparamA
\def\mycondspace{&\\}
\def\condtoprule{\toprule}
}

\usepackage[utf8]{inputenc}
\usepackage[nocompress]{cite}
\usepackage{amsmath,amssymb}
\usepackage{xcolor}
\usepackage{mathabx}
\usepackage{pxfonts}

\def\doubleparam{
\def\lnsp{1.0}
\def\taband{}
\def\invtaband{&}
\def\endtab{}
\def\mylnbk{\\}
\def\mylnbksp{\\ \vspace{\myvspace}}
\def\myarrow{\drsh}
\def\myemptycell{}
\def\mycondspace{\vspace{2mm}}
\def\condtoprule{}
}

\singleparam
\def\mytab#1{\begin{tabular}{#1}}

\doubleparam
\def\mytab#1{}

\usepackage{url,multirow,booktabs,tabularx,cancel,graphicx}

\usepackage{setspace}
\usepackage[ruled]{algorithm}
\usepackage[noend]{algorithmic}

\DeclareFontFamily{OT1}{pzc}{}
\DeclareFontShape{OT1}{pzc}{m}{it}{<-> s * [1.200] pzcmi7t}{}
\DeclareMathAlphabet{\mathpzc}{OT1}{pzc}{m}{it}

\newtheorem{theorem}{Theorem}
\newtheorem{lemma}{Lemma}
\newtheorem{proposition}{Proposition}
\newtheorem{observation}{Observation}
\newtheorem{reduction}{Reduction Bound}

\newcommand\chain{\mathcal H}
\newcommand{\Tch}{T_{\!\square}}
\newcommand{\T}{T_{\!\circ}}
\newcommand{\Tred}[1]{\T^{#1}}

\newcommand{\leco}[4]{(#1,#2,#3,#4)}

\DeclareMathOperator\INV{INV}
\newcommand{\dinv}{d_{\scriptscriptstyle\INV}^{}}
\newcommand{\dinvid}{d_{\scriptscriptstyle\INV}^{\scriptscriptstyle id}}

\newcommand{\tinvid}{\tau^*}
\newcommand\optcost[1]{\tinvid\!\!=\!#1}

\newcommand{\sepsymb}{\!\Leftrightarrow\!}

\newcommand\compsymb[1]{\overline{#1}}
\newcommand\bigT{\mathcal{P}}
\newcommand\tagset{\textsc{tag}}
\newcommand\subtree{\Upsilon}
\newcommand\subtreeof[1]{\subtree(#1)}
\newcommand\compsubtree{\compsymb{\subtree}}
\newcommand\compsubtreeof[1]{\compsubtree\!\left(#1\right)}
\newcommand\extsubtree{\subtree^{+\!}}
\newcommand\extsubtreeof[1]{\extsubtree(#1)}
\newcommand\coroot{\!\!\bowtie\!\!}

\newcommand\mateat{\!\vdash\!\!}
\newcommand\gmate[2]{#1\mateat\extsubtreeof{#2}}
\newcommand\mate[3]{\gmate{#2}{#3}}
\newcommand\qcoroot{\smallsmile}

\newcommand\tl[1]{#1^t}
\newcommand\hd[1]{#1^h}
\newcommand\rev[1]{\bar{#1}}

\newcommand\bvertex{\phi}

\newcommand\ga{\alpha}
\newcommand\gb{\beta}

\renewcommand\gg{G}
\newcommand\ext{\gamma}
\renewcommand\aa{A}
\newcommand\bb{B}

\renewcommand\notag{\emptyset}
\newcommand\cc{\mathcal K}

\newcommand\anysymb{\mathrm{T}}
\newcommand\asymb{\mathrm{A}}
\newcommand\bsymb{\mathrm{B}}
\newcommand\absymb{\asymb\bsymb}
\newcommand\noasymb{\phantom{\asymb}}
\newcommand\nobsymb{\phantom{\bsymb}}

\makeatletter
\renewcommand{\boxed}[1]{\text{\fboxsep=0.005em\fbox{\m@th$\displaystyle\:\!#1$}}}
\makeatother

\newcommand\abset{\subtree_{\!\absymb}}
\newcommand\aset{\subtree_{\!\asymb}}
\newcommand\bset{\subtree_{\!\bsymb}}
\newcommand\uset{\subtree_{\!\notag}}

\newcommand{\la}{L_{\asymb}}
\newcommand{\lb}{L_{\bsymb}}
\newcommand{\lab}{L_{\absymb}}
\newcommand{\lc}{L_{\notag}}

\newcommand{\ra}{\ell_{\!\asymb}}
\newcommand{\rb}{\ell_{\!\bsymb}}
\newcommand{\rab}{\ell_{\!\absymb}}
\newcommand{\rc}{\ell_{\!\notag}}

\newcommand\costonepath[1]{\overset{\centerdot}{#1}}
\newcommand\shcostonepath[1]{\overset{\,\centerdot}{#1}}
\newcommand\costtwopath{\overset{\centerdot\,\centerdot}{\trav}}

\newcommand\trav{\multimapboth}
\newcommand\semitrav{\multimapinv}
\newcommand{\travone}{\!\!\costonepath{\trav}\,\!\!}
\newcommand{\travtwo}{\,\!\!\costtwopath\,\!\!}
\newcommand{\semitravone}{\,\!\!\shcostonepath{\semitrav}\,\!\!}

\newcommand\ctype{\ovoid}

\newcommand\run{\Lambda}
\newcommand\srun{\lambda}

\newcommand\myand{\ $\wedge$\ ~}
\newcommand\myor{\ $\vee$\ ~}

\def\triangleSubtree#1{\begin{picture}(0,0)
  \thicklines
      \qbezier( 16,-24)(8,-12)(0,0)
      \qbezier(-16,-24)(-8,-12)(0,0)
      \qbezier(-16,-24)(0,-24)(16,-24)
  \put(00,-15){\makebox(0,0){#1}}
\end{picture}}

\def\squarevertex{\begin{picture}(0,0)
  \put(0,0){\makebox(0,0){\LARGE $\blacksquare$}}
\end{picture}}

\def\colorTagA#1{\begin{picture}(0,0)
  \put(0,0){\makebox(0,0){\color{#1}\scalebox{.55}{\boldmath$\asymb$}}}
\end{picture}}

\def\tagA{\colorTagA{black}}

\def\wTagA{\colorTagA{black}}

\def\colorTagAB#1{\begin{picture}(0,0)
  \put(0,0){\makebox(0,0){\color{#1}\scalebox{.55}{\boldmath$\absymb$}}}
\end{picture}}

\def\tagAB{\colorTagAB{black}}

\def\wTagAB{\colorTagAB{black}}

\def\colorTagB#1{\begin{picture}(0,0)
  \put(0,0){\makebox(0,0){\color{#1}\scalebox{.55}{\boldmath$\bsymb$}}}
\end{picture}}

\def\tagB{\colorTagB{black}}

\def\wTagB{\colorTagB{black}}

\def\lfsize{\footnotesize}

\def\bottomlabel#1{\begin{picture}(0,0)
  \put(0,-13){\makebox(0,0){\lfsize#1}}
\end{picture}}

\def\toplabel#1{\begin{picture}(0,0)
  \put(0,13){\makebox(0,0){\lfsize#1}}
\end{picture}}

\def\cornerlabel#1{\begin{picture}(0,0)
  \put(11,7){\makebox(0,0){\lfsize#1}}
\end{picture}}

\def\rightlabel#1{\begin{picture}(0,0)
  \put(13,0){\makebox(0,0){\lfsize#1}}
\end{picture}}

\def\leftlabel#1{\begin{picture}(0,0)
  \put(-13,0){\makebox(0,0){\lfsize#1}}
\end{picture}}

\def\badvertex{\begin{picture}(0,0)
  \put(0,0){\color{white}\circle*{14}}
  \put(0,0){\circle{14}}
\end{picture}}

\def\goodvertex{\begin{picture}(0,0)
  \put(0,0){\circle*{14}}
\end{picture}}

\def\goodtagvertex{\begin{picture}(0,0)
  \put(0,0){\goodvertex}
  \put(0,0){\makebox(0,0){\color{white}\rule{7pt}{6pt}}}
\end{picture}}

\def\badABvertex{\begin{picture}(0,0)
  \put(0,0){\badvertex}
  \put(0,0){\tagAB}
\end{picture}}

\def\badAvertex{\begin{picture}(0,0)
  \put(0,0){\badvertex}
  \put(0,0){\tagA}
\end{picture}}

\def\badBvertex{\begin{picture}(0,0)
  \put(0,0){\badvertex}
  \put(0,0){\tagB}
\end{picture}}

\def\goodABvertex{\begin{picture}(0,0)
  \put(0,0){\goodtagvertex}
  \put(0,0){\wTagAB}
\end{picture}}

\def\goodAvertex{\begin{picture}(0,0)
  \put(0,0){\goodtagvertex}
  \put(0,0){\wTagA}
\end{picture}}

\def\goodBvertex{\begin{picture}(0,0)
  \put(0,0){\goodtagvertex}
  \put(0,0){\wTagB}
\end{picture}}

\def\Leaf#1{\begin{picture}(0,0)
              \put(0,0){#1}
            \end{picture}}

\def\UnTree#1#2{\begin{picture}(0,0)
  \put(0,0){\line(0,-1){20}}
  \put(0,0){#1}
  \put(0,-20){#2}
\end{picture}}

\def\UnTreeLeft#1#2{\begin{picture}(0,0)
  \put(0,0){\line(-1,0){20}}
  \put(0,0){#1}
  \put(-20,0){#2}
\end{picture}}

\def\UnTreeRight#1#2{\begin{picture}(0,0)
  \put(0,0){\line(1,0){20}}
  \put(0,0){#1}
  \put(20,0){#2}
\end{picture}}

\def\ParTreesNarrow#1#2{\begin{picture}(0,0)
  \put(-15,0){\line(1,0){30}}
  \put(-15,0){#1}
  \put(15,0){#2}
\end{picture}}

\def\ParTrees#1#2{\begin{picture}(0,0)
  \put(-25,0){\line(1,0){50}}
  \put(-25,0){#1}
  \put(25,0){#2}
\end{picture}}

\def\ParTreesWide#1#2{\begin{picture}(0,0)
  \put(-40,0){\line(1,0){80}}
  \put(-40,0){#1}
  \put(40,0){#2}
\end{picture}}

\def\ThreeParTreesCustom#1#2#3{\begin{picture}(0,0)
  \put(-100,0){\line(1,0){70}}
  \put(-40,0){\line(1,0){130}}
  \put(-100,0){#1}
  \put(-40,0){#2}
  \put( 90,0){#3}
\end{picture}}

\def\ThreeParTrees#1#2#3{\begin{picture}(0,0)
  \put(-50,0){\line(1,0){50}}
  \put(0,0){\line(1,0){50}}
  \put(-50,0){#1}
  \put(0,0){#2}
  \put(50,0){#3}
\end{picture}}

\def\PenTree#1#2#3#4#5#6{\begin{picture}(0,0)
\put(0,0){\line(-3,-1){50}}
  \put(0,0){\line(-5,-3){30}}
  \put(0,0){\line(0,-1){20}}
  \put(0,0){\line(5,-3){30}}
\put(0,0){\line(3,-1){50}}
  \put(0,0){#1}
  \put(-50,-20){#2}
  \put(-25,-20){#3}
  \put(0,-20){#4}
  \put(25,-20){#5}
  \put(50,-20){#6}
\end{picture}}

\def\QuaTreeWide#1#2#3#4#5{\begin{picture}(0,0)
  \thinlines
  \qbezier( 0,0)(-30,-10)(-60,-20)
  \qbezier( 0,0)(-10,-10)(-20,-20)
  \qbezier( 0,0)(10,-10)(20,-20)
  \qbezier( 0,0)(30,-10)(60,-20)
    \put(0,0){#1}
  \put(-60,-20){#2}
  \put(-20,-20){#3}
  \put(20,-20){#4}
  \put(60,-20){#5}
\end{picture}}

\def\QuaTree#1#2#3#4#5{\begin{picture}(0,0)
\put(0,0){\line(-3,-2){30}}
  \put(0,0){\line(-1,-2){10}}
  \put(0,0){\line(1,-2){10}}
  \put(0,0){\line(3,-2){30}}
    \put(0,0){#1}
  \put(-30,-20){#2}
  \put(-10,-20){#3}
  \put(10,-20){#4}
  \put(30,-20){#5}
\end{picture}}

\def\TriTree#1#2#3#4{\begin{picture}(0,0)
  \put(0,0){\line(-4,-3){20}}
  \put(0,0){\line(4,-3){20}}
  \put(0,0){\line(0,-1){20}}
  \put(0,0){#1}
  \put(-20,-20){#2}
  \put(0,-20){#3}
  \put(20,-20){#4}
\end{picture}}

\def\TriTreeWide#1#2#3#4{\begin{picture}(0,0)
  \put(0,0){\line(-2,-1){30}}
  \put(0,0){\line(2,-1){30}}
  \put(0,0){\line(0,-1){20}}
  \put(0,0){#1}
  \put(-30,-20){#2}
  \put(0,-20){#3}
  \put(30,-20){#4}
\end{picture}}

\def\TriTreeRight#1#2#3#4{\begin{picture}(0,0)
  \put(0,0){\line(0,-1){20}}
  \put(0,0){\line(4,-3){20}}
  \put(0,0){\line(5,-2){40}}
  \put(0,0){#1}
  \put(0,-20){#2}
  \put(20,-20){#3}
  \put(40,-20){#4}
\end{picture}}

\def\TriTreeExtraWide#1#2#3#4{\begin{picture}(0,0)
  \put(0,0){\line(-3,-1){50}}
  \put(0,0){\line(3,-1){50}}
  \put(0,0){\line(0,-1){20}}
  \put(0,0){#1}
  \put(-50,-20){#2}
  \put(0,-20){#3}
  \put(50,-20){#4}
\end{picture}}

\def\BiTree#1#2#3{\begin{picture}(0,0)
  \put(0,0){\line(-1,-1){20}}
  \put(0,0){\line(1,-1){20}}
  \put(0,0){#1}
  \put(-20,-20){#2}
  \put(20,-20){#3}
\end{picture}}

\def\BiTreeLeft#1#2#3{\begin{picture}(0,0)
  \put(0,0){\line(-4,-3){20}}
  \put(0,0){\line(0,-1){20}}
  \put(0,0){#1}
  \put(-20,-20){#2}
  \put(0,-20){#3}
\end{picture}}

\def\BiTreeRight#1#2#3{\begin{picture}(0,0)
  \put(0,0){\line(0,-1){20}}
  \put(0,0){\line(4,-3){20}}
  \put(0,0){#1}
  \put(0,-20){#2}
  \put(20,-20){#3}
\end{picture}}

\def\BiTreeNarrow#1#2#3{\begin{picture}(0,0)
  \put(0,0){\line(-1,-2){10}}
  \put(0,0){\line(1,-2){10}}
  \put(0,0){#1}
  \put(-10,-20){#2}
  \put(10,-20){#3}
\end{picture}}

\def\BiTreeWide#1#2#3{\begin{picture}(0,0)
  \put(0,0){\line(-3,-2){30}}
  \put(0,0){\line(3,-2){30}}
  \put(0,0){#1}
  \put(-30,-20){#2}
  \put(30,-20){#3}
\end{picture}}

\begin{document}

\title{Computing the Inversion-Indel Distance}

\author{Eyla~Willing, Jens~Stoye and Mar\'ilia~D.~V.~Braga
\thanks{E. Willing, J. Stoye and M. D. V. Braga are with the Genome Informatics group, Faculty of Technology and Center for Biotechnology (CeBiTec), Bielefeld University, Germany}}

\maketitle

\begin{abstract}
The inversion distance, that is the distance between two unichromosomal genomes with the same content allowing only inversions of DNA segments, can be exactly computed
thanks to a pioneering approach of Hannenhalli and Pevzner from 1995.
In 2000, El-Mabrouk extended the inversion model to perform the comparison of unichromosomal genomes with unequal contents, combining inversions with insertions and deletions (indels) of DNA segments, giving rise to the inversion-indel distance.
However, only a heuristic was provided for its computation.
In 2005, Yancopoulos, Attie and Friedberg started a new branch of research by introducing the generic double cut and join (DCJ) operation, that can represent several genome rearrangements (including inversions).
In 2006, Bergeron, Mixtacki and Stoye showed that the DCJ distance can be computed in linear time with a very simple procedure.
As a consequence, in 2010 we gave a linear-time algorithm to compute the DCJ-indel distance.
This result allowed the inversion-indel model to be revisited from another angle.
In 2013, 
we could show that, when the diagram that represents the relation between the two compared genomes has no {\em bad components}, the inversion-indel distance is equal to the DCJ-indel distance.
In the present work we complete the study of the inversion-indel distance by giving the first algorithm to compute it exactly even in the presence of bad components.
\end{abstract}


\section{Introduction}

Genome rearrangement is a branch of comparative genomics that analyses the arrangement of two or more genomes on an abstract level, in which a genome is represented as one (unichromosomal) or more (multichromosomal) ordered sequences of oriented DNA fragments (\emph{markers}).
The structural rearrangements of chromosomal segments, such as inversions, translocations and transpositions, besides fusions and fissions of chromosomes, can change the location (with respect to the chromosome), the order and the orientation of markers.
A natural measure of comparison is then the \emph{distance}, that counts the minimum number of rearrangements required to transform one given genome into another one.
	
One of the first proposed 
models aimed to compare unichromosomal genomes (e.g. of bacteria), considering only inversions of contiguous markers.
Hannenhalli and Pevzner~\cite{HAN-PEV-1999} devised a polynomial time algorithm to compute the inversion distance when each marker occurs exactly once in each genome, based on a diagram that represents the relation between the two given genomes and is composed of cycles only.
While for most of the instances based on real data the inversion distance can be obtained by simply counting the number of cycles in this diagram, the complete analysis of the problem requires the computation of other parameters -- called \emph{bad components} --
that can be derived from the 
\emph{relational diagram} by a procedure that is considerably technical.
	
The inversion model was then extended to a 
multichromosomal model including, besides inversions, also fusions and
fissions of chromosomes, and translocations, still allowing, with a
technically even more complex procedure, the exact computation of the distance
in polynomial time 
when each marker occurs exactly once in each genome~\cite{HAN-PEV-1995}.

A relaxation on the genome content allows markers that occur
in one genome but not in the other, and \emph{vice-versa}. 
A marker of this type
can be added to or removed from a genome by content-modifying operations
called \emph{insertions} and \emph{deletions} (\emph{indels} collectively).
Combining structural rearrangements and indels is non-trivial if 
contiguous markers can be inserted or deleted in a single event.
Almost 20 years ago, El-Mabrouk~\cite{ELM-2000} studied the
inversion model with this relaxation to include indels. 
However, specially due to its technical complexity, the author could give only a heuristic to compute the inversion-indel distance.
     
The field of genome rearrangement received a new boost when
Yancopoulos, Attie and Friedberg~\cite{YAN-ATT-FRI-2005} presented a generic
model, based on the \emph{double cut-and-join (DCJ)} operation, that
allows inversions, translocations, fissions and fusions, and also
excisions and integrations of circular chromosomes. It was shown that the
DCJ distance can be computed with the help of a surprisingly
simple, linear time procedure when each maker occurs exactly once in each
genome~\cite{BER-MIX-STO-2006b}.
The relaxation of the genomes to include indels, besides DCJ operations,
increased the technical aspects of the model, but
immediately resulted in a linear time exact
algorithm to compute the DCJ-indel
distance~\cite{YAN-FRI-2009,BRA-WIL-STO-2011}.
    
The potential of using the DCJ model to revisit the inversion model in a
more efficient way was first demonstrated
in~\cite{BER-MIX-STO-2009}.
More recently, the DCJ-indel model was used to give the first exact
algorithm to compute the inversion-indel distance in the absence of
bad components~\cite{WIL-ZAC-BRA-STO-2013}.
In the present work, we complete the study of the inversion-indel model by
giving a quadratic time algorithm to exactly compute the respective
distance even in the presence of bad components. 
Observe that, since inversions are intrachromosomal rearrangements, the
inversion-indel model is restricted to unichromosomal genomes.
Our approach is developed considering circular chromosomes, but can be easily generalized to linear chromosomes\footnote{Linear chromosomes can be artificially ``circularized'' by the well known technique of \emph{capping} its two extremities. Fixing the capping of one chromosome, there are only two possible ways of capping the other chromosome. The choice that gives the smaller inversion-indel distance gives the overall result.}.

This paper is organized as follows. In Section~\ref{sec:preliminaries}, we show how the relational 
diagram, with its good and bad components, allows the study of the inversion-indel model and gives a lower bound to the inversion-indel distance. In Section~\ref{sec:tree} we introduce the {\it tagged component tree}, a structure that helps us to handle
bad components. In Section~\ref{sec:results} we present our main results, effectively showing how to use the tagged component tree to compute the inversion-indel distance. This requires a lengthy, but
finite analysis of cases of bad component \emph{cutting} and \emph{merging}. 
In our approach we first \emph{safely} reduce the tagged component tree to a \emph{residual tree} with 2 to 10 leaves.
Then we enumerate the optimal covers of all possible residual trees. Each possible residual tree corresponds to a \emph{topology} with a certain \emph{leaf composition}. Our enumeration includes 56 \emph{leaf compositions}, each one corresponding to an average of 3 topologies.
We present some key examples of the case analysis, 
explaining the properties that need to be considered. The full
enumeration is then given in the appendix.
Section~\ref{sec:conclusion} concludes with a short summary.

\section{Preliminaries}\label{sec:preliminaries}

Each marker in a chromosome is an oriented DNA fragment.
The representation of a marker~$g$ in a chromosome~$\ga$ can be
the symbol~$g$, if it is read in direct orientation in~$\ga$, or the
symbol~$\rev{g}$, if it is read in reverse orientation.
In the present work, duplicated markers are not allowed. 
We represent a circular chromosome~$\ga$ by a string~$s$, obtained by the concatenation of all symbols
in~$\ga$, read in any of the two directions (since $\ga$ is circular, we can start to read it at any marker).

Given two circular chromosomes~$\ga$ and~$\gb$, possibly with
unequal contents, let~$\gg$,~$\aa$ and~$\bb$
be three disjoint sets, such that~$\gg$ is the set of {\em common markers} that
occur once in~$\ga$ and once in~$\gb$,~$\aa$ is the set of markers that occur only
in~$\ga$, and~$\bb$ is the set of markers that occur only in~$\gb$. The markers in
sets~$\aa$ and~$\bb$ are also called {\em unique markers}.
For example, if we have~$\ga=(ab\bar x\bar cyd)$ and~$\gb=(asbuvcd)$,
then~$\gg=\{a,b,c,d\}$,~$\aa=\{x,y\}$ and~$\bb=\{s,u,v\}$.

\subsection{The inversion-indel model}

An \emph{inversion} is the operation
that cuts a circular chromosome at two different positions, creating two linear
segments with four open ends, and joins these open ends such that in the new
circular chromosome the relative orientation of the linear
segments is inverted.
Consider, for example, an inversion applied to chromosome~$\ga=(ab\bar x\bar cyd)$, that cuts before and after~$\bar x\bar c$, creating the
segments~$\bullet ydab\bullet$ and
$\bullet \bar x\bar c\bullet$, where the symbol~$\bullet$ represents
the open ends. If we then join the first with the third and the second with the fourth
open end, we obtain
$\ga'=(abcxyd)$, in which the orientation
of $\bar x\bar c$ is inverted with respect to the longer segment.

Since the chromosomes can have unequal marker sets, we also need to consider {\em insertions} and
{\em deletions} of blocks of contiguous markers~\cite{ELM-2000,YAN-FRI-2009,BRA-WIL-STO-2011}.
We refer to insertions and deletions collectively as \emph{indels}.
Indels have two restrictions~\cite{BRA-WIL-STO-2011}: (i) markers of~$\gg$ cannot be deleted; and (ii)
an insertion cannot produce duplicated
markers.
We illustrate an indel
with the following example: the deletion of markers~$xy$ from chromosome~$\ga'=(abcxyd)$ results
in~$\ga''=(abcd)$.

Given two circular chromosomes $\ga$ and $\gb$, one can transform or
\emph{sort} $\ga$ into $\gb$ with inversions and indels.
We are particularly interested in
computing the \emph{inversion-indel distance}, that is denoted by $\dinvid(\ga,\gb)$
and corresponds to the minimum number of inversions and indels required to sort
$\ga$ into $\gb$.
Observe that, if~$|\gg| \leq 1$, the problem of sorting~$\ga$ into~$\gb$
becomes trivial: we simply delete at once the unique content of the chromosome
of~$\ga$ and insert at once, in the proper orientation, the unique content of the
chromosome of~$\gb$. Due to this fact, we assume in this work that~$|\gg| \geq 2$.

\subsection{Relational diagram}

Given two circular chromosomes~$\ga$ and~$\gb$, their {\em relational
diagram}~\cite{BRA-2013}, denoted by~$R(\ga,\gb)$, is a particular view of the {\em master graph}~\cite{FRI-DAR-YAN-2008} and shows the elements of chromosome~$\ga$
in an upper horizontal line and the elements of chromosome~$\gb$ in a lower
horizontal line.
We denote the two extremities of each marker $g \in \gg$ by~$\tl{g}$ (tail) and~$\hd{g}$ (head).
For each extremity of $g$ the diagram $R(\ga,\gb)$ has a vertex in the upper line and a vertex in the lower line.
Clearly, each line (that corresponds to one of the two chromosomes) has $2|\gg|$ vertices,
and its vertices are distributed following the same order of the corresponding chromosome.
Since the chromosomes are circular, we have to choose one marker~$a \in \gg$ from which we start to read
both chromosomes, s.t.\ in both lines the leftmost vertex is $a^h$ and the rightmost is $a^t$.
Then, for each marker $g \in \gg$, we connect the upper and the lower vertices
that represent $\tl{g}$ by a dotted edge.
Similarly, we connect the upper and the lower vertices that represent $\hd{g}$
by a dotted edge.

Moreover, for each integer $i$ from $1$ to $|\gg|$, let~$\ext_i^1$ and~$\ext_i^2$ be
the upper vertices (analogously lower vertices) at positions $2i-1$ and $2i$ of
the corresponding line of the diagram.
We connect the upper vertices (analogously lower vertices)~$\ext_i^1$ and~$\ext_i^2$
by an \emph{$\ga$-edge} (analogously \emph{$\gb$-edge}) labeled by~$\ell$,
which is the substring composed of the markers of chromosome $\ga$ (analogously chromosome $\gb$)
that are between the extremities represented by~$\ext_i^1$ and~$\ext_i^2$.
We say that~$\ext_i^1$ and~$\ext_i^2$ are {\em $\gg$-adjacent}, that is, they represent extremities of
occurrences of markers from $\gg$ in chromosome $\ga$ (analogously $\gb$), so that in-between only markers
from $\aa$ (analogously $\bb$) can appear. In other words, the label~$\ell$ contains no marker of~$\gg$.
When the label of an $\ga$-edge (or a $\gb$-edge) is empty, the edge is
said to be {\em clean}, otherwise it is said to be {\em labeled}.

Each vertex is now connected to one dotted edge and either to one $\ga$-edge
(for the upper vertices) or to one $\gb$-edge (for the lower vertices), thus the
degree of each vertex is two and the diagram is simply a collection of cycles.
Each cycle alternates a pair ($\ga$-edge, dotted edge) with a pair 
($\gb$-edge, dotted edge), consequently the length of each cycle is a multiple
of 4.
A cycle that contains at least one labeled edge is said to be labeled, otherwise
the cycle is said to be clean.
The cycles of~$R(\ga,\gb)$ containing only
two dotted edges (and one $\ga$-edge and one $\gb$-edge)
are called {\em $2$-cycles} and are said to be {\em inversion-sorted}.
Longer cycles are {\em inversion-unsorted} and have to be reduced, by applying
inversions, to $2$-cycles. This
procedure is called {\em inversion-sorting} of~$\ga$ into~$\gb$: when the relational diagram of $\ga$ and $\gb$ is composed exclusively of $2$-cycles, the markers from $\gg$ appear exactly in the same relative order and relative orientations in $\ga$  and $\gb$~\cite{HAN-PEV-1999}.

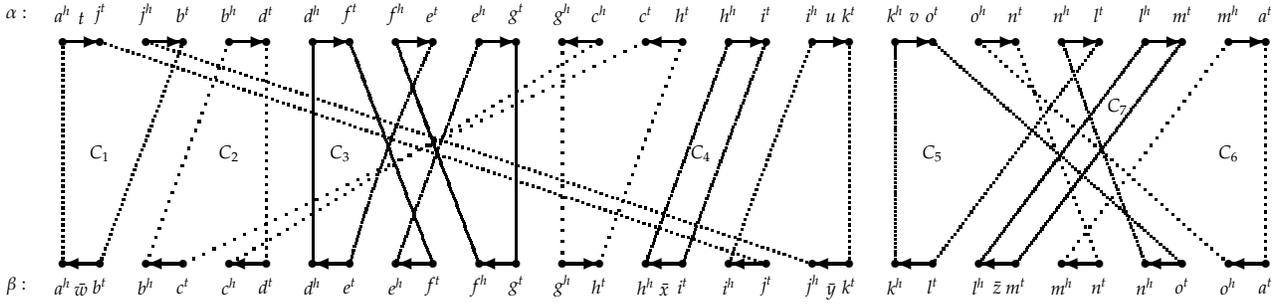
\begin{figure*}
  \centering
	\scriptsize
	\setlength{\unitlength}{.7pt}
    \begin{picture}(700,150)
    
        \put( 50, 70){\makebox(0,0){$C_1$}}
        \put(120, 70){\makebox(0,0){$C_2$}}
        \put(180, 70){\makebox(0,0){$C_3$}}
        \put(375, 70){\makebox(0,0){$C_4$}}
        \put(500, 70){\makebox(0,0){$C_5$}}
        \put(600, 95){\makebox(0,0){$C_7$}}
        \put(660, 70){\makebox(0,0){$C_6$}}

        

        \thicklines
        \qbezier[40]( 30,130)( 30, 70)( 30, 10)
        \qbezier[120]( 50,130)(230, 70)(410, 10)
        \qbezier[120]( 75,130)(255, 70)(435, 10)
        \qbezier[45]( 95,130)( 72, 70)( 50, 10)
        \qbezier[25](120,130)( 97, 70)( 75, 10)
        \qbezier[28](140,130)(140, 70)(140, 10)
        \qbezier[85](165,130)(165, 70)(165, 10)
        \qbezier[95](185,130)(207, 70)(230, 10)
        \qbezier[95](210,130)(232, 70)(255, 10)
        \qbezier[65](230,130)(202, 70)(185, 10)
        \qbezier[65](255,130)(232, 70)(210, 10)
        \qbezier[85](275,130)(275, 70)(275, 10)
        \qbezier[20](300,130)(300, 70)(300, 10)
        \qbezier[45](320,130)(220, 70)(120, 10)
        \qbezier[45](345,130)(220, 70)( 95, 10)
        \qbezier[25](365,130)(342, 70)(320, 10)
        \qbezier[65](390,130)(367, 70)(345, 10)
        \qbezier[65](410,130)(387, 70)(365, 10)
        \qbezier[45](435,130)(412, 70)(390, 10)
        \qbezier[40](455,130)(455, 70)(455, 10)
        \qbezier[55](480,130)(480, 70)(480, 10)
        \qbezier[80](500,130)(567, 70)(635, 10)
        \qbezier[45](525,130)(592, 70)(660, 10)
        \qbezier[30](545,130)(567, 70)(590, 10)
        \qbezier[60](570,130)(592, 70)(615, 10)
        \qbezier[60](590,130)(545, 70)(500, 10)
        \qbezier[95](615,130)(570, 70)(525, 10)
        \qbezier[95](635,130)(590, 70)(545, 10)
        \qbezier[30](660,130)(615, 70)(570, 10)
        \qbezier[30](680,130)(680, 70)(680, 10)

        \put( 30,130){\circle*{5}}
        \put( 30,130){\vector( 1,0){20}}
        \put( 50,130){\circle*{5}}
        \put( 75,130){\circle*{5}}
        \put( 75,130){\vector( 1,0){20}}
        \put( 95,130){\circle*{5}}
        \put(120,130){\circle*{5}}
        \put(120,130){\vector( 1,0){20}}
        \put(140,130){\circle*{5}}
        \put(165,130){\circle*{5}}
        \put(165,130){\vector( 1,0){20}}
        \put(185,130){\circle*{5}}
        \put(210,130){\circle*{5}}
        \put(210,130){\vector( 1,0){20}}
        \put(230,130){\circle*{5}}
        \put(255,130){\circle*{5}}
        \put(255,130){\vector( 1,0){20}}
        \put(275,130){\circle*{5}}
        \put(300,130){\circle*{5}}
        \put(320,130){\vector(-1,0){20}}
        \put(320,130){\circle*{5}}
        \put(345,130){\circle*{5}}
        \put(365,130){\vector(-1,0){20}}
        \put(365,130){\circle*{5}}
        \put(390,130){\circle*{5}}
        \put(390,130){\vector( 1,0){20}}
        \put(410,130){\circle*{5}}
        \put(435,130){\circle*{5}}
        \put(435,130){\vector( 1,0){20}}
        \put(455,130){\circle*{5}}
        \put(480,130){\circle*{5}}
        \put(480,130){\vector( 1,0){20}}
        \put(500,130){\circle*{5}}
        \put(525,130){\circle*{5}}
        \put(525,130){\vector( 1,0){20}}
        \put(545,130){\circle*{5}}
        \put(570,130){\circle*{5}}
        \put(570,130){\vector( 1,0){20}}
        \put(590,130){\circle*{5}}
        \put(615,130){\circle*{5}}
        \put(615,130){\vector( 1,0){20}}
        \put(635,130){\circle*{5}}
        \put(660,130){\circle*{5}}
        \put(660,130){\vector( 1,0){20}}
        \put(680,130){\circle*{5}}

        \put(  5,145){\makebox(0,0){$\ga:$}}

        \put( 30,145){\makebox(0,0){$\hd{a}$}}
        \put( 40,143){\makebox(0,0){$t$}}
        \put( 50,145){\makebox(0,0){$\tl{j}$}}
        \put( 75,145){\makebox(0,0){$\hd{j}$}}
        \put( 95,145){\makebox(0,0){$\tl{b}$}}
        \put(120,145){\makebox(0,0){$\hd{b}$}}
        \put(140,145){\makebox(0,0){$\tl{d}$}}
        \put(165,145){\makebox(0,0){$\hd{d}$}}
        \put(185,145){\makebox(0,0){$\tl{f}$}}
        \put(210,145){\makebox(0,0){$\hd{f}$}}
        \put(230,145){\makebox(0,0){$\tl{e}$}}
        \put(255,145){\makebox(0,0){$\hd{e}$}}
        \put(275,145){\makebox(0,0){$\tl{g}$}}
        \put(300,145){\makebox(0,0){$\hd{g}$}}
        \put(320,145){\makebox(0,0){$\hd{c}$}}
        \put(345,145){\makebox(0,0){$\tl{c}$}}
        \put(365,145){\makebox(0,0){$\tl{h}$}}
        \put(390,145){\makebox(0,0){$\hd{h}$}}
        \put(410,145){\makebox(0,0){$\tl{i}$}}
        \put(435,145){\makebox(0,0){$\hd{i}$}}
        \put(445,143){\makebox(0,0){$u$}}
        \put(455,145){\makebox(0,0){$\tl{k}$}}
        \put(480,145){\makebox(0,0){$\hd{k}$}}
        \put(490,143){\makebox(0,0){$v$}}
        \put(500,145){\makebox(0,0){$\tl{o}$}}
        \put(525,145){\makebox(0,0){$\hd{o}$}}
        \put(545,145){\makebox(0,0){$\tl{n}$}}
        \put(570,145){\makebox(0,0){$\hd{n}$}}
        \put(590,145){\makebox(0,0){$\tl{l}$}}
        \put(615,145){\makebox(0,0){$\hd{l}$}}
        \put(635,145){\makebox(0,0){$\tl{m}$}}
        \put(660,145){\makebox(0,0){$\hd{m}$}}
        \put(680,145){\makebox(0,0){$\tl{a}$}}

        \put( 30,10){\circle*{5}}
        \put( 50,10){\vector(-1,0){20}}
        \put( 50,10){\circle*{5}}
        \put( 75,10){\circle*{5}}
        \put( 95,10){\vector(-1,0){20}}
        \put( 95,10){\circle*{5}}
        \put(120,10){\circle*{5}}
        \put(140,10){\vector(-1,0){20}}
        \put(140,10){\circle*{5}}
        \put(165,10){\circle*{5}}
        \put(185,10){\vector(-1,0){20}}
        \put(185,10){\circle*{5}}
        \put(210,10){\circle*{5}}
        \put(230,10){\vector(-1,0){20}}
        \put(230,10){\circle*{5}}
        \put(255,10){\circle*{5}}
        \put(275,10){\vector(-1,0){20}}
        \put(275,10){\circle*{5}}
        \put(300,10){\circle*{5}}
        \put(300,10){\vector( 1,0){20}}
        \put(320,10){\circle*{5}}
        \put(345,10){\circle*{5}}
        \put(365,10){\vector(-1,0){20}}
        \put(365,10){\circle*{5}}
        \put(390,10){\circle*{5}}
        \put(410,10){\vector(-1,0){20}}
        \put(410,10){\circle*{5}}
        \put(435,10){\circle*{5}}
        \put(455,10){\vector(-1,0){20}}
        \put(455,10){\circle*{5}}
        \put(480,10){\circle*{5}}
        \put(500,10){\vector(-1,0){20}}
        \put(500,10){\circle*{5}}
        \put(525,10){\circle*{5}}
        \put(545,10){\vector(-1,0){20}}
        \put(545,10){\circle*{5}}
        \put(570,10){\circle*{5}}
        \put(590,10){\vector(-1,0){20}}
        \put(590,10){\circle*{5}}
        \put(615,10){\circle*{5}}
        \put(635,10){\vector(-1,0){20}}
        \put(635,10){\circle*{5}}
        \put(660,10){\circle*{5}}
        \put(680,10){\vector(-1,0){20}}
        \put(680,10){\circle*{5}}
        
        \put( 5,-2){\makebox(0,0){$\gb:$}}
        \put( 30,-2){\makebox(0,0){$\hd{a}$}}
        \put(40,-4){\makebox(0,0){$\bar w$}}
        \put( 50,-2){\makebox(0,0){$\tl{b}$}}
        \put( 75,-2){\makebox(0,0){$\hd{b}$}}
        \put( 95,-2){\makebox(0,0){$\tl{c}$}}
        \put(120,-2){\makebox(0,0){$\hd{c}$}}
        \put(140,-2){\makebox(0,0){$\tl{d}$}}
        \put(165,-2){\makebox(0,0){$\hd{d}$}}
        \put(185,-2){\makebox(0,0){$\tl{e}$}}
        \put(210,-2){\makebox(0,0){$\hd{e}$}}
        \put(230,-2){\makebox(0,0){$\tl{f}$}}
        \put(255,-2){\makebox(0,0){$\hd{f}$}}
        \put(275,-2){\makebox(0,0){$\tl{g}$}}
        \put(300,-2){\makebox(0,0){$\hd{g}$}}
        \put(320,-2){\makebox(0,0){$\tl{h}$}}
        \put(345,-2){\makebox(0,0){$\hd{h}$}}
        \put(355,-4){\makebox(0,0){$\bar x$}}
        \put(365,-2){\makebox(0,0){$\tl{i}$}}
        \put(390,-2){\makebox(0,0){$\hd{i}$}}
        \put(410,-2){\makebox(0,0){$\tl{j}$}}
        \put(435,-2){\makebox(0,0){$\hd{j}$}}
        \put(445,-5){\makebox(0,0){$\bar y$}}
        \put(455,-2){\makebox(0,0){$\tl{k}$}}
        \put(480,-2){\makebox(0,0){$\hd{k}$}}
        \put(500,-2){\makebox(0,0){$\tl{l}$}}
        \put(525,-2){\makebox(0,0){$\hd{l}$}}
        \put(535,-3){\makebox(0,0){$\bar z$}}
        \put(545,-2){\makebox(0,0){$\tl{m}$}}
        \put(570,-2){\makebox(0,0){$\hd{m}$}}
        \put(590,-2){\makebox(0,0){$\tl{n}$}}
        \put(615,-2){\makebox(0,0){$\hd{n}$}}
        \put(635,-2){\makebox(0,0){$\tl{o}$}}
        \put(660,-2){\makebox(0,0){$\hd{o}$}}
        \put(680,-2){\makebox(0,0){$\tl{a}$}}
      \end{picture}

    \caption{For chromosomes~$\ga=(atjbdfeg\bar chiukvonlm)$
  and~$\gb=(awbcdefghxijyklzmno)$, with
  sets of common and unique markers $\gg=\{a,b,c,d,e,f,g,h,i,j,k,l,m,n,o\}$, $\aa=\{t,u,v\}$ and
  $\bb=\{w,x,y,z\}$, the relational diagram has the five inversion-unsorted cycles $C_1$, $C_2$, $C_3$, $C_5$ and $C_6$ and the two inversion-sorted cycles $C_4$ and $C_7$.
  Cycles $C_1$, $C_4$, $C_5$ and
  $C_7$ are labeled and cycles $C_2$, $C_3$ and $C_6$ are clean.
  }
  \label{fig:rel-graph}
\end{figure*}

By walking through each cycle of the relational diagram, arbitrarily in one of
the two possible directions, we assign an orientation to each $\ga$-edge and to each
$\gb$-edge. 
We represent the labels according to the assigned direction instead of taking a simple left-to-right orientation
for each edge, in order to avoid any ambiguity.
Figure~\ref{fig:rel-graph} shows a relational diagram of two chromosomes~$\ga$ and~$\gb$.


\subsubsection{Bad and good cycles, bad and good components}\label{sec:bad-x-good}

The \emph{relative} orientations of $\ga$-edges and $\gb$-edges within one cycle
are useful for classifying different types of inversions.
In the following, without loss of
generality, we will refer to operations applied to $\ga$-edges of~$R(\ga,\gb)$, but
a symmetric analysis could be done using $\gb$-edges.
Given an inversion $\rho$, the \emph{inversion cost} of $\rho$, denoted by
$\|\rho\|$, depends on the effect of $\rho$ on the number of cycles in~$R(\ga,\gb)$ as
follows~\cite{HAN-PEV-1999}:
\begin{itemize}
  \item a \emph{split inversion} $\rho$ has $\|\rho\|=0$: it is  applied to two $\ga$-edges that are in the same cycle and have opposite relative orientations, splitting the original cycle into two (increasing the number of cycles in~$R(\ga,\gb)$ by one);
  
  \item a \emph{neutral inversion} $\rho$ has $\|\rho\|=1$: it is applied to two 
$\ga$-edges that are in the same cycle with the same relative orientation,
changing the internal organization of the original cycle without splitting it;
 
  \item a \emph{joint inversion} $\rho$ has $\|\rho\|=2$: it is applied to $\ga$-edges from two distinct
  cycles $C_1$ and $C_2$, joining $C_1$ and $C_2$ into one cycle
  (decreasing the number of cycles in~$R(\ga,\gb)$ by one).
  
\end{itemize}

If a cycle $C$ does not have a pair of $\ga$-edges with opposite orientations, it cannot be split by any inversion applied to its $\ga$-edges and is
called a {\em bad cycle}. Otherwise the cycle $C$ is said to be {\em good}.
Two distinct cycles $C$ and $C'$ are said to be {\em interleaving} when in the relational
diagram there is at least one $\ga$-edge of $C$ between two $\ga$-edges of $C'$
and at least one $\ga$-edge of $C'$ between two $\ga$-edges of $C$.
An {\em interleaving path} connecting two cycles $C$ and $C'$ is
defined as a sequence of cycles $C_1,C_2,\ldots,C_k$ such that $C_1=C$,
$C_k=C'$ and $C_i$ and $C_{i+1}$ are interleaving for all $i$, $1\leq i<k$.
An {\em interleaving component} or simply {\em component} is then
a maximal set of cycles $\cc$ where each $C \in \cc$ is connected
by an interleaving path to any other $C' \in \cc$.

Components can be of three types. The first type is a $2$-cycle, that can never interleave with any other
cycle and is called a \emph{trivial component}. The other two types are
components of inversion-unsorted cycles. 
A non-trivial component that contains at least one good cycle can be sorted with split inversions and is called a \emph{good component}, otherwise it is called a {\em bad component}~\cite{HAN-PEV-1999}.
The relational diagram represented in Figure~\ref{fig:rel-graph} has six
components:
one good (the cycle $C_2$), two trivial (the cycles $C_4$ and $C_7$) and three bad
(each of the cycles $C_1$ and $C_3$ and interleaving cycles $C_5$ and $C_6$).
Let $\dinv(\ga,\gb)$ be the minimum number of inversions required to transform
$R(\ga,\gb)$ into a collection of inversion-sorted $2$-cycles.  
It is well known that when $R(\ga,\gb)$ has no bad components, this transformation can be done by applying
split inversions only~\cite{HAN-PEV-1999}. Since the maximum number of cycles in $R(\ga,\gb)$ is equal
to $|\gg|$, we have:

\begin{lemma}[adapted 
from~\cite{HAN-PEV-1999,MEI-WAL-DIA-2000}]\label{lem:dcjinvSort} For two
circular chromosomes~$\ga$ and~$\gb$, such that $R(\ga,\gb)$ has no bad
component, $\dinv(\ga,\gb)=|\gg|-c$, where $c$ is the number of cycles in $R(\ga,\gb)$.
\end{lemma}

\subsubsection{Runs, indel-potential and inversion-indel distance lower bound}

Given two
chromosomes~$\ga$ and~$\gb$ and a cycle~$C$ of~$R(\ga,\gb)$, a \emph{run} is a
maximal subpath of~$C$, in which the first and the last edges are
labeled and all labeled edges belong to the same
chromosome~\cite{BRA-WIL-STO-2011}. A run with labels in chromosome~$\ga$ is also called
an $\aa$-run, and a run with labels in chromosome~$\gb$ is called a~$\bb$-run. Observe that, since we start counting runs in any border between an $\aa$- and a $\bb$-run, a cycle can have either 0, or 1, or an
even number of runs. We
denote by $\run(C)$ the number of runs in cycle~$C$. 
As an example, note that the cycle $C_1$ represented in Figure~\ref{fig:rel-graph} has 2 runs
($\{\hd{a}t\tl{j},\tl{j}\hd{i},\hd{i}u\tl{k}\}$ is an $\aa$-run, while $\{\tl{k}\bar y\hd{j},\hd{j}\tl{b}, \tl{b}\widebar w\hd{a}\}$ is a $\bb$-run).
While doing the inversion-sorting of good components with split inversions
only, an entire run can be accumulated into a single edge of the relational
diagram~\cite{BRA-WIL-STO-2011,WIL-ZAC-BRA-STO-2013}. The number $\run(C)$ is then an upper bound for the contribution of cycle $C$ in the final number of indels obtained while doing the inversion-sorting of $R(\ga,\gb)$ with split inversions only. The exact value of this contribution is the \emph{indel-potential} of a cycle~$C$, that is denoted by~$\srun(C)$ and can be directly derived from~$\run(C)$ as follows:

\begin{proposition}[adapted from~\cite{BRA-WIL-STO-2011}]\label{prop:indel-pot}
Given two circular chromosomes~$\ga$ and~$\gb$, the
indel-potential of a cycle~$C$ of~$R(\ga,\gb)$ is given by

\[
\srun(C) = 
\begin{cases}
\run(C)\:, & \mbox{ if $\run(C) \in \{0,1,2\}$}\:, \\[2mm]
\frac{\run(C)}{2}+1\:, & \mbox{ if $\run(C) \in \{4,6,8,\ldots\}$}\:.
\end{cases}
\] 
\end{proposition}

Let $\srun_0$ and $\srun_1$ be,
respectively, the sum of the indel-potentials for the components of the
relational diagram before and after an inversion $\rho$.
The \emph{indel-cost} of $\rho$ is then $\Delta \srun(\rho) = \srun_1
- \srun_0$, and the inversion-indel cost of $\rho$ is defined as
$\Delta d(\rho) = \|\rho\| + \Delta \srun(\rho)$.
If $R(\ga,\gb)$ has good cycles, it is always possible to find a split
inversion $\rho$ such that $\rho$ does not increase the overall
indel-potential of $R(\ga,\gb)$, that is, $\Delta
\srun(\rho)=0$ and, consequently, $\Delta
d(\rho)=0$~\cite{WIL-ZAC-BRA-STO-2013}.
Moreover, it has been shown that using neutral or joint inversions one can never achieve $\Delta d < 0$. Therefore, we have the following result:
 
\begin{theorem}[from~\cite{BRA-WIL-STO-2011,WIL-ZAC-BRA-STO-2013}]\label{thm:inv-indel-distance-good}
Given two circular
chromosomes $\ga$ and $\gb$, such that $R(\ga,\gb)$ has~$c$ cycles and no bad component, we
have $$\dinvid(\ga,\gb) = |\gg| - c + \!\!\!\! \sum_{C \in R(\ga,\gb)} \!\!\!\!\!\!
\srun(C).$$
\end{theorem}

\section{Handling bad components with the tagged component tree}\label{sec:tree}

When $R(\ga,\gb)$ has bad components, it requires additional neutral and joint inversions to be sorted. Therefore, in this case the formula given in Theorem~\ref{thm:inv-indel-distance-good} is only a lower bound for the inversion-indel distance.
Denoting by $\tinvid(\ga,\gb)$ the extra cost
of optimally combining neutral and joint inversions to sort the bad components of
$R(\ga,\gb)$ and given that~$c$ is the number of cycles in $R(\ga,\gb)$, we can express the complete formula for the inversion-indel distance as:

$$\dinvid(\ga,\gb) = |\gg| - c +
\!\!\!\! \sum_{C \in R(\ga,\gb)} \!\!\!\!\!\! \srun(C) + \tinvid(\ga,\gb).$$ 

\subsection{Neutral and joint inversions sort bad components}
For computing $\tinvid(\ga,\gb)$ it is necessary to examine the possible values
of the inversion-indel cost $\Delta d$ for neutral and joint inversions.
In the remainder of this section we will recall results
from~\cite{HAN-PEV-1999,WIL-ZAC-BRA-STO-2013}, explaining the approach of
sorting bad components with neutral and joint inversions that minimize indels.
Before that, we need to introduce the following notation.
A cycle with no run is represented by $\ctype$.
A cycle with exactly one
$\aa$-run (respectively one $\bb$-run) is represented by $\ctype_{\!\asymb}$ (respectively $\ctype_{\bsymb}$).
Similarly, a cycle with $2i$ runs is represented by $\ctype_{\!\absymb}^{~\!i}$.
For simplicity, we
denote $\ctype_{\!\absymb}^{~\!1}$ by $\ctype_{\!\absymb}$.

\subsubsection{Neutral inversions cut bad components}

Let $\cc$ be a bad component and let $C$ be any (bad) cycle in $\cc$.
A neutral inversion, applied to $C$, turns $\cc$ into a good
component~\cite{HAN-PEV-1999}.
This type of inversion is said to be a {\em cut} of the bad component.
If $R(\ga,\gb)$ has a bad cycle of type $\ctype_{\!\absymb}^{~\!i}$ with $i\geq 2$, it is always 
possible to apply a neutral inversion $\rho$ such that we obtain a good cycle of
type $\ctype_{\absymb}^{~\!\!i-1}$. In other words, two $\aa$-runs and two
$\bb$-runs are merged by $\rho$, implying $\Delta\srun(\rho)=-1$
and, consequently, $\Delta d(\rho)=0$.
For all other types of bad cycles, it is not possible to decrease
the overall indel-potential with neutral inversions~\cite{WIL-ZAC-BRA-STO-2013}.

\begin{observation}\label{ass:costless-neutral}
Since it is always possible to apply a costless neutral
inversion on a cycle of type $\ctype_{\!\absymb}^{~\!i}$ with $i \geq 2$,
we can assume that $R(\ga,\gb)$ has no cycle of this type, that is, we can assume that any cycle in
$R(\ga,\gb)$ has at most 2 runs. In this setting, the most efficient way to cut a
bad component is by applying a neutral inversion $\rho$ with $\Delta d(\rho)= 1$.
\end{observation}

\subsubsection{Joint inversions merge bad components}

Let $\rho$ be a joint inversion applied to two distinct cycles $C_1$ and $C_2$,
resulting in a single cycle $C$. Observe that the cycle $C$ is always good, even
when both $C_1$ and $C_2$ are bad. 
Moreover, if $C_1$ and $C_2$ belong to two distinct components $\cc_1$ and
$\cc_2$, these components are \emph{merged}
into a single good component $\cc$ that contains the good cycle $C$~\cite{HAN-PEV-1999}.
Actually, an inversion $\rho$ has the potential of merging even more than two components at
once, as we will describe in the following.
Without loss of generality, let us assume that, if $R(\ga,\gb)$ has at least two components,
the first and the last $\ga$-edges of $R(\ga,\gb)$ belong to two distinct
components.


Let $\cc_1$, $\cc_2$ and $\cc_3$ be three distinct components in $R(\ga,\gb)$ such that
if we take the rightmost $\ga$-edge of $\cc_1$ and look at the following
$\ga$-edges one by one, we find an edge of $\cc_3$ before finding an edge
of $\cc_2$.
In the same way, if we take the rightmost $\ga$-edge of $\cc_2$ and look at the
following $\ga$-edges one by one, we find an edge of $\cc_3$ before
finding an edge of $\cc_1$.
The component $\cc_3$ is then said to {\em separate}~$\cc_1$ and~$\cc_2$.
(In Figure~\ref{fig:rel-graph} the good component~$\{C_1\}$ separates the good component~$\{C_2\}$, the bad component~$\{C_3\}$ and the trivial component~$\{C_4\}$ from the bad component~$\{C_5,C_6\}$ and the trivial component~$\{C_7\}$.
Similarly, the bad component~$\{C_5,C_6\}$ separates~$\{C_7\}$ from~$\{C_1\}$, $\{C_2\}$, $\{C_3\}$ and $\{C_4\}$.)
By joining with an inversion $\rho$ two cycles $C_1$ and $C_2$, that belong
to two distinct components $\cc_1$ and $\cc_2$ respectively, we merge not only
the components $\cc_1$ and $\cc_2$, but also all components that separate $\cc_1$ and $\cc_2$, into a single component $\cc$.
Even when all merged components are bad, the new component $\cc$ is always
good~\cite{HAN-PEV-1999}.
A joint inversion can then be used to merge two or more bad components.
Recall that any joint inversion $\rho$ has $\|\rho\|=2$, therefore $\Delta d(\rho) = 2
+ \Delta \srun(\rho)$.
Table~\ref{tab:merging-cost} enumerates the inversion-indel costs of the
different types of joint inversions.

\begin{table}
\caption{Inversion-indel costs of distinct types of joint inversions (any joint inversion $\rho$ has inversion cost $\|\rho\|=2$, 
a cycle with any number of runs is represented by $\ctype_{*}$)}
\label{tab:merging-cost}
\medskip
\centering
\begin{tabular}{cccc}
\hline
{\bf sources} & {\bf resultant} &
\boldmath$\Delta \srun(\rho)$  & \boldmath$\Delta d(\rho)$\\[1mm]
\hline
$\ctype + \ctype_{*}$ & $\ctype_{*}$ & 0 & 2\\
$\ctype_{\!\asymb} + \ctype_{\bsymb}$ & $\ctype_{\!\absymb}$ & 0 & 2\\[2mm]
$\ctype_{\!\asymb} + \ctype_{\!\asymb}$ & $\ctype_{\!\asymb}$ & -1 & 1\\[1mm]
$\ctype_{\bsymb} + \ctype_{\bsymb}$ & $\ctype_{\bsymb}$ & -1 & 1\\[1mm]
$\ctype_{\!\asymb} + \ctype_{\!\absymb}^{~\!i}$ & $\ctype_{\!\absymb}^{~\!i}$ & -1 & 1\\[1mm]
$\ctype_{\bsymb} + \ctype_{\!\absymb}^{~\!i}$ & $\ctype_{\!\absymb}^{~\!i}$ & -1 & 1\\[2mm]
$\ctype_{\!\absymb}^{~\!i} + \ctype_{\!\absymb}^{~\!i'}$ & $\ctype_{\!\absymb}^{\!(i+i'\!-1)}$ & -2
& 0\\[1mm]
\hline
\end{tabular}
\endtab\\

\end{table}

\begin{observation}\label{ass:costless-joint}
Since it is always possible to apply a costless joint
inversion on two cycles of type $\ctype_{\!\absymb}^{~\!i}$, we assume that
$R(\ga,\gb)$ has at most one cycle of this type and that any joint inversion
$\rho$ merging components of $R(\ga,\gb)$ has $\Delta d(\rho) \in \{1,2\}$.
\end{observation}

Observations~\ref{ass:costless-neutral}
and~\ref{ass:costless-joint} imply that we can assume that
$R(\ga,\gb)$ has at most one cycle of type
$\ctype_{\aa\bb}$ and no cycle of type $\ctype_{\aa\bb}^{~i}$ with $i \geq 2$. This assures that a cut of a
bad component costs always 1 and that merging bad components costs either 1 or 2.

\subsection{Tagged component tree}

The component separation described above can be identified by an alternative structure, that represents the \emph{chaining} and \emph{nesting} relationships of the components and helps us to find the best way of cutting and merging bad components of
$R(\ga,\gb)$. 

\subsubsection{Chained tagged component tree}

Let a \emph{chain} be a sequence of components $\cc_1$, $\cc_2$, ..., $\cc_n$ for which the rightmost $\ga$-edge of $\cc_i$ is succeeded by the leftmost $\ga$-edge of $\cc_{i+1}$, for $1 \leq i < n$. A chain is \emph{maximal} when it cannot be extended to the left nor to the right.
A maximal chain $\chain$ is \emph{nested} in a component $\cc$ when the leftmost $\ga$-edge of $\chain$ is preceded by an $\ga$-edge of $\cc$ and the rightmost $\ga$-edge of $\chain$ is succeeded by an $\ga$-edge of $\cc$.

We can then build the \emph{chained tagged component tree} $\Tch(\ga,\gb)$ as follows~\cite{BER-MIX-STO-2005,BER-MIX-STO-2009,WIL-ZAC-BRA-STO-2013}:

\begin{enumerate}
\item Each component $\cc$ from $R(\ga,\gb)$ is represented by a \emph{round node} $v$. If $\cc$ is a bad component, then $v$ is a \emph{bad node}, drawn
in white. Otherwise $v$ is a \emph{good node}, drawn in black. A
good node can represent a trivial or a good component. Each round node $v$ has a tag set $\tagset(v)$, initially empty, that can receive at most two tags:
(i) the tag {\boldmath$\asymb$}
is added to $\tagset(v)$ if at least one cycle of the component $\cc$ has an
$\aa$-run; and (ii) the tag {\boldmath$\bsymb$}
is added to $\tagset(v)$ if at least one cycle of $\cc$ has a $\bb$-run.

\item Each maximal chain is represented
  by a \emph{square node} whose children are the round nodes
  that represent the components of this chain. A square node is either the root
  or a child of the 
  component in which this chain is nested.

\end{enumerate}

The chained tagged component tree $\Tch(\ga,\gb)$ is the same component tree as introduced in~\cite{BER-MIX-STO-2005}, except that here we add the tag sets. This tree has the following important property.
Let $P$ be a path connecting two distinct round nodes $u_1$ and $u_2$ in $\Tch(\ga,\gb)$. The round nodes in $P \backslash \{u_1, u_2\}$ correspond exactly to the components that separate $u_1$ and $u_2$ in $R(\ga,\gb)$.

\subsubsection{Max-flower contraction}

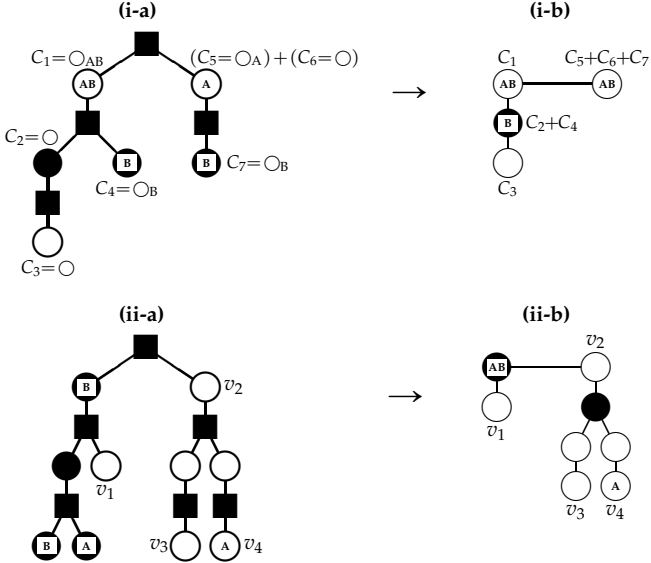
\begin{figure}[ht]
  \centering
	\scriptsize
	\lfsize
	\setlength{\unitlength}{0.75pt}

	\begin{tabular}{ccc}
	\bf (i-a)~~~~~~~~~~~~~ & &\bf (i-b)\\
	\begin{picture}(170,130)
		\thicklines
		\put(60,120){
		  \BiTreeWide{\squarevertex}
		   {\UnTree{\badABvertex\toplabel{\scriptsize$C_1\!\!=\!\ctype_{\absymb}~~~~~~$}}
		     {\BiTree{\squarevertex}
		       {\UnTree{\goodvertex\toplabel{\scriptsize$C_2\!\!=\!\ctype~~~~~$}}
		          {\UnTree{\squarevertex}{\badvertex\bottomlabel{\scriptsize$C_3\!\!=\!\ctype$}}}
		       }
		       {\goodBvertex\bottomlabel{\scriptsize$C_4\!\!=\!\ctype_{\bsymb}$}}
		     }
		    }
  		    {\UnTree{\badAvertex\toplabel{\scriptsize$~~~~~~~~~~~~~~~~~~~~~(C_5\!\!=\!\ctype_{\asymb})\!+\!(C_6\!\!=\!\ctype)$}}
		         {\UnTree{\squarevertex}
		           {\goodBvertex\rightlabel{\scriptsize$~~~~~~~~C_7\!\!=\!\ctype_{\bsymb}$}}
		         }
		    }
         }
        \end{picture} &
        \begin{picture}(20,130)
		\put(10,94){\makebox(0,0){\Large \boldmath$\rightarrow$}}
		\end{picture} &
		\begin{picture}(90,130)
		\put(45,100){
		  \ParTrees
		    {\UnTree{\badABvertex\toplabel{\scriptsize$C_1$}}{\UnTree{\goodBvertex\rightlabel{~~~~~\scriptsize$C_2\!\!+\!\!C_4$}}{\badvertex\bottomlabel{\scriptsize$C_3$}}}}
		    {\Leaf{\badABvertex\toplabel{\scriptsize$C_5\!\!+\!\!C_6\!\!+\!\!C_7$}}}
		 }
    \end{picture}
\end{tabular} 
\mylnbksp
	\begin{tabular}{ccc}
	\bf (ii-a)~~~~~~~~~~~ & & \bf (ii-b)\\
	    \begin{picture}(170,130)
		\thicklines
		\put(65,120)
		 {\BiTreeWide{\squarevertex}
		  {\UnTree{\goodBvertex}
		   {\BiTreeNarrow{\squarevertex}
		    {\UnTree
		      {\goodvertex}
		      {\BiTreeNarrow{\squarevertex}
		                {\goodBvertex}
		                {\goodAvertex}}
		    }
		    {\badvertex\bottomlabel{$v_1$}}}}
		 {\UnTree
		   {\badvertex\rightlabel{\;$v_2$}}
		   {\BiTreeNarrow{\squarevertex}
		      {\UnTree{\badvertex}
		        {\UnTree{\squarevertex}{\badvertex\leftlabel{$v_3$}}}}
		      {\UnTree{\badvertex}
		        {\UnTree{\squarevertex}{\badAvertex\rightlabel{\;$v_4$}}}}
		  }}
		}
		\end{picture} &
        \begin{picture}(20,130)
		\put(10,94){\makebox(0,0){\Large \boldmath$\rightarrow$}}
		\end{picture} &
		\begin{picture}(90,130)
		\put(45,110)
		   {\ParTrees
		     {\UnTree{\goodABvertex}{\badvertex\bottomlabel{$v_1$}}}
		     {\UnTree{\badvertex\toplabel{$v_2$}}
		      {\BiTreeNarrow
		       {\goodvertex}
		       {\UnTree
		         {\badvertex}
		         {\badvertex\bottomlabel{$v_3$}}
		       }
		       {\UnTree{\badvertex}{\badAvertex\bottomlabel{$v_4$}}}
		     }
		    }
		  }
		
    \end{picture}
	\end{tabular}
\vspace{-0.3cm}
\caption{
(i-a) The chained tagged component tree $\Tch(\ga,\gb)$ derived from
the relational diagram represented in Figure~\ref{fig:rel-graph}, with three bad (white) and three
good (black) nodes and their respective tags.
(i-b) The tree~$\T(\ga,\gb)$ obtained from $\Tch(\ga,\gb)$. 
(ii-a,b) Another example illustrating the max-flower contraction.
}
\label{fig:tree}
\end{figure}

Let a \emph{max-flower} of a chained tagged component tree be a maximal connected subgraph composed of good and/or square nodes only. Suppose we have a max-flower $f$ of a chained tagged component tree $T$. The \emph{contraction} of $f$ in $T$ consists of two steps: 
\begin{enumerate}
\item Replace $f$ by a
single good round node $g$, such that $g$ is connected to all bad nodes connected to
$f$ and $\tagset(g)$ is the union of tag sets of all nodes from $f$ (assuming that square nodes have empty tag sets);  
\item If $g$ has an empty tag set and exactly two neighbors $b_1$ and $b_2$, $g$ is removed from the tree and $b_1$ is directly connected to $b_2$; otherwise, if $g$ is a leaf and $b$ is the bad node connected to $g$, the tag set $\tagset(g)$ is added to~$\tagset(b)$ and $g$ is removed from the tree.
\end{enumerate}
After repeating this procedure for all (disjoint) max-flowers we obtain an unrooted \emph{tagged component tree} $T'$, that is said to be \emph{flower-contracted}, that is, the tree $T$' has no square nodes, no good leaves, no good nodes of degree 2 with an empty tag set, and each remaining max-flower of $T'$ corresponds to a single good node.
As we will see in the next subsection, the computation of $\tinvid(\ga,\gb)$ is derived from the analysis of an unrooted tagged component tree that is exactly the flower-contracted version of the chained tagged component tree $\Tch(\ga,\gb)$ and is denoted by $\T(\ga,\gb)$. See examples of its construction in Figure~\ref{fig:tree}. 


\vspace{5mm}

\subsection{Cover and optimal cover}

The diagram $R(\ga,\gb)$ can have any number of good and bad cycles of types
$\ctype$, $\ctype_\aa$ and $\ctype_\bb$. Due to Observations~\ref{ass:costless-neutral}
and~\ref{ass:costless-joint}, we assume that $R(\ga,\gb)$ has at most one cycle of type
$\ctype_{\aa\bb}$ and no cycle of type $\ctype_{\aa\bb}^{~i}$ with $i \geq 2$. Therefore, the cost of cutting a component is always one, independently of
the number of tags of the corresponding node in $\T(\ga,\gb)$ and --
even if a node of $\T(\ga,\gb)$ has both tags -- at most one of them can be
``used'' in each joint inversion merging components, that costs either 1 or 2.

Similar to \cite{BER-MIX-STO-2005}, in our analysis we focus on paths covering the bad nodes of $\T(\ga,\gb)$. 
We denote by $\tau(P)$ the cost of a path $P$, that can be of two types. 
The path $P$ is \emph{short} if it contains a single bad node. It corresponds to the cut of the component represented by its bad node and has therefore cost $\tau(P)=1$.
On the other hand, the path $P$ is \emph{long} if it contains at least two bad nodes. It
corresponds to the merging of all components represented by its nodes, by a
joint inversion applied to two cycles, each of the two belonging to one of the
two endpoints of the path.
The cost of $P$ is therefore either $\tau(P)=1$, if its endpoints share at least
one tag (\emph{indel-saving} path), or $\tau(P)=2$ otherwise
(\emph{indel-neutral} path).

A {\em cover} of $\T(\ga,\gb)$ is defined as a
set of paths $\bigT$ such that each bad
node of $\T(\ga,\gb)$ is contained in at least one path $P \in \bigT$.
The cost of $\bigT$ is given by $\tau(\bigT)=\sum_{P \in \bigT} \tau(P)$. 
If $\tau(\bigT)$ is minimum, $\bigT$ is 
an \emph{optimal cover} and
$\tau(\bigT)=\tau(\T(\ga,\gb))=\tinvid(\ga,\gb)$~\cite{BER-MIX-STO-2009,WIL-ZAC-BRA-STO-2013}.

The reason behind contracting the tree to get one with the minimum number of good nodes and whose leaves are bad nodes is  simply to get a more compact tree, while preserving the same bad nodes and tags of the original tree. In other words, since the flower-contractions preserve in $\T(\ga,\gb)$ all the tags from $\Tch(\ga,\gb)$ in the same relative positions with respect to the bad nodes, we have $\tau(\T(\ga,\gb)) = \tau(\Tch(\ga,\gb))$.

\subsubsection{Covering a tree with traversals}

Given a tree $T$ with $\ell$ leaves, a \emph{leaf-branch} of $T$ is: (i)~either the complete tree $T$, if~$\ell \leq 2$; (ii) or, if $\ell \geq 3$, 
a maximal path $u_1,u_2,\ldots,u_k$,
such that $u_1$ is a leaf of $T$ and, for $i=2,\ldots,k$, the degree of internal node $u_i$ in $T$ is two. 
The second case includes a leaf-branch consisting of a leaf only, if that is directly connected to a branching node of $T$.

A path $P$ whose both endpoints are leaves is called a \emph{traversal}.
Observe that, if $\ell \ge 2$ is the number of leaves of a tagged component tree $\T(\ga,\gb)$, the minimum number of traversals required to cover all $\ell$ leaves is $\lceil \frac{\ell}{2} \rceil$. 
Indeed, any unrooted tree with at least two leaves can be completely covered with this minimum number of traversals. If the number of leaves $\ell$ is even, the traversals can be obtained with a very simple procedure (Algorithm~\ref{alg:covering-traversals}, from \cite{ERD-SOU-STO-2011}). If $\ell$ is odd, it is necessary to remove one leaf-branch from the tree, run Algorithm~\ref{alg:covering-traversals} and add one extra traversal connecting the removed leaf-branch and any other leaf.

\begin{algorithm}
\caption{$\textsc{CoverTreeWithTraversals}$}
\label{alg:covering-traversals}
\algsetup{indent=2em}
\linespread{1.2}
\small
\begin{algorithmic}
  \REQUIRE unrooted tree $T$ with $2n$ leaves
  \ENSURE 
  set $\mathcal P$ of $n$ traversals covering all nodes of $T$
 \STATE ~
 \STATE Considering an embedding of $T$ into the plane, 
        enumerate the leaves from $1$ to $2n$ in circular order;
 \STATE $\mathcal P = \emptyset$;
 \FOR {$i=1$ \TO $n$}
 \STATE $\mathcal P = \mathcal P \cup \{\text{path connecting leaves } i \text{ and } i+n\}$; 
 \ENDFOR
  \STATE Return $\mathcal P$;
 \end{algorithmic}
 \end{algorithm}

The fact that Algorithm~\ref{alg:covering-traversals} covers the complete tree proves the following theorem.

\begin{theorem}[Adapted from \cite{ERD-SOU-STO-2011}] \label{th:balanced-vertex}
Given a tree $T$ with $2n$ leaves, it is always possible to find a vertex $v$ in $T$, such that there is a cover of $T$ composed of $n$ traversals, all of whom containing the vertex $v$. Such a vertex is called a \emph{balanced vertex} of $T$.
\end{theorem}
\begin{IEEEproof}[Proof given in \cite{ERD-SOU-STO-2011}]
Considering any two paths in the set returned by Algorithm~\ref{alg:covering-traversals}, their endpoints alternate along the circle which contains the leaves in increasing order. Therefore these two paths clearly intersect each other. If in a tree a set of paths does not contain two disjoint paths, then all the paths share a common vertex $v$ (a proof is given in~\cite{GYA-LEH-1970}). And because these paths connect $v$ to the leaves, they cover all edges of the tree.
\end{IEEEproof}

\subsubsection{The simplest instances give a lower and an upper bound}

A node of the tagged component tree $\T(\ga,\gb)$ whose tag set is empty is called \emph{clean}, otherwise it is \emph{tagged}.
We will now give solutions for the simplest instances of the problem of computing the cost $\tinvid(\ga,\gb)$ of an optimal cover of $\T(\ga,\gb)$, that are: (i) the case in which all leaves are tagged and share at least one tag and (ii) the case in which all leaves are clean. These two cases correspond, respectively, to tight lower and upper bounds for $\tinvid(\ga,\gb)$ and are based on optimal covers composed of traversals and at most one short path. 

\def\stateOneTag{
Let all $\ell\geq2$ leaves of a tagged component tree $\T(\ga,\gb)$ share at least one tag. The cost of any optimal cover of $\T(\ga,\gb)$ is $\tinvid(\ga,\gb)=\left\lceil\frac{\ell}{2}\right\rceil$.
}
\def\proofOneTag{
Any traversal is indel-saving (costs 1) because the leaves of $\T(\ga,\gb)$ share a tag.
As we have already explained before, with Algorithm~\ref{alg:covering-traversals} we can find $\left\lceil\frac{\ell}{2}\right\rceil$ traversals covering $\T(\ga,\gb)$ completely.
}

\begin{theorem}\label{th:one-tag}
\stateOneTag
\end{theorem}
\begin{IEEEproof}
\proofOneTag
\end{IEEEproof}

When the tree has clean leaves, the length of the leaf-branches can influence the cost of an optimal cover.
A leaf-branch of the tree~$\T(\ga,\gb)$ is \emph{long} if it contains at least two bad nodes, otherwise it is \emph{short}. 

\def\stateNoTag{
Let all $\ell\geq 2$ leaves of a tagged component tree $\T(\ga,\gb)$ be clean. The cost of any optimal cover of $\T(\ga,\gb)$ is 
$$\tinvid(\ga,\gb) = \left\{\begin{array}{ll}
    \ell+1&\mbox{if $\ell$ is odd and all leaf-branches }\mylnbk 
         \invtaband  \mbox{are long (``fortress''~\cite{HAN-PEV-1999}),}\\[2mm]
    \ell  &\mbox{otherwise.}
  \end{array}\right.$$
}
\def\proofNoTag{
Any traversal is indel-neutral (costs 2) because the leaves of $\T(\ga,\gb)$ are clean. 
When $\ell$ is even, with Algorithm~\ref{alg:covering-traversals} we can cover the tree with a 
set of $\frac{\ell}{2}$ traversals, giving an optimal cost of $\ell$ in this case.
When $\ell$ is odd, let $S$ be a short leaf-branch of $\T(\ga,\gb)$, if there is any, otherwise any of the long leaf-branches.
The subtree obtained from $\T(\ga,\gb)$ by removing the leaf-branch $S$ can be covered by a 
set of $\frac{\ell-1}{2}$ traversals obtained wit Algorithm~\ref{alg:covering-traversals}.
Then, if $S$ is a short leaf-branch, it can be covered by a short path, giving an optimal cost of $\ell - 1 + 1 = \ell$.
Otherwise, $S$ is a long leaf-branch and we have the classical situation of a ``fortress''~\cite{HAN-PEV-1999}, in which $S$ can only be covered by a long path with cost 2, resulting in the optimal cost of $\ell-1 + 2 = \ell+1$.
}
    \begin{theorem}\label{th:no-tag}
\stateNoTag
\end{theorem}
\begin{IEEEproof}
\proofNoTag
\end{IEEEproof}


\section{Obtaining an optimal cover of trees with at least two leaves that share no tag}\label{sec:results}

In this section we will 
describe our approach of finding an optimal
cover of all intermediate and more intrincate instances of the tagged component tree. 

Given a path
$P$, a \emph{$P$-reduction} $\Tred{-P}(\ga,\gb)$ of $\T(\ga,\gb)$ results in a smaller
tagged component tree as follows.
Let $T'$ be the tagged tree obtained from $\T(\ga,\gb)$ by replacing each bad node $b$ of $P$ by a good node $g$, such that $\tagset(g)=\tagset(b)$.
Then $\Tred{-P}(\ga,\gb)$ is the flower-contracted version of
$T'$. A $P$-reduction is said to be \emph{safe} if  $\tau(\T(\ga,\gb))=\tau(\Tred{-P}(\ga,\gb)) + \tau(P)$, otherwise it is \emph{unsafe}.
We will build optimal covers by first ``simplifying'' the tagged component tree with safe reductions.

From now on we will denote the three possible non-empty tag sets $\{\asymb\}$,
$\{\bsymb\}$ and $\{\asymb,\bsymb\}$, respectively, simply by $\asymb$, $\bsymb$ and $\absymb$.
We can then refer to
$\asymb$-, $\bsymb$-, $\notag$- or $\absymb$-nodes or -leaves, depending on the respective set of tags.
Tagged component trees and their optimal covers will be shown in figures with the following notation: 

\begin{itemize}
\item Leaves will be usually labeled as described: (i) $a$, $a_1$, $a_2$, etc., will be labels of $\asymb$-leaves; (ii) $b$, $b_1$, $b_2$, etc., will be labels of $\bsymb$-leaves; (iii) $x$, $x_1$, $x_2$, etc., will be labels of $\absymb$-leaves; and (iv) $s$ is the label of the solo leaf (defined in Section~\ref{sec:solo}), while $c$, $c_1$, $c_2$, etc., will be labels of other clean leaves.
\item In the given optimal covers, 
traversals will be represented by the symbol~$\trav$, semi-traversals (defined in Section~\ref{sec:mates}) will be represented by the symbol~$\semitrav$ and the dots above these symbols will indicate the cost of each path.
\end{itemize}



\subsection{Properties of tagged component tree topologies and their covers}

We need to describe the possible topologies of tagged component trees in order to devise our approach of simplifying the tree and finding optimal covers.

\subsubsection{Subtrees and partition subtrees}
    Let ${\mathcal N}$ be a set of nodes of $\T(\ga,\gb)$. Then $\subtreeof{\mathcal N}$ denotes the subtree induced by ${\mathcal N}$, i.~e., the smallest connected subtree of
    $\T(\ga,\gb)$ containing all the nodes from ${\mathcal N}$. 
    The leaves of $\T(\ga,\gb)$ can be partitioned into four disjoint classes
    $\la$, $\lb$, $\lc$ and $\lab$, that are, respectively,
    the sets of $\asymb$-leaves, $\bsymb$-leaves, $\notag$-leaves and $\absymb$-leaves of $\T(\ga,\gb)$.
    A \emph{partition subtree} is then a subtree $\subtreeof{\mathcal L}$ where ${\mathcal L}$ is the union of one to four leaf classes.
    In general, the topology and connections between the possible partition subtrees
  drive our approach of finding an optimal cover for the tree $\T(\ga,\gb)$.
    The four partition subtrees
    $\subtreeof{\la}$, $\subtreeof{\lb}$, 
    $\subtreeof{\lc}$ and $\subtreeof{\lab}$ are called \emph{canonical} subtrees of $\T(\ga,\gb)$. While canonical subtree $\subtreeof{\lc}$ is said to be \emph{clean}, subtrees $\subtreeof{\la}$, $\subtreeof{\lb}$ and $\subtreeof{\lab}$ are said to be \emph{tagged}.
 
    

   \subsubsection{Complementary subtree}  
    Let $\subtreeof{\mathcal L}$ be a partition subtree. 
    The \emph{complementary subtree} of 
    $\subtreeof{\mathcal L}$, denoted by $\compsubtreeof{\mathcal L}$, is the partition
    subtree $\subtreeof{\mathcal L'}$, where ${\mathcal L'}$ is the set composed of all leaves of $\T(\ga,\gb)$ except those from
    ${\mathcal L}$. 
    Complementary subtree is a reciprocal concept, thus~$\compsubtreeof{\mathcal L'}$ is the subtree~$\subtreeof{\mathcal L}$ itself.
    For example, considering the canonical subtree $\subtreeof{\la}$, its complementary subtree is $\compsubtreeof{\la}=\subtreeof{\lb\cup\lab\cup\lc}$.
    Another example is the partition subtree $\subtreeof{\lb\cup\lc}$, whose complementary subtree is $\compsubtreeof{\lb \cup \lc}=\subtreeof{\la\cup\lab}$. 
    Note that $\compsubtreeof{\mathcal L}$ and $\T(\ga,\gb)\backslash\subtreeof{\mathcal L}$ can be distinct and, while the first is necessarily a tree, the second can be a forest.

   \subsubsection{Links and bad links}
   
     If $\subtreeof{{\mathcal L}_1}$ and $\subtreeof{{\mathcal L}_2}$ are two non-empty and disjoint partition subtrees, there is exactly one path connecting $\subtreeof{\mathcal L_1}$ to
    $\subtreeof{\mathcal L_2}$ in $\T(\ga,\gb)$.
    This path is called the \emph{link} between $\subtreeof{\mathcal L_1}$ and
    $\subtreeof{\mathcal L_2}$. A link is a \emph{bad link} when it contains at
    least one bad node, otherwise it is a \emph{good link}.
    A bad link that
    contains two or more bad nodes is called \emph{long bad link}.
    Otherwise, if it contains a single bad node, it is called \emph{short bad
    link}. 
    
   \subsubsection{Co-rooted and separated partition subtrees}
    
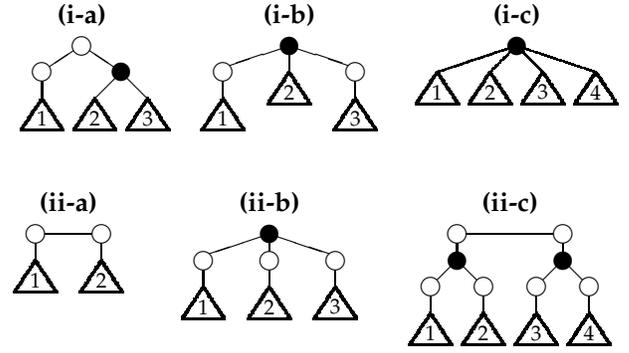
\begin{figure}[ht]
        \setlength{\unitlength}{0.5pt}
    \centering
    \begin{tabular}{c@{~~~~~~}c@{~~~~~~}c}
    {\bf (i-a)} & {\bf (i-b)} & {\bf (i-c)}\\
    \begin{picture}(100,100)
		\put(50,90){\BiTreeWide
		    {\badvertex}
		    {\UnTree{\badvertex}{\triangleSubtree{\footnotesize$1$}}}
		    {\BiTree{\goodvertex}{\triangleSubtree{\footnotesize$2$}}{\triangleSubtree{\footnotesize$3$}}}
		    }
    \end{picture}&
    \begin{picture}(130,100)
		\put(65,90){\TriTreeExtraWide
		    {\goodvertex}
		    {\UnTree{\badvertex}{\triangleSubtree{\footnotesize$1$}}}
		    {\triangleSubtree{\footnotesize$2$}}
		    {\UnTree{\badvertex}{\triangleSubtree{\footnotesize$3$}}}
		  }
    \end{picture}&
    \begin{picture}(130,100)
		\put(65,90){\QuaTreeWide
		    {\goodvertex}
		    {\triangleSubtree{\footnotesize$1$}}
		    {\triangleSubtree{\footnotesize$2$}}
		    {\triangleSubtree{\footnotesize$3$}}
		    {\triangleSubtree{\footnotesize$4$}}
		  }  
    \end{picture}
    \end{tabular} 
    \mylnbksp
    \begin{tabular}{c@{~~~~~~}c@{~~~~~~}c}
    {\bf (ii-a)} & {\bf (ii-b)} & {\bf (ii-c)}\\
        \begin{picture}(90,100)
		\put(45,90){\ParTrees
		    {\UnTree{\badvertex}{\triangleSubtree{\footnotesize$1$}}}
		    {\UnTree{\badvertex}{\triangleSubtree{\footnotesize$2$}}}
		    }
\end{picture} &
    \begin{picture}(130,100)
		\put(65,90){\TriTreeExtraWide
		    {\goodvertex}
		    {\UnTree{\badvertex}{\triangleSubtree{\footnotesize$1$}}}
		    {\UnTree{\badvertex}{\triangleSubtree{\footnotesize$2$}}}
		    {\UnTree{\badvertex}{\triangleSubtree{\footnotesize$3$}}}
		    }
\end{picture} &
    \begin{picture}(150,100)
		\put(75,90){\ParTreesWide
		    {\UnTree{\badvertex}{\BiTree{\goodvertex}{\UnTree{\badvertex}{\triangleSubtree{\footnotesize$1$}}}{\UnTree{\badvertex}{\triangleSubtree{\footnotesize$2$}}}}}
		    {\UnTree{\badvertex}{\BiTree{\goodvertex}{\UnTree{\badvertex}{\triangleSubtree{\footnotesize$3$}}}{\UnTree{\badvertex}{\triangleSubtree{\footnotesize$4$}}}}}
		    }
\end{picture}
\end{tabular}
\\
    \caption{Each of the subtrees numbered 1-4 is one distinct canonical subtree in these illustrations of different types of topology. In (i-a) subtrees~2 and~3 are co-rooted and separated from subtree~1. In (i-b) subtrees~1 and~3 are isolated, while subtree~2 is non-isolated. In (i-c) we have a fully co-rooted tagged component tree with the four canonical subtrees. In the second half we have fully separated tagged component trees with (ii-a) two, (ii-b) three, or (ii-c) four canonical subtrees.
    }
    \label{fig:topologies}
    \end{figure}

    Two non-empty partition subtrees $\subtreeof{\mathcal L_1}$ and $\subtreeof{\mathcal L_2}$ are said to be \emph{separated} when they are
    disjoint and connected by a bad link. Otherwise, they are \emph{co-rooted}.
    A partition subtree $\subtreeof{\mathcal L}$ is said to be
    \emph{isolated} if it is separated from its complementary subtree $\compsubtreeof{\mathcal L}$, and 
    \emph{non-isolated} if it is co-rooted with its complementary subtree $\compsubtreeof{\mathcal L}$. 
    If each canonical subtree is non-isolated and, in addition, the partition subtree $\subtreeof{\la\!\!\cup\!\lb}$ is non-isolated, then $\T(\ga,\gb)$ is said to be \emph{fully
    co-rooted}.
    At the other extreme, if each canonical subtree $\subtreeof{L}$ is isolated and, in addition, for a tree with four leaf types, either the subtree $\subtreeof{\la\!\!\cup\!\lb}$, or the subtree $\subtreeof{\la \!\!\cup\!\lc}$, or the subtree $\subtreeof{\la\!\!\cup\!\lab}$ is isolated, then $\T(\ga,\gb)$ is said
    to be \emph{fully separated}. Examples of various topologies are given in Figure~\ref{fig:topologies}.
    
\subsubsection{In-, out- and semi-traversals}      
    Let $u_1$ and $u_2$ be two leaves of~$\T(\ga,\gb)$. 
If both~$u_1$ and~$u_2$ are in the same canonical subtree~$\subtreeof{L}$, the traversal between~$u_1$ and~$u_2$ is said to be an \emph{in-traversal} (of the canonical
subtree~$\subtreeof{L}$). 
On the other hand, if~$u_1$ and~$u_2$ are in distinct canonical subtrees~$\subtreeof{L_1}$ and~$\subtreeof{L_2}$, the traversal between~$u_1$ and~$u_2$ is said to be an \emph{out-traversal} (between the
canonical subtrees~$\subtreeof{L_1}$ and~$\subtreeof{L_2}$). Note that in-traversals of tagged canonical subtrees are indel-saving (cost~1), while in-traversals of the clean canonical subtree are indel-neutral (cost~2). Out-traversals are usually indel-neutral, except in the case of one subtree being~$\subtreeof{\lab}$ and the other subtree being either~$\subtreeof{\la}$ or~$\subtreeof{\lb}$, when they are indel-saving. 

Now let $u_1$ be a leaf of a canonical subtree~$\subtreeof{L}$ and~$u_2$ be an internal node of~$\T(\ga,\gb)$ that is not in~$\subtreeof{L}$. The path connecting~$u_1$ and~$u_2$ is called a \emph{semi-traversal}, that can be indel-saving or indel-neutral depending on whether~$u_1$ and~$u_2$ share a tag or not.


\subsubsection{Tag mates}\label{sec:mates}

Let a tagged canonical subtree $\subtreeof{L}$ be separated from a partition subtree $\subtreeof{\mathcal L}$ by a bad link~$P$. 
As we explain in the following, the bad link~$P$ can be possibly covered by an
indel-saving out-traversal or by an indel-saving semi-traversal.

\begin{figure}[ht]
\setlength{\unitlength}{0.75pt}
\centering
\lfsize
\begin{tabular}{c@{~}c@{~}c@{~}c} 
{\bf (i) trivial mate} & {\bf (ii) tag mate} & {\bf (iii) tag mate} & {\bf (iv) no mate}\\[1em]
\begin{picture}(80,120)
  \put(25,110){\ParTreesNarrow
		        {\UnTree{\badvertex}
		          {\UnTree{\badvertex}{\badAvertex\bottomlabel{$a$}}}}
		        {\BiTreeNarrow{\goodvertex}
		          {\UnTree{\badvertex}{\badvertex\bottomlabel{$c_1$}}}
		          {\BiTreeRight{\badvertex}
		           {\badvertex\bottomlabel{$c_2$}}
		           {\badABvertex\bottomlabel{$x$}}
		          }
		        }
		      }
  \put(40,40){\makebox(0,0){$\optcost{3}$}}
  \put(40,20){\makebox(0,0){$a_{\;\!} \travone x$,}}
  \put(40, 5){\makebox(0,0){$c_{1\!} \travtwo c_2$}}
\end{picture} &
\begin{picture}(80,120)
  \put(35,110){\ParTreesNarrow
                {\UnTree{\badvertex\toplabel{$p$}}
	  	          {\UnTree{\badvertex}{\badAvertex\bottomlabel{$a$}}}}
                {\BiTreeNarrow{\goodABvertex\toplabel{$m$}}
		          {\UnTree{\badvertex}{\badvertex\bottomlabel{$c_1$}}}
		          {\UnTree{\badvertex}{\badvertex\bottomlabel{$c_2$}}}}} 
  \put(40,40){\makebox(0,0){$\optcost{3}$}}
  \put(40,20){\makebox(0,0){$a \semitravone m$,}}
  \put(40, 5){\makebox(0,0){$c_{1\!} \travtwo c_2$}}
\end{picture} &
\begin{picture}(80,120)
  \put(35,110){\ParTreesNarrow
                {\UnTree{\badABvertex\toplabel{$p\!=\!m$}}
		          {\UnTree{\badvertex}{\badAvertex\bottomlabel{$a$}}}}
		        {\BiTreeNarrow{\goodvertex}
		          {\UnTree{\badvertex}{\badvertex\bottomlabel{$c_1$}}}
		          {\UnTree{\badvertex}{\badvertex\bottomlabel{$c_2$}}}}}
  \put(40,40){\makebox(0,0){$\optcost{3}$}}
  \put(40,20){\makebox(0,0){$a \semitravone m$,}}
  \put(40, 5){\makebox(0,0){$c_{1\!} \travtwo c_2$}}
\end{picture} &
\begin{picture}(80,120)
		\put(40,110){\ParTreesNarrow
		   {\UnTree{\badvertex\toplabel{$p$}}
		    {\UnTree{\badABvertex\leftlabel{$\tilde{m}$}
		    }{\badAvertex\bottomlabel{$a$}}}}
	       {\BiTreeNarrow{\goodvertex}
		       {\UnTree{\badvertex}{\badvertex\bottomlabel{$c_1$}}}
		       {\UnTree{\badvertex}{\badvertex\bottomlabel{$c_2$}}}
		      }
		    }
		\put(40,40){\makebox(0,0){$\optcost{4}$}}
		\put(40,20){\makebox(0,0){$a_{1\!} \travtwo c_1$,}}
		\put(40, 5){\makebox(0,0){$c_{1\!} \travtwo c_2$}}
\end{picture} 
\end{tabular} 
\caption{\label{fig:mate}Tag mates are illustrated in four examples.
In~(i)~the $\absymb$-leaf $x$ is a trivial $\asymb$-mate of $\subtreeof{\la}$ at $\subtreeof{\lc \cup \lab}$. 
In examples (ii) and (iii), whose optimal covers have cost~3, there is no trivial tag mate, but the internal $\absymb$-node $m$ is an~$\asymb$-mate of $\subtreeof{\la}$ at~$\subtreeof{\lc\!\cup\!\{p\}}$.
In example (iv) the internal $\absymb$-node~$\tilde{m}$ 
is not 
at $\subtreeof{\lc\cup\{p\}}$ and an optimal cover has cost~4.}
\end{figure}
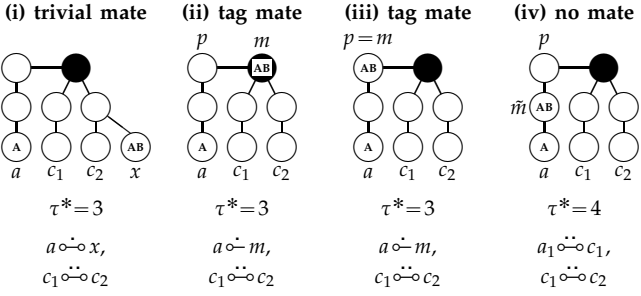

First suppose that at least one leaf of the subtree $\subtreeof{\mathcal L}$ shares a tag with the leaves of $\subtreeof{L}$.
In this case, the bad link $P$ can be
covered by an indel-saving out-traversal $P'$ (with $\tau(P')=1$) and we say that subtree $\subtreeof{\mathcal L}$ contains a \emph{trivial tag mate} of $\subtreeof{L}$.
An example of trivial $\asymb$-mate is given in Figure~\ref{fig:mate}~(i).
	
Now suppose that subtree $\subtreeof{\mathcal L}$ has no trivial tag mate of $\subtreeof{L}$.
In this case $\subtreeof{\mathcal L}$ might have some internal node that shares a tag with the leaves of $\subtreeof{L}$, 
therefore it can still be possible to cover the bad link $P$ by an indel-saving semi-traversal.
More precisely, let the \emph{extended subtree} $\extsubtreeof{\mathcal L}$ be $\subtreeof{{\mathcal L} \cup \{p\}}$, where $p$ is the bad node of the bad link $P$ that is closest to $\subtreeof{\mathcal L}$.
If the tag set of at least one node~$m$ of~$\extsubtreeof{\mathcal L}$ shares a tag with the leaves of~$\subtreeof{L}$, the node~$m$ is said to be a \emph{non-trivial tag mate} of~$\subtreeof{L}$ at~$\extsubtreeof{\mathcal L}$.
The fact that $\extsubtreeof{\mathcal L}$ contains a non-trivial tag mate of~$\subtreeof{L}$ is denoted by~$\mate{\mathcal L}{\anysymb}{\mathcal L}$, where~$\anysymb$ is either the tag~$\asymb$ or the tag~$\bsymb$. Examples of non-trivial tag mates are
$\mate{\lb}{\bsymb}{\la\!\cup\lc}$ and
$\mate{\lb\cup\lab}{\asymb}{\lc}$, the latter shown in Figure~\ref{fig:mate}~(ii)-(iii).

Observe that, if $\mathcal L$ includes the class~$\lab$, it is unnecessary to verify whether the extended subtree~$\extsubtreeof{\mathcal L}$ contains a non-trivial tag mate, because in this case it obviously contains a trivial one.
	
\subsection{Procedure of computing an optimal cover}

For any canonical subtree $\subtreeof{L}$, the cost of an in-traversal of $\subtreeof{L}$ is at most the same as the cost of an out-traversal between $\subtreeof{L}$ and any other canonical subtree. This suggests an approach of searching optimal covers maximizing in-traversals. 
We need to be careful, however, and also allow out-traversals for covering bad links and possibly unpaired leaves. 


Given a tree $\T(\ga,\gb)$ with at least two types of leaves that share no tag, 
we will search for an optimal cover with a three-step procedure.
The first step consists of determining whether the \emph{solo leaf} exists: it is a leaf of the clean subtree that must be kept by any reduction.
The second step consists of safely reducing the sizes of the four subtrees $\subtreeof{\la}$,
$\subtreeof{\lb}$, $\subtreeof{\lc}$ and $\subtreeof{\lab}$ of $\T(\ga,\gb)$ with 
in-traversals,
until we reach a certain minimum number of leaves per canonical subtree, obtaining a \emph{residual tree}. The third step is then finding an optimal cover of the residual tree,
that requires a lengthy enumeration of all possible cases.

\subsubsection{Solo leaf}\label{sec:solo}

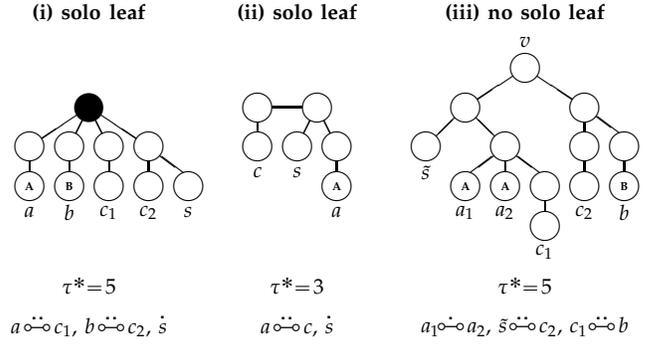
\begin{figure}[ht]
\setlength{\unitlength}{0.75pt}
\centering
\lfsize
\begin{tabular}{c@{~~~~}c@{~~~~}c}
	{\bf (i) solo leaf}& {\bf (ii) solo leaf}& {\bf (iii) no solo leaf}\\[1em]
\begin{picture}(90,145)
		\put(45,115){\QuaTree{\goodvertex}
		    {\UnTree{\badvertex}{\badAvertex\bottomlabel{$a$}}}
		    {\UnTree{\badvertex}{\badBvertex\bottomlabel{$b$}}}
		    {\UnTree{\badvertex}{\badvertex\bottomlabel{$c_1$}}}
		    {\BiTreeRight{\badvertex}{\badvertex\bottomlabel{$c_2$}}{\badvertex\bottomlabel{$s$}}}
		    }
		\put(45,25){\makebox(0,0){$\optcost{5}$}}
		\put(45, 5){\makebox(0,0){$a \travtwo c_1$, $b \travtwo c_2$, $\costonepath{s}$}}
\end{picture} &
\begin{picture}(90,145)
		\put(40,115){\ParTreesNarrow
		       {\UnTree{\badvertex}{\badvertex\bottomlabel{$c$}}}
		       {\BiTreeNarrow{\badvertex}
		        {\badvertex\bottomlabel{$s$}}
		        {\UnTree{\badvertex}{\badAvertex\bottomlabel{$a$}}}
		       }
		    }
  \put(45,25){\makebox(0,0){$\optcost{3}$}}
  \put(45,5){\makebox(0,0){$a \travtwo c$, $\costonepath{s}$}}
\end{picture} &
 \begin{picture}(110,145) 
		\put(55,135){\BiTreeWide{\badvertex\toplabel{$v$}}
		      {\BiTree{\badvertex}
		        {\badvertex\bottomlabel{$\tilde{s}$}}
		        {\TriTree{\badvertex}
   		          {\badAvertex\bottomlabel{$a_1$}}
   		          {\badAvertex\bottomlabel{$a_2$}}
   		          {\UnTree{\badvertex}{\badvertex\bottomlabel{$c_1$}}}
   		        }
   		       }
   		       {\BiTreeRight{\badvertex}
		        {\UnTree{\badvertex}{\badvertex\bottomlabel{$c_2$}}}
                {\UnTree{\badvertex}{\badBvertex\bottomlabel{$b$}}}
		       }
		  } 
  \put(55,25){\makebox(0,0){$\optcost{5}$}}
  \put(55,5){\makebox(0,0){$a_{1\!} \travone a_2$, $\tilde{s}_{\!} \travtwo c_2$, $c_{1\!} \travtwo b$}}
\end{picture}
\end{tabular}
\\
\caption{\label{fig:solo}The concept of solo leaf is illustrated with three examples. In (i) the solo leaf $s$ is in a short leaf-branch of $\subtreeof{\lc}$. In (ii) the solo leaf $s$ is in a short leaf-branch of $\T(\ga,\gb)$ that is a long leaf-branch of $\subtreeof{\lc}$. In (iii), although $\tilde{s}$ is in a short leaf-branch of $\T(\ga,\gb)$ and even of~$\subtreeof{\lc}$, no cover including a short path with $\tilde{s}$ is optimal.}
\end{figure}

When reducing the clean subtree, we need to be careful with a particularity of the topology that can influence the cost of an optimal cover. 
Let the tree $\T(\ga,\gb)$ with at least three leaves have a clean leaf $s$ in a short leaf-branch and let $\Tred{-s}(\ga,\gb)$ be the tree obtained by pruning $s$ and performing a flower-contraction in $\T(\ga,\gb)$. Obviously, the tree $\Tred{-s}(\ga,\gb)$ has exactly one leaf less than $\T(\ga,\gb)$. If
$\tau(\T(\ga,\gb))=\tau(\Tred{-s}(\ga,\gb)) + 1$, then $s$ is said to be a \emph{solo leaf}. In other words, the clean leaf $s$ is a solo leaf if 
there is some optimal cover of $\T(\ga,\gb)$ that includes a short path covering $s$. See examples in Figure~\ref{fig:solo}~(i) and~(ii).

The solo leaf is the last leaf to be paired when we search for an optimal cover of $\T(\ga,\gb)$. 
In some sense, it represents the absence of a ``fortress''~\cite{HAN-PEV-1999}: if, instead, the last unpaired clean leaf would not be a solo leaf but part of a long leaf-branch, it could only be covered together with its branch by an indel-neutral path, whose cost is~2.

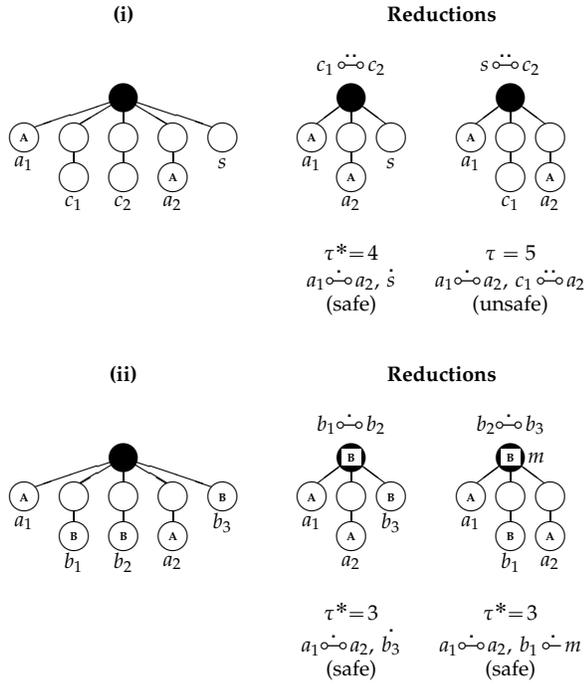
\begin{figure}[ht]
	\setlength{\unitlength}{0.75pt}
	\centering
	\lfsize
    \begin{tabular}{c@{~~~~}c@{~~}c}
    {\bf (i)} & \multicolumn{2}{c}{\bf Reductions} \\
    \hfill\\
    & {\bf $c_1 \travtwo c_2$} & {\bf $s \travtwo c_2$} \\
    	\begin{picture}(130,80)
		\put(65,70){\PenTree{\goodvertex}
		    {\badAvertex\bottomlabel{$a_1$}}
		    {\UnTree{\badvertex}{\badvertex\bottomlabel{$c_1$}}}
		    {\UnTree{\badvertex}{\badvertex\bottomlabel{$c_2$}}}
		    {\UnTree{\badvertex}{\badAvertex\bottomlabel{$a_2$}}}
		    {\badvertex\bottomlabel{$s$}}
		 }
		\end{picture} 
		&
		\begin{picture}(70,80)
		\put(35,70){\TriTree{\goodvertex}
		    {\badAvertex\bottomlabel{$a_1$}}
		    {\UnTree{\badvertex}{\badAvertex\bottomlabel{$a_2$}}}
		    {\badvertex\bottomlabel{$s$}}
		}
		\end{picture} 
		&
		\begin{picture}(70,80)
		\put(35,70){\TriTree{\goodvertex}
		    {\badAvertex\bottomlabel{$a_1$}}
		    {\UnTree{\badvertex}{\badvertex\bottomlabel{$c_1$}}}
		    {\UnTree{\badvertex}{\badAvertex\bottomlabel{$a_2$}}}
		}
		\end{picture}\\
		& $\optcost{4}$ & $\tau = 5$\\
		& $a_1 \travone a_2$, $\costonepath{s}$ &
		$a_1 \travone a_2$, $c_1 \travtwo a_2$\\
		& (safe) & (unsafe)\\
		\hfill\\
		\hfill\\
    {\bf (ii)} & \multicolumn{2}{c}{\bf Reductions} \\
    \hfill\\
    & {\bf $b_1 \travone b_2$} & {\bf $b_2 \travone b_3$} \\
    	\begin{picture}(130,80)
		\put(65,70){\PenTree{\goodvertex}
		    {\badAvertex\bottomlabel{$a_1$}}
		    {\UnTree{\badvertex}{\badBvertex\bottomlabel{$b_1$}}}
		    {\UnTree{\badvertex}{\badBvertex\bottomlabel{$b_2$}}}
		    {\UnTree{\badvertex}{\badAvertex\bottomlabel{$a_2$}}}
		    {\badBvertex\bottomlabel{$b_3$}}
		 }
		\end{picture} 
		&
		\begin{picture}(70,80)
		\put(35,70){\TriTree{\goodBvertex}
		    {\badAvertex\bottomlabel{$a_1$}}
		    {\UnTree{\badvertex}{\badAvertex\bottomlabel{$a_2$}}}
		    {\badBvertex\bottomlabel{$b_3$}}
		}
		\end{picture} 
		&
		\begin{picture}(70,80)
		\put(35,70){\TriTree{\goodBvertex\rightlabel{$m$}}
		    {\badAvertex\bottomlabel{$a_1$}}
		    {\UnTree{\badvertex}{\badBvertex\bottomlabel{$b_1$}}}
		    {\UnTree{\badvertex}{\badAvertex\bottomlabel{$a_2$}}}
		}
		\end{picture}\\
		& $\optcost{3}$ & $\optcost{3}$\\
		& $a_1 \travone a_2$, $\costonepath{b_3}$ &
		$a_1 \travone a_2$, $b_1 \semitravone m$\\
		& (safe) & (safe)\\
		\end{tabular}
		\caption{In (i), subtrees $\subtreeof{\la}$ and $\subtreeof{\lc}$ are co-rooted and the tree has a solo leaf. After the safe ($c_1,c_2)$-reduction, the reduced tree keeps the solo leaf and can be covered with cost 2. On the other hand, after the unsafe ($s,c_2$)-reduction, the reduced tree can be covered only with cost~3.
		In (ii), subtrees $\subtreeof{\la}$ and $\subtreeof{\lb}$ are co-rooted. After the safe ($b_1,b_2$)-reduction, the reduced tree can be covered with cost 2, including a short path covering leaf~$b_3$. If instead we do the ($b_2,b_3$)-reduction, subtrees~$\subtreeof{\la}$ and $\subtreeof{\lb}$ are separated in the reduced tree. However, the flower-contracted vertex $m$ is a tag mate of leaf $b_1$. Therefore, this reduction is also safe. In other words, there is no advantage in keeping the short leaf-branch ($b_3$).
		}
	\label{fig:reduction1}
	\end{figure}

In contrast to clean leaves in short leaf-branches of $\T(\ga,\gb)$, tagged leaves in short leaf-branches need no special treatment
for the following reasons:
\begin{itemize}
\item If originally a tagged canonical subtree is composed of a single leaf, it is indeed important to identify whether its leaf is in a long or in a short leaf-branch, since the latter allows it to be covered by a short path. But
the single leaf being in a short leaf-branch is equivalent to the corresponding subtree being non-isolated, a condition that can be easily verified.
\item If a tagged canonical subtree $\subtreeof{L}$ has at least two leaves,
there is no advantage of any leaf being in a short leaf-branch: in this case it could possibly be covered by a short path, but if it would be in a long leaf-branch, the whole branch could be covered by an indel-saving path, whose cost is the same as that of a short path.
\end{itemize}
The difference between keeping a clean leaf or a  tagged leaf in a short leaf-branch is illustrated in Figure~\ref{fig:reduction1}.



A tagged component tree $\T(\ga,\gb)$ can have more than one clean short leaf-branch, but in our approach at most one is covered by a short path: even if some optimal cover includes two short paths covering two clean short leaf-branches, there is another optimal cover with the same cost covering the two leaf-branches by an in-traversal of $\subtreeof{\lc}$. Therefore, if the tree has clean short leaf-branches, it suffices to characterize at most one as solo leaf.

However, even if $\T(\ga,\gb)$ has a single clean short leaf-branch, the leaf of this branch is not necessarily a solo leaf, as shown in Figure~\ref{fig:solo}~(iii). Actually the procedure of searching for a solo leaf is quite tricky and depends on various aspects of the distinct topologies of a tagged component tree. The only approach we could devise so far tests all possible hypotheses (each clean short leaf-branch of $\T(\ga,\gb)$ as the possible solo leaf) and takes one that results in the smallest cover cost. This exhaustive search dominates the complexity of our whole procedure for computing the distance, as we will demonstrate in Section~\ref{sec:complexity}.

\subsubsection{Reducing tagged component trees with balanced in-traversals}\label{sec:redu}

Given a canonical subtree $\subtreeof{L}$ with $|L| \geq 4$, a \emph{balanced in-traversal} of $\subtreeof{L}$ is defined as follows:
\begin{itemize}
\item If $|L|$ is even, any traversal from the set returned by Algorithm~\ref{alg:covering-traversals} is a balanced in-traversal.
\item If $|L|$ is odd, let $P$ be any leaf-branch of $\subtreeof{L}$. A balanced in-traversal of $\subtreeof{L}$ is a balanced in-traversal of~$\subtreeof{L}\backslash P$.
\end{itemize}

Let a path $P$ be an in-traversal of some canonical subtree $\subtreeof{L}$. The corresponding $P$-reduction costs~2 if~$L=\lc$ and~1 otherwise. It is called a \emph{balanced reduction} if $P$ is a balanced in-traversal.
And it is said to be \emph{preserving} if $\Tred{-P}(\ga,\gb)$ and $\T(\ga,\gb)$ have the same topology with respect to
partition subtrees, bad links and solo leaf,
and if in $\Tred{-P}(\ga,\gb)$ subtree $\subtreeof{L}$ has exactly two leaves less than in $\T(\ga,\gb)$. 

\def\statePreserving{
Let $\subtreeof{L}$ be a tagged canonical subtree of $\T(\ga,\gb)$ with $|L| \geq 4$. Any balanced $P$-reduction of $\subtreeof{L}$ is preserving.
}
\def\proofPreserving{
Since the in-traversal $P$ is balanced, it ``crosses'' the subtree $\subtreeof{L}$. This guarantees that the $P$-reduction does not create a new bad leaf, nor a new bad link, nor it increases the length of any remaining leaf-branch.  Furthermore, the flower-contraction after the reduction preserves all tags, therefore we do not lose any tag mate. 
}
\begin{lemma}\label{lem:preserving} 
\statePreserving
\end{lemma}
\begin{IEEEproof}
\proofPreserving
\end{IEEEproof}

\def\stateCleanPreserving{
For any tagged component tree $\T(\ga,\gb)$ with~$|\lc| \geq 4$, a preserving $P$-reduction of $\subtreeof{\lc}$ is any balanced $P$-reduction such that, if $\T(\ga,\gb)$ has a solo leaf~$s$, $P$ does not cover $s$. Such a reduction is called \emph{solo-safe reduction}. If $\T(\ga,\gb)$ has no solo leaf, any balanced reduction of $\subtreeof{\lc}$ is a solo-safe reduction.
}
\def\proofCleanPreserving{
Here we can repeat the arguments of the proof of Lemma~\ref{lem:preserving}. Furthermore, since $P$ does not cover the solo leaf, we do not lose it.
}

\begin{lemma}\label{lem:clean-preserving} 
\stateCleanPreserving
\end{lemma}
\begin{IEEEproof}
\proofCleanPreserving
\end{IEEEproof}


A preserving balanced reduction is very often safe. Indeed, a solo-safe reduction of $\subtreeof{\lc}$ down to two leaves is always safe. The same is true for any
balanced reduction of tagged canonical subtrees $\subtreeof{\la}$ and $\subtreeof{\lb}$ down to two leaves and for any
balanced reduction of tagged canonical subtree $\subtreeof{\lab}$ down to three leaves. Actually, only in some particular cases a balanced reduction of $\subtreeof{\lab}$ from four to two leaves is unsafe. These facts will be demonstrated by the Reduction Bounds stated later in this section. We will also demonstrate that, although a canonical subtree  $\subtreeof{L}$ with only three leaves has no balanced in-traversal, a reduction 
to a single leaf can be safe, even if it is non-preserving (as we showed in Figure~\ref{fig:reduction1}).

We can now analyse the topologies of each type of tagged component tree and set the lower bounds for safe reductions in each canonical subtree. Since a reduction removes two leaves from the tree, for each case analysis we need to consider whether the number of leaves in the canonical subtree being reduced is even or odd. We start by setting the bounds for canonical subtrees $\subtreeof{\la}$ and $\subtreeof{\lb}$:

\def\stateCanAorCanB{
Each of the canonical subtrees $\subtreeof{\la}$ and $\subtreeof{\lb}$ can be safely reduced to one or two leaves in any tagged component tree.
}
\def\proofCanAorCanB{
Without loss of generality, we will do the case analsis considering only $\subtreeof{\la}$.
First suppose we have a tagged component tree with two non-empty canonical subtrees $\subtreeof{\la}$ and $\subtreeof{L}$. Then suppose $\subtreeof{\la}$ has three leaves and all of them are endpoints of out-traversals to canonical subtree $\subtreeof{L}$. In this case we could replace two out-traversals between $\subtreeof{\la}$ and $\subtreeof{L}$ by an in-traversal of $\subtreeof{\la}$ and an in-traversal of $\subtreeof{L}$. Note that $L$ can be either $\lb$ or $\lc$, but in both cases this replacement would lead to a smaller cost than the original traversals.
Now suppose we have a tagged component tree with three non-empty canonical subtrees $\subtreeof{\la}$, $\subtreeof{L_1}$ and $\subtreeof{L_2}$. Then suppose $\subtreeof{\la}$ has three leaves and two of them are endpoints of out-traversals to canonical subtree $\subtreeof{L_1}$, while the third leaf of $\subtreeof{\la}$ is endpoint of an out-traversal to canonical subtree $\subtreeof{L_2}$. In this case we could replace an out-traversal between $\subtreeof{\la}$ and $\subtreeof{L_1}$ and the out-traversal between $\subtreeof{\la}$ and $\subtreeof{L_2}$ by an in-traversal of $\subtreeof{\la}$ and an out-traversal between $\subtreeof{L_1}$ and $\subtreeof{L_2}$. By assigning the possible canonical subtrees to $L_1$ and $L_2$, one can verify that this replacement would lead to at most the same cost as the original traversals. (A similar analysis can be done for the case in which the tagged component tree has four non-empty canonical subtrees.) Clearly, if a reduction from 3 to 1 is possible, then a reduction from 4 to 2 is also possible.
}

\begin{proposition}\label{prop:red-a-and-b}
\stateCanAorCanB
\end{proposition}
\IEEEproof In Appendix 1.


Let $\ell_t$ be the number of leaves in canonical subtree~$\subtreeof{L_t}$. 
The \emph{leaf composition} of the tree $\T(\ga,\gb)$ is given by the values $\ra$, $\rb$, $\rc$  and $\rab$.

Due to the symmetry of costs for $\asymb$-nodes and for $\bsymb$-nodes in covering paths, we assume without loss of generality that after the reduction we obtain a residual tree in which $\ra\geq\rb$. To achieve this, if $\subtreeof{\lb}$ is non-empty and if $\subtreeof{\la}$ is either empty or has an odd number of leaves, we can simply ``swap'' $\asymb$-nodes and $\bsymb$-nodes. Now, for each type of tagged component tree, the bounds of 
each canonical subtree
will be defined in the five Reduction Bounds listed below and whose proofs are given in Appendix~1.


\vspace{2mm}


\def\stateTwoCan{{\bf Two non-empty canonical subtrees \boldmath$\subtreeof{L_1}$ and \boldmath$\subtreeof{L_2}$ such that \boldmath$L_1$ and \boldmath$L_2$ share no tag:} 
Let $\ell_1=|L_1|$ and $\ell_2=|L_2|$. Without loss of generality, assume that $\subtreeof{L_1}$ is tagged. Given that $w \in \{1, 2\}$ is the cost of an in-traversal of $\subtreeof{L_2}$ (that can be tagged or clean),
the number of leaves in both subtrees $\subtreeof{L_1}$ and $\subtreeof{L_2}$ can be safely reduced to one or two, with cost
$ \left\lfloor (\ell_1 - 1)/2 \right\rfloor + w\left\lfloor (\ell_2 - 1)/2 \right\rfloor$.}

\begin{reduction}\label{red:two-can}
\stateTwoCan
\end{reduction}

\vspace{3mm}




\def\stateThreeCanABC{{\bf Three non-empty canonical subtrees \boldmath$\subtreeof{\la}$, \boldmath$\subtreeof{\lb}$, and \boldmath$\subtreeof{\lc}$:}
\begin{enumerate}
\item If $\ra$ and $\rb$ are odd, and given that $n = \max \{ \rc, 2\}$,
the number of leaves in subtrees $\subtreeof{\la}$ and $\subtreeof{\lb}$ can be safely reduced to one, and the number of leaves in subtree $\subtreeof{\lc}$ can be safely reduced to two or three, with cost 
$(\ra - 1)/2 + (\rb - 1)/2 + 2\left\lfloor (n - 2)/2 \right\rfloor$. 
\item Otherwise $\ra$ is even, the number of leaves in subtree $\subtreeof{\la}$ can be safely reduced to two and the number of leaves in subtrees $\subtreeof{\lb}$ and $\subtreeof{\lc}$ can be safely reduced to one or two, with cost $(\ra - 2)/2 + \lfloor (\rb - 1)/2 \rfloor + 2\lfloor (\rc - 1)/2 \rfloor$.
\end{enumerate}}

\begin{reduction}\label{red:three-can-a-b-c}
\stateThreeCanABC
\end{reduction}

\vspace{3mm}

\def\stateThreeCanABAB{{\bf Three non-empty canonical subtrees \boldmath$\subtreeof{\la}$, \boldmath$\subtreeof{\lb}$, and \boldmath$\subtreeof{\lab}$:}
\begin{enumerate}
\item If both $\ra$ and $\rb$ are odd, the number of leaves in subtrees $\subtreeof{\la}$ and $\subtreeof{\lb}$ can be safely reduced to one and the number of leaves in subtree $\subtreeof{\lab}$ can be safely reduced to one or two, with cost 
$ (\ra - 1)/2 + (\rb - 1)/2 + \lfloor (\rab - 1)/2 \rfloor$. 
\item Otherwise, $\ra$ is even. If $\rb$ is odd and given that $n = \max \{\rab, 2\}$, the number of leaves in subtrees $\subtreeof{\la}$ can be safely reduced to two, the number of leaves in $\subtreeof{\lb}$ can be safely reduced to one and the number of leaves in subtree $\subtreeof{\lab}$ can be safely reduced to two or three, with cost 
$(\ra - 2)/2 + (\rb - 1)/2 + \left\lfloor (n - 2)/2 \right\rfloor$. 
\item Finally, both $\ra$ and $\rb$ are even and $n = \max \{\rab, 3\}$. The number of leaves in subtrees $\subtreeof{\la}$ and $\subtreeof{\lb}$ can be safely reduced to two, and the number of leaves in subtree $\subtreeof{\lab}$ can be safely reduced to three or four, with cost 
$(\ra - 2)/2 + (\rb - 2)/2 + \left\lfloor (n - 3)/2 \right\rfloor$.
\end{enumerate}}

\begin{reduction}\label{red:three-can-a-b-ab}
\stateThreeCanABAB
\end{reduction}

\vspace{3mm}

\def\stateThreeCanACAB{{\bf Three non-empty canonical subtrees \boldmath$\subtreeof{\la}$, \boldmath$\subtreeof{\lc}$, and \boldmath$\subtreeof{\lab}$:}
\begin{enumerate}
\item If at least one value among $\ra$ and $\rc$ is odd, the number of leaves in each subtree can be safely reduced to one or two, with cost 
$\lfloor (\ra - 1)/2 \rfloor + 2\lfloor (\rc - 1)/2 \rfloor + \lfloor (\rab - 1)/2 \rfloor$.
\item  Otherwise, both $\ra$ and $\rc$ are even and $n = \max \{\rab, 2\}$. Then the number of leaves in subtrees $\subtreeof{\la}$ and $\subtreeof{\lc}$ can be safely reduced to two, and the number of leaves in subtree $\subtreeof{\lab}$ can be safely reduced to two or three, with cost 
$(\ra - 2)/2 + (\rc - 2) + \left\lfloor (n - 2)/2 \right\rfloor$.
\end{enumerate}}

\begin{reduction}\label{red:three-can-a-c-ab}
\stateThreeCanACAB
\end{reduction}

\vspace{3mm}

\def\stateFourCan{{\bf Four non-empty canonical subtrees:}
\begin{enumerate}
\item If all $\ra$, $\rb$ and $\rc$ are odd, the number of leaves in subtrees $\subtreeof{\la}$, $\subtreeof{\lb}$ and $\subtreeof{\lc}$ can be safely reduced to one, and the number of leaves in subtree $\subtreeof{\lab}$ can be safely reduced to one or two, with cost 
$(\ra - 1)/2 + (\rb - 1)/2 + (\rc - 1) + \lfloor (\rab - 1)/2 \rfloor$.
\item Otherwise, if $\rb$ is odd and among $\ra$ and $\rc$ one value is odd and the other is even, we set $n = \max \{ \rab, 2\}$. The number of leaves in subtree $\subtreeof{\lb}$ can be safely reduced to one, the number of leaves in subtrees $\subtreeof{\la}$ and $\subtreeof{\lc}$ can be safely reduced to one or two, and the number of leaves in subtree 
$\subtreeof{\lab}$ can be safely reduced to two or three, with cost 
$\lfloor (\ra - 1)/2 \rfloor + (\rb - 1)/2 + 2\lfloor (\rc - 1)/2 \rfloor + \left\lfloor (n - 2)/2 \right\rfloor$.
\item Finally, if $\ra$ is even and at least one value among $\rb$ and $\rc$ is also even, we set $n = \max \{ \rab, 3\}$.
The number of leaves in subtree $\subtreeof{\la}$ can be safely reduced to two, the number of leaves in subtrees $\subtreeof{\lb}$ and $\subtreeof{\lc}$ can be safely reduced to one or two, and the number of leaves in subtree 
$\subtreeof{\lab}$ can be safely reduced to three or four, with cost 
$(\ra - 2)/2 + \lfloor (\rb - 1)/2 \rfloor + 2\lfloor (\rc - 1)/2 \rfloor + \left\lfloor (n - 3)/2 \right\rfloor$.
\end{enumerate}}

\begin{reduction}\label{red:four-can}
\stateFourCan
\end{reduction}

\vspace{2mm}

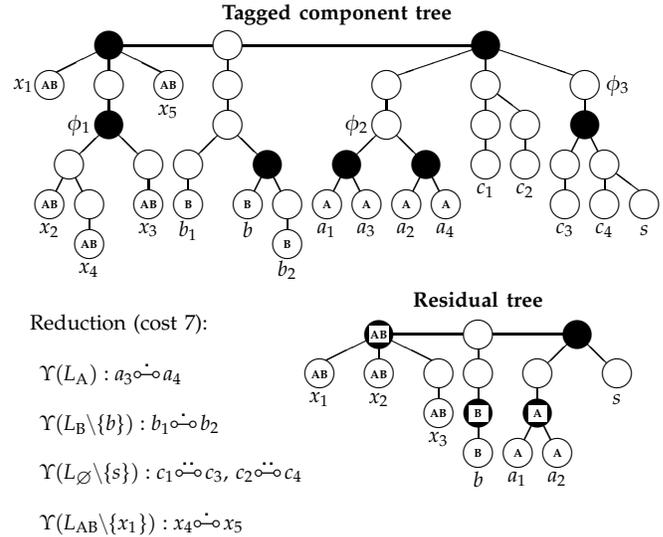
\begin{figure}[ht]
	\setlength{\unitlength}{0.75pt}
	\centering
	\lfsize
    \begin{tabular}{c}
    {\bf Tagged component tree}\\
    	\begin{picture}(320,130)
		\put(145,120){\ThreeParTreesCustom
		    {\TriTreeWide{\goodvertex}
		      {\badABvertex\leftlabel{$x_1$}}
		      {\UnTree{\badvertex}
		        {\BiTree{\goodvertex\leftlabel{$\bvertex_1$~}}
		          {\BiTreeNarrow{\badvertex}
		            {\badABvertex\bottomlabel{$x_2$}}
		            {\UnTree{\badvertex}{\badABvertex\bottomlabel{$x_4$}}}
		          }
		          {\UnTree{\badvertex}{\badABvertex\bottomlabel{$x_3$}}}
		        } 
		      }     
		      {\badABvertex\bottomlabel{$x_5$}}
		    }
		    {\UnTree{\badvertex}
		      {\UnTree{\badvertex}
		       {\BiTree{\badvertex}
		        {\UnTree{\badvertex}{\badBvertex\bottomlabel{$b_1$}}}
		         {\BiTreeNarrow{\goodvertex}
		            {\badBvertex\bottomlabel{$b$}}
		            {\UnTree{\badvertex}{\badBvertex\bottomlabel{$b_2$}}}
		         }
		       }
		      }
		    }
		    {\TriTreeExtraWide{\goodvertex}
		       {\UnTree{\badvertex}
		         {\BiTree{\badvertex\leftlabel{~$\bvertex_2$~~}}
		           {\BiTreeNarrow{\goodvertex}
		             {\badAvertex\bottomlabel{$a_1$}}
		             {\badAvertex\bottomlabel{$a_3$}}
		           }
		           {\BiTreeNarrow{\goodvertex}
		             {\badAvertex\bottomlabel{$a_2$}}
		             {\badAvertex\bottomlabel{$a_4$}}
		           }
		         }     
		       }
		       {\BiTreeRight{\badvertex}
		           {\UnTree{\badvertex}{\badvertex\bottomlabel{$c_1$}}}
		           {\UnTree{\badvertex}{\badvertex\bottomlabel{$c_2$}}}
		       }
		       {\UnTree{\badvertex\rightlabel{~~$\bvertex_3$}}
		           {\BiTreeNarrow{\goodvertex}
		             {\UnTree{\badvertex}{\badvertex\bottomlabel{$c_3$}}}
		             {\BiTreeRight{\badvertex}
		               {\badvertex\bottomlabel{$c_4$}}
		               {\badvertex\bottomlabel{$s$}}
		             }
		           }
		         }
		       }
		   }
		\end{picture} 
		\end{tabular} 
		\begin{tabular}{c@{~~~}c}
          & {\bf Residual tree} \\
		\begin{picture}(120,130)
		\put(50,125){\makebox(0,0){\lfsize Reduction (cost $7$):}}
		\put(10,100){\makebox(0,0)[l]{\lfsize$\subtreeof{\la}: a_3 \travone a_4$}}
		\put(10,75){\makebox(0,0)[l]{\lfsize$\subtreeof{\lb\backslash\{b\}}: b_1 \travone b_2$}}
		
		\put(10,50){\makebox(0,0)[l]{\lfsize$\subtreeof{\lc\backslash\{s\}}: c_{1\!} \travtwo c_3$, $c_{2\!} \travtwo c_4$}}
		\put(10,25){\makebox(0,0)[l]{\lfsize$\subtreeof{\lab\backslash\{x_1\}}: x_4 \travone x_5$}}
		\end{picture} &
		\begin{picture}(200,130)
		\put(100,120){\ThreeParTrees
		    {\TriTreeWide{\goodABvertex}
		      {\badABvertex\bottomlabel{$x_1$}}
		      {\badABvertex\bottomlabel{$x_2$}}
		      {\UnTree{\badvertex}{\badABvertex\bottomlabel{$x_3$}}}
		    }
		    {\UnTree{\badvertex}
		      {\UnTree{\badvertex}
		        {\UnTree{\goodBvertex}
		          {\badBvertex\bottomlabel{$b$}}
		        }
		      }
		    }
		    {\BiTree{\goodvertex}
		      {\UnTree{\badvertex}
		        {\BiTreeNarrow{\goodAvertex}
		          {\badAvertex\bottomlabel{$a_1$}}
		          {\badAvertex\bottomlabel{$a_2$}}
		        }
		      }
		      {\badvertex\bottomlabel{$s$}}
		     }
		   }
		\end{picture}
		\end{tabular}
		\caption{Execution of Reduction Bound~\ref{red:four-can} (case 2) for a tree with $\ra=4$, $\rb=3$, $\rc=5$ and $\rab=5$. The leaf $s$ is the solo leaf and vertices $\bvertex_1$, $\bvertex_2$ and $\bvertex_3$ are, respectively, balanced vertices in subtrees $\subtreeof{\lab\backslash\{x_1\}}$, $\subtreeof{\la}$ and $\subtreeof{\lc\backslash\{s\}}$. 
		The obtained residual tree has $\ra=2$, $\rb=1$, $\rc=1$ and $\rab=3$. Observe that, before and after the reduction, subtrees $\subtreeof{\la}$ and $\subtreeof{\lb}$ are both isolated, while $\subtreeof{\lc}$ and $\subtreeof{\lab}$ are both non-isolated.
		}
	\label{fig:reduction3}
	\end{figure}


Reduction Bounds~\ref{red:two-can} to~\ref{red:four-can} jointly define an algorithm to find the residual tree of any tagged component tree (Algorithm~\ref{alg:residual} in Appendix~1). Once we have the resulting residual tagged component tree (see an example of an execution of Reduction Bound~\ref{red:four-can} - case 2 - in   Figure~\ref{fig:reduction3}), its optimal cover can be obtained from the lengthy enumeration of all possible residual trees, given in the remainder of this section.   


\subsubsection{Covering all residual trees}

A \emph{residual leaf composition} can be represented by a quadruple~$\leco \ra\rb\rc\rab$ of four non-negative values that always has $\ra$ in the first position, $\rb$ in the second position, $\rc$ in the third position and $\rab$ in the fourth position.
Recall that, due to the symmetry of costs for $\asymb$-nodes and for $\bsymb$-nodes in
covering paths, we assume without loss of generality that $\ra\geq\rb$.
According to the results presented so far, we have 56 residual leaf compositions distributed in three groups.

\vspace{2mm}

\begin{figure}[ht]
	\setlength{\unitlength}{0.7pt}
	\lfsize
	\centering
	\mytab{c@{~~~~~~~~~~~}c}
	   \begin{tabular}{c@{~~~~~~~~~~~}c} 
    {\bf (i) co-rooted} & {\bf (ii) short link}\\  
	\begin{picture}(90,125)
		\put(45,115){\QuaTree{\goodvertex}
		       {\UnTree{\badvertex}{\badAvertex\bottomlabel{$a_1$}}}
		       {\UnTree{\badvertex}{\badAvertex\bottomlabel{$a_2$}}}
		       {\UnTree{\badvertex}{\badBvertex\bottomlabel{$b_1$}}}
		       {\UnTree{\badvertex}{\badBvertex\bottomlabel{$b_2$}}}
		    }
		\put(45,25){\makebox(0,0){$\optcost{2}$}}
		\put(45, 5){\makebox(0,0){$a_1 \travone a_2$, $b_1 \travone b_2$}}
		\end{picture} &
	\begin{picture}(90,125)
		\put(45,115){\BiTree{\badvertex\bottomlabel{$u$}}
		      {\BiTreeNarrow{\goodvertex}
		       {\UnTree{\badvertex}{\badAvertex\bottomlabel{$a_1$}}}
		       {\UnTree{\badvertex}{\badAvertex\bottomlabel{$a_2$}}}
		      }
		      {\BiTreeNarrow{\goodvertex}
		       {\UnTree{\badvertex}{\badBvertex\bottomlabel{$b_1$}}}
		       {\UnTree{\badvertex}{\badBvertex\bottomlabel{$b_2$}}}
		      }
		    }
		\put(45,25){\makebox(0,0){$\optcost{3}$}}
		\put(45, 5){\makebox(0,0){$a_1 \travone a_2$, $b_1 \travone b_2$, $\costonepath{u}$}}
		\end{picture}
	\end{tabular} \taband \mylnbksp
	\begin{tabular}{c@{~~~~~~~~~~~}c}
	{\bf (iii) tag mate} & {\bf (iv) long link}\\
	\begin{picture}(90,125)
		\put(45,115){\ParTrees
		    {\UnTree{\badvertex}
		      {\BiTreeNarrow{\goodvertex}
		       {\UnTree{\badvertex}{\badAvertex\bottomlabel{$a_1$}}}
		       {\UnTree{\badvertex}{\badAvertex\bottomlabel{$a_2$}}}
		      }
		     }
		    {\UnTree{\badAvertex\rightlabel{$m$}}
		      {\BiTreeNarrow{\goodvertex}
		       {\UnTree{\badvertex}{\badBvertex\bottomlabel{$b_1$}}}
		       {\UnTree{\badvertex}{\badBvertex\bottomlabel{$b_2$}}}
		      }
		     }
		    }
		\put(45,25){\makebox(0,0){$\optcost{3}$}}
		\put(45, 5){\makebox(0,0){$a_1 \travone a_2$, $a_{1\!} \semitravone m$, $b_1 \travone b_2$}}
		\end{picture} & 
\begin{picture}(90,125)
		\put(45,115){\ParTrees
		    {\UnTree{\badvertex}
		      {\BiTreeNarrow{\goodvertex}
		       {\UnTree{\badvertex}{\badAvertex\bottomlabel{$a_1$}}}
		       {\UnTree{\badvertex}{\badAvertex\bottomlabel{$a_2$}}}
		      }
		     }
		    {\UnTree{\badvertex}
		      {\BiTreeNarrow{\goodvertex}
		       {\UnTree{\badvertex}{\badBvertex\bottomlabel{$b_1$}}}
		       {\UnTree{\badvertex}{\badBvertex\bottomlabel{$b_2$}}}
		      }
		     }
		    }
		\put(45,25){\makebox(0,0){$\optcost{4}$}}
		\put(45, 5){\makebox(0,0){$a_{1\!} \travtwo b_1$, $a_{2\!} \travtwo b_2$}}
		\end{picture}\\[1em]
\end{tabular}
\endtab
	\caption{Possible topologies of leaf composition $\leco 2200$.
	A symmetric case of (iii) 
	is omitted.}
	\label{fig:leco2200}
	\end{figure}
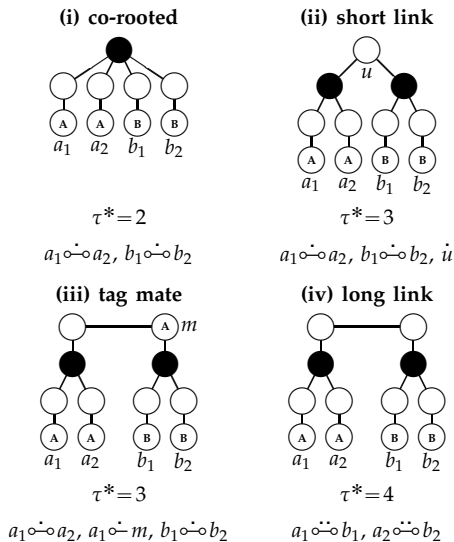

{\bf Group 1:} Residual trees with exactly two non-empty canonical subtrees $\subtreeof{L_1}$ and $\subtreeof{L_2}$, such that the tag sets of the classes $L_1$ and $L_2$ share no tag, defined in Reduction Bound~\ref{red:two-can}. 
This group includes 11 residual leaf compositions listed in Table~\ref{tab:residual1}. The complete enumeration is given in Appendix~2.
As an example of Group 1, we give the topologies and respective optimal covers of leaf composition $\leco 2200$ in Figure~\ref{fig:leco2200}.

\begin{table}[ht]
\caption{All 11 residual leaf compositions of Group 1}
\label{tab:residual1}
\centering
\begin{tabular}{ccc}
\toprule
\multicolumn{3}{c}{Reduction Bound \ref{red:two-can}}\\
\midrule
$\leco\ra\rb00$ & $\leco\ra0\rc0$ &
$\leco00\rc\rab$\\
 \midrule
 $\leco 1100$ & $\leco 1010$ & $\leco 0011$\\
 & $\leco 1020$ & $\leco 0012$\\
 $\leco 2100$ & $\leco 2010$ & $\leco 0021$\\
 $\leco 2200$ & $\leco 2020$ & $\leco 0022$\\
 \bottomrule
\end{tabular}
\end{table}

\vspace{2mm}



\begin{figure}[ht]
	\setlength{\unitlength}{0.7pt}
	\centering
	\lfsize
		
    \begin{tabular}{c}
    {\bf Non-reducible topology of leaf composition \boldmath$\leco 1130$:}\\
    {\bf both \boldmath$\subtreeof{\la}$ and $\subtreeof{\lb}$ isolated, solo leaf, no tag mates}\\
	\begin{picture}(100,125)
		\put(50,115){\BiTreeWide{\badvertex}
		       {\TriTree{\goodvertex}
		         {\badvertex\bottomlabel{$s$}}
		         {\UnTree{\badvertex}{\badAvertex\bottomlabel{$a$}}}
		         {\UnTree{\badvertex}{\badBvertex\bottomlabel{$b$}}}
		       }
		       {\BiTreeNarrow{\goodvertex}
		        {\UnTree{\badvertex}{\badvertex\bottomlabel{$c_1$}}}
		        {\UnTree{\badvertex}{\badvertex\bottomlabel{$c_2$}}}
		       }
		    }
		\put(50,25){\makebox(0,0){$\optcost{5}$}}
		\put(50, 5){\makebox(0,0){$\costonepath{s}$, $a \travtwo c_1$, $b \travtwo c_2$}}
		\end{picture}
	   \end{tabular}\\
	   
	\vspace{3mm}   
	 
	{\bf Topologies of leaf composition \boldmath$\leco 1130$ reducible to \boldmath$\leco 1110$}\\
	\vspace{2mm}
    
		\begin{tabular}{c@{~~}c@{~~}c}
	    \multicolumn{3}{c}{\bf (i) \boldmath$\subtreeof{\la}$ non-isolated}\\
		\begin{picture}(150,105)
		\put(75,95){\PenTree{\goodvertex}
		         {\UnTree{\badvertex}{\badvertex\bottomlabel{$c$}}}
		         {\UnTree{\badvertex}{\badBvertex\bottomlabel{$b$}}}
		         {\Leaf{\badAvertex\bottomlabel{$a$}}}
		         {\UnTree{\badvertex}{\badvertex\bottomlabel{$c_1$}}}
		         {\UnTree{\badvertex}{\badvertex\bottomlabel{$c_2$}}}
		    }
		\put(75,25){\makebox(0,0){$\optcost{5}$}}
		\put(75, 5){\makebox(0,0){$c_{1\!} \travtwo c_2$}}
		\put(135, 5){\makebox(0,0){+ residual:}}
		\end{picture} &
		\begin{picture}(20,105)
		\put(15,75){\makebox(0,0){$\rightarrow$}}
		\end{picture} &
		\begin{picture}(100,105)
		\put(50,95){\TriTree{\goodvertex}
		         {\UnTree{\badvertex}{\badvertex\bottomlabel{$c$}}}
		         {\UnTree{\badvertex}{\badBvertex\bottomlabel{$b$}}}
		         {\Leaf{\badAvertex\bottomlabel{$a$}}}
		    }
		\put(50,25){\makebox(0,0){$\optcost{3}$}}
		\put(50, 5){\makebox(0,0){$\costonepath{a}$, $b \travtwo c$}}
		\end{picture}
		\end{tabular}\\ 
		\mycondspace
		\begin{tabular}{c@{~~}c@{~~}c}
		\multicolumn{3}{c}{\bf (ii) tag mate}\\
		\begin{picture}(150,125)
		\put(75,115){\BiTreeWide{\badAvertex\bottomlabel{$m$}}
		       {\BiTreeNarrow{\goodvertex}
		         {\UnTree{\badvertex}{\badAvertex\bottomlabel{$a$}}}
		         {\UnTree{\badvertex}{\badBvertex\bottomlabel{$b$}}}
		       }
		       {\TriTree{\goodvertex}
		        {\badvertex\bottomlabel{$c$}}
		        {\UnTree{\badvertex}{\badvertex\bottomlabel{$c_1$}}}
		        {\UnTree{\badvertex}{\badvertex\bottomlabel{$c_2$}}}
		       }
		    }
		\put(75,25){\makebox(0,0){$\optcost{5}$}}
		\put(75, 5){\makebox(0,0){$c_{1\!} \travtwo c_2$}}
		\put(135, 5){\makebox(0,0){+ residual:}}
		\end{picture}& 
		\begin{picture}(20,125)
		\put(15,75){\makebox(0,0){$\rightarrow$}}
		\end{picture} &
		\begin{picture}(100,125)
		\put(50,115){\TriTree{\goodvertex}
		         {\UnTree{\badvertex}{\badAvertex\bottomlabel{$a$}}}
		         {\UnTree{\badvertex}{\badBvertex\bottomlabel{$b$}}}
		         {\UnTree{\badvertex\cornerlabel{$m$}\tagA}{\badvertex\bottomlabel{$c$}}}
		    }
		\put(50,25){\makebox(0,0){$\optcost{3}$}}
		\put(50, 5){\makebox(0,0){$b \travtwo c$, $a_{\!} \semitravone m$}}
		\end{picture}
		\end{tabular} \mylnbksp
		\begin{tabular}{c@{~~}c@{~~}c}
		\multicolumn{3}{c}{\bf (iii) both $\subtreeof{\la}$ and $\subtreeof{\lb}$ isolated, no solo leaf, no mates}\\
		\begin{picture}(150,125)
		\put(75,115){\BiTreeWide{\badvertex}
		       {\TriTree{\goodvertex}
 		         {\UnTree{\badvertex}{\badvertex\bottomlabel{$c_1$}}}
		         {\UnTree{\badvertex}{\badAvertex\bottomlabel{$a$}}}
		         {\UnTree{\badvertex}{\badBvertex\bottomlabel{$b$}}}
		       }
		       {\BiTreeNarrow{\goodvertex}
		        {\UnTree{\badvertex}{\badvertex\bottomlabel{$c$}}}
		        {\UnTree{\badvertex}{\badvertex\bottomlabel{$c_2$}}}
		       }
		    }
		\put(75,25){\makebox(0,0){$\optcost{6}$}}
		\put(75, 5){\makebox(0,0){$c_{1\!} \travtwo c_2$}}
		\put(135, 5){\makebox(0,0){+ residual:}}
		\end{picture} &
		\begin{picture}(20,125)
		\put(15,75){\makebox(0,0){$\rightarrow$}}
		\end{picture} &
       \begin{picture}(100,125)
		\put(50,95){\TriTree{\goodvertex}
		         {\UnTree{\badvertex}{\badAvertex\bottomlabel{$a$}}}
		         {\UnTree{\badvertex}{\badBvertex\bottomlabel{$b$}}}
		         {\UnTree{\badvertex}{\badvertex\bottomlabel{$c$}}}
		    }
		\put(50,25){\makebox(0,0){$\optcost{4}$}}
		\put(50, 5){\makebox(0,0){$a \travtwo c$, $b \travtwo c$}}
		\end{picture}
	\end{tabular}

	\caption{
	Leaf composition $\leco 1130$ has a single non-reducible topology (with 
	 both $\subtreeof{\la}$ and $\subtreeof{\lb}$ isolated, solo leaf and no tag mates)
 and
	three reducible topologies (symmetric cases of (i) and (ii) are omitted).
	}
	\label{fig:leco1130}
	\end{figure}
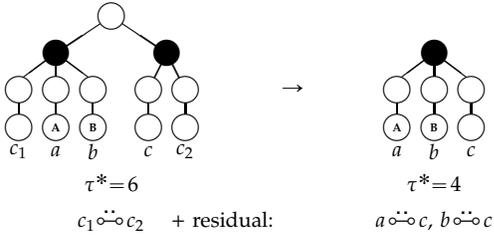

{\bf Group 2:} Residual trees with exactly three non-empty canonical subtrees, 
forming three subgroups, defined by Reduction Bounds~\ref{red:three-can-a-b-c} to~\ref{red:three-can-a-c-ab}.
This group includes 25 residual leaf compositions listed in Table~\ref{tab:residual2}. The complete enumeration is given in Appendix~2.
In particular, in this group we have the residual leaf composition $\leco 1130$ (Reduction Bound~\ref{red:three-can-a-b-c}, first case) and also residual leaf compositions with three or four leaves in subtree $\subtreeof{\lab}$ (cases of Reduction Bounds~\ref{red:three-can-a-b-ab} and~\ref{red:three-can-a-c-ab}).
Some topologies of these leaf compositions can be further reduced by an extra in-traversal and are then said to be \emph{reducible}.
As an example of Group 2, we give the topologies and respective optimal covers of leaf composition $\leco 1130$ and its 
reduction $\leco 1110$ in Figure~\ref{fig:leco1130}.


\begin{table}[ht]
\caption{All 25 residual leaf compositions of Group 2 
}
\label{tab:residual2}
\centering
\small
\mytab{c|c}
\begin{tabular}{c|cc}
\toprule
 \multicolumn{3}{c}{Reduction Bound~\ref{red:three-can-a-b-c}}\\
 \midrule
  \multicolumn{1}{c}{Case 1} & \multicolumn{2}{c}{Case 2}\\ 
 $\leco11\rc0$ & $\leco21\rc0$ & $\leco22\rc0$\\
 \midrule
$\leco 1110$ & $\leco 2110$ & $\leco 2210$\\
$\leco 1120$ & $\leco 2120$ & $\leco 2220$\\
$\leco 1130$ & &\\
\myemptycell
\bottomrule
\end{tabular} \taband \mylnbksp
\begin{tabular}{c|c|c}
\toprule
\multicolumn{3}{c}{Reduction Bound~\ref{red:three-can-a-b-ab}}\\
\midrule
\multicolumn{1}{c}{Case 1} & \multicolumn{1}{c}{Case 2} & \multicolumn{1}{c}{Case 3}\\
$\leco110\rab$ & $\leco210\rab$ & $\leco220\rab$\\
\midrule
$\leco 1101$ & $\leco 2101$ & $\leco 2201$\\
$\leco 1102$ & $\leco 2102$ & $\leco 2202$\\
& $\leco 2103$ & $\leco 2203$\\
& & $\leco 2204$\\
\bottomrule
\end{tabular} 
\endtab\\

\vspace{\myvspace}

\begin{tabular}{ccc|c}
\toprule
\multicolumn{4}{c}{Reduction Bound~\ref{red:three-can-a-c-ab}}\\
\midrule
\multicolumn{3}{c}{Case 1} & \multicolumn{1}{c}{Case 2}\\
 $\leco101\rab$ & $\leco102\rab$ & $\leco201\rab$ & $\leco202\rab$\\
 \midrule
  $\leco 1011$ & $\leco 1021$ & $\leco 2011$ & $\leco 2021$\\
  $\leco 1012$ & $\leco 1022$ & $\leco 2012$ & $\leco 2022$\\
  & & & $\leco 2023$ \\
\bottomrule
\end{tabular}
\end{table}

{\bf Group 3:} Residual trees with four non-empty canonical subtrees, defined by Reduction Bound \ref{red:four-can}, whose cases~2 and~3 include residual leaf compositions with three or four leaves in subtree $\subtreeof{\lab}$, containing reducible topologies.
In total, this group includes 20 residual leaf compositions listed in Table~\ref{tab:residual3}. The complete enumeration is given in Appendix~2.
As an example, we consider the leaf composition \leco 2223. The non-reducible topologies and respective optimal covers are shown in Figure~\ref{fig:leco2223nR}, and the ones reducible to \leco 2221 are shown in Figure~\ref{fig:leco2223R}, both in Appendix~2.

\begin{table}[ht]
\caption{All 20 residual leaf compositions of Group 3}
\label{tab:residual3}
\centering
\vspace{-2mm}
\small

\begin{tabular}{c}
\toprule
~~~~~~~~~~~~~Reduction Bound \ref{red:four-can}~~~~~~~~~~~~~
\end{tabular}\\

\vspace{\myvspace}

\mytab{c|c}
\begin{tabular}{c|cc}
\toprule
\multicolumn{1}{c}{Case 1} & \multicolumn{2}{c}{Case 2}\\ 
$\leco111\rab$ & $\leco112\rab$ & $\leco211\rab$\\
\midrule
\leco 1111 &\leco 1121 &\leco 2111\\
\leco 1112 &\leco 1122 &\leco 2112\\
 &\leco 1123 &\leco 2113\\
\myemptycell
\bottomrule
\end{tabular} \taband \mylnbksp
\begin{tabular}{ccc}
\condtoprule
\multicolumn{3}{c}{Case 3}\\
$\leco212\rab$ & $\leco221\rab$ & $\leco222\rab$ \\
\midrule
\leco 2121 &\leco 2211 &\leco 2221\\
\leco 2122 &\leco 2212 &\leco 2222\\
\leco 2123 &\leco 2213 &\leco 2223\\
\leco 2124 &\leco 2214 &\leco 2224\\
\bottomrule
\end{tabular}
\endtab
\end{table}

\subsubsection{Proof of correctness and complexity}\label{sec:complexity}

The correctness of our approach is a consequence of the correctness of Reduction Bounds~\ref{red:two-can} to~\ref{red:four-can} and the enumeration of all possible residual trees, as stated in the following theorem.


\def\stateCorrectness{
Given a tagged component tree $\T(\ga,\gb)$, let $\tau$ be the sum of the cost of applying the appropriate Reduction Bound (among \ref{red:two-can}-\ref{red:four-can}) to $\T(\ga,\gb)$ and the cost of covering the obtained residual tree $\Tred{R}(\ga,\gb)$, that can be found in the full enumeration (in Appendix~2). The value $\tau$ is optimal, that is, $\tau = \tinvid(\ga,\gb)$.
}
\def\proofCorrectness{
Reduction Bounds~\ref{red:two-can} to~\ref{red:four-can} (whose proofs are given in 
Appendix~1) give the number of safe balanced in-traversals that we can use to reduce each canonical subtree. For the residual trees, the exhaustive enumeration of all topologies guarantees that we can always find an optimal cover.

The following considerations show that the topology description that we consider contains all necessary properties to the search of optimal covers.

Since the cost of an in-traversal of  canonical subtree $\subtreeof{L}$ is at most the same as the cost of an out-traversal between $\subtreeof{L}$ and any other canonical subtree, in the absence of bad links and if each canonical subtree has an even number of leaves, it is clear that with in-traversals only we cover the tree optimally. 

Otherwise, bad links, tag mates and solo leaf are sufficient properties to drive our search for optimal covers:
\begin{itemize}
\item Note that, if long bad links exist, one can find an optimal cover by attempting to cover them by only three types of paths: (1) indel-saving out-traversals, (2) indel-saving semi-traversals (with tag mates) and (3) indel-neutral out-traversals. (Another possibility would be to have \emph{bridges}, that are paths connecting two internal nodes. Indeed, an optimal cover can contain a bridge, but in this case there is always an alternative optimal cover composed of traversals, semi-traversals and at most one short path - see Proposition~\ref{prop:no-bridge} 
in Appendix~1.)
Therefore, besides the leaves, tag mates are the only nodes we have to consider to cover bad links.
\item As explained in Section~\ref{sec:solo}, if there are unpaired leaves, only in the clean subtree we need to be careful and keep the solo leaf, that is in a short leaf-branch of $\T(\ga,\gb)$, to be the last leaf to be paired. For tagged leaves, it makes no difference whether the leaf is in a short or in a long leaf-branch, unless originally the corresponding subtree is composed of a single leaf (in this case, the leaf being in a short leaf-branch is equivalent to the corresponding subtree being non-isolated, a condition that is considered when we enumerate the covers of all possible residual trees).
\end{itemize}
}

\begin{theorem}\label{th:correctness}
\stateCorrectness
\end{theorem}
\begin{IEEEproof}
\proofCorrectness
\end{IEEEproof}


With respect to the complexity, the critical point is finding solo-safe reductions of the clean subtree, which results in a quadratic procedure.

\def\stateComplexity{
The time and space complexity of our procedure of computing the inversion-indel distance is quadratic in the number of markers $|\gg|+|\aa|+|\bb|$.
}
\begin{lemma}\label{lem:complexity}
\stateComplexity
\end{lemma}
\begin{IEEEproof}
In Appendix 1.
\end{IEEEproof}

\section{Conclusion}\label{sec:conclusion}

In this work we have completed the study of the inversion-indel distance 
between two chromosomes $\ga$ and $\gb$ with unequal contents, by
presenting an approach to compute the extra cost $\tinvid(\ga,\gb)$
of sorting bad components of the relational diagram $R(\ga,\gb)$.
We gave easy-to-compute lower and upper bounds. For the intermediate instances, our approach
is based on the composition of several properties of tagged component trees and the enumeration of optimal covers of a long list of minimal residual trees that can be obtained by reducing them.  
The whole procedure can be done in quadratic time and space on the number of markers $|\gg|+|\aa|+|\bb|$.

In future work, it seems now possible to address the more general problem of computing the multichromosomal \emph{genomic distance}~\cite{HAN-PEV-1995} with insertions and deletions using techniques similar to those used in this paper combined with the more general component tree presented in~\cite{BER-MIX-STO-2009}.

\bibliographystyle{IEEEtran}
\bibliography{invindel}

\clearpage
\onecolumn
\setcounter{page}{1}


\singleparamA
\def\mytab#1{\begin{tabular}{#1}}

\setstretch{1.25}



\section*{Appendix 1: Omitted Proofs and Algorithms}

~ 

\subsection*{Proofs}






The bounds given by Proposition~\ref{prop:red-a-and-b} and Reduction Bounds~\ref{red:two-can} to~\ref{red:four-can} were set assuming the worst case of fully separated trees.
The reason is that,
if a fully separated tree $T_1$ and non fully separated tree $T_2$ have the same number of leaves per canonical subtree, in $T_1$ we might need to keep more leaves for covering bad links. In other words, if, together with the reduction, some amount of remaining leaves is enough for giving an optimal cover of $T_1$, the same number is clearly enough for giving an optimal cover of $T_2$.

\vspace{2mm}

\noindent 
{\it Proposition \ref{prop:red-a-and-b}:}
\stateCanAorCanB
\begin{IEEEproof}
\proofCanAorCanB
\end{IEEEproof}

\vspace{2mm}

\noindent 
{\it Reduction Bound \ref{red:two-can}:}
\stateTwoCan
\begin{IEEEproof}
Two leaves from each canonical subtree are sufficient to cover the possible bad link between them and compensate an unpaired leaf from the other subtree. 
It is then clear that, if $\ell_1 \geq 4$ (respect. $\ell_2 \geq 4$), it can be reduced to $2$ or $3$ by preserving reductions (Lemma~\ref{lem:preserving}) that are safe. 
The question now is whether $\ell_1$ and/or $\ell_2$ can still be reduced from three to one.
If one subtree has only one leaf, it is clear that we can reduce the other from three leaves to one. If both subtrees have three leaves, it is clear that both can be reduced to one leaf. If one subtree has three leaves and the other has two, we have the following possibilities: 

(i) $L_1$ is either $\la$ or $\lab$ and $L_2= \lc$: in this case we have one out-traversal to cover the bad link. Suppose we have a cover that contains another out-traversal. This cover would have a cost of five. However, if instead we do an in-traversal in each canonical subtree, we also have a cost of five, therefore this reduction is safe.

(ii) $L_1 = \la$ and $L_2 = \lb$: the reduction in this case was proven to be safe by Proposition~\ref{prop:red-a-and-b}.
\end{IEEEproof}

\vspace{2mm}

The following proposition will be used to prove the correctness of Reduction Bounds~\ref{red:three-can-a-b-c} to~\ref{red:three-can-a-c-ab}.

\vspace{2mm}

\begin{proposition}\label{prop:cover-of-three}
The bad links of a fully separated tagged component tree $\T(\ga,\gb)$ with three non-empty canonical subtrees $\subtreeof{L_1}$, $\subtreeof{L_2}$ and $\subtreeof{L_3}$ can be covered by two paths, one connecting $\subtreeof{L_1}$ and $\subtreeof{L_2}$, and the other connecting $\subtreeof{L_1}$ and $\subtreeof{L_3}$.
\end{proposition}
\begin{IEEEproof}
 We simply refer to Figure~\ref{fig:topologies} (ii-b) of the paper. It is clear that if we pick any of the three subtrees, say subtree no. 1, a pair with a path between subtrees~1 and~2 and a path between subtrees~1 and~3 covers all bad links of the tagged component tree.
\end{IEEEproof}

\vspace{2mm}
{\it Reduction Bound \ref{red:three-can-a-b-c}:}
\stateThreeCanABC
\begin{IEEEproof}
Proposition~\ref{prop:red-a-and-b} guarantees that $\subtreeof{\la}$ and $\subtreeof{\lb}$ can be safely reduced to one or two leaves. Since all bad links can be covered by two out-traversals (Proposition \ref{prop:cover-of-three}), we need to reserve two leaves from $\subtreeof{\lc}$ to cover the bad links, and two other leaves from $\subtreeof{\lc}$ to possibly compensate unpaired leaves in the other two subtrees, meaning that the number of leaves in $\subtreeof{\lc}$ can be safely reduced to four or five. Further reductions depend on the following case analysis. 

Case 1: Note that, if $\ra$ is odd and $\rb$ is odd, they can be both reduced to one and we do not need to compensate unpaired leaves. Then the number of leaves in $\subtreeof{\lc}$ can be safely reduced to two or three. Let us examine whether we could still reduce from three to one: if both tagged subtrees are isolated and the tree has the solo leaf, such a reduction could either create a new leaf or transform the solo leaf into a leaf in a long leaf-branch, being then unsafe. Therefore in this case the bounds for the clean subtree are two or three.

Case 2: Now let $\ra$ be even. 
Suppose that we have five leaves in $\subtreeof{\lc}$, two leaves in $\subtreeof{\lb}$
and four out-traversals involving $\subtreeof{\lc}$. We could then replace, at the same cost, two of these out-traversals by an in-traversal of $\subtreeof{\lc}$ and an out-traversal between $\subtreeof{\la}$ and $\subtreeof{\lb}$. But then we would have a triangle covering the bad links, therefore we could replace one other out-traversal between $\subtreeof{\lc}$ and $\subtreeof{\la}$ by an in-traversal of $\subtreeof{\lc}$ (covering its unpaired leaf) and an in-traversal of $\subtreeof{\la}$, at the same cost. (A similar analysis could be done for the case of a single leaf in $\subtreeof{\lb}$.) Consequently, if the number of leaves in $\subtreeof{\lc}$ is odd, it can be safely reduced to one. Clearly, if the number of leaves in $\subtreeof{\lc}$ is even, it can be safely reduced to two.
\end{IEEEproof}

\vspace{2mm}
{\it Reduction Bound \ref{red:three-can-a-b-ab}:}
\stateThreeCanABAB
\begin{IEEEproof}
Proposition~\ref{prop:red-a-and-b} guarantees that $\subtreeof{\la}$ and $\subtreeof{\lb}$ can be safely reduced to one or two leaves. Since all bad links can be covered by two out-traversals (Proposition~\ref{prop:cover-of-three}), we need to reserve two leaves from $\subtreeof{\lab}$ to cover the bad links and two other leaves from $\subtreeof{\lab}$ to possibly compensate unpaired leaves in the other two subtrees, meaning that the number of leaves in $\subtreeof{\lab}$ can be safely reduced to four or five. Further reductions depend on the following case analysis. 

Case 1: Note that, if $\ra$ is odd and $\rb$ is odd, they can be both reduced to one and we do not need to compensate unpaired leaves. Then number of leaves in $\subtreeof{\lab}$ can be safely reduced to two or three. But actually we could still reduce from three to one: suppose the unpaired (third) leaf of subtree $\subtreeof{\lab}$ is covered by an out-traversal to one of the other two subtrees. At the same cost we could replace it by an in-traversal of $\subtreeof{\lab}$. Therefore in this case $\subtreeof{\lab}$ can be safely reduced to one or two.

Case 2: Now let $\ra$ be even, that is reduced to two, and $\rb$ be odd, that is reduced to one. The number of leaves in $\subtreeof{\lab}$ can be then safely reduced to three or four. But actually we could still reduce from four to two: suppose the unpaired (fourth) leaf of subtree $\subtreeof{\lab}$ is covered by an out-traversal to one of the other two subtrees. At the same cost we could replace it by an in-traversal of $\subtreeof{\lab}$. Therefore in this case $\subtreeof{\lab}$ can be safely reduced to two or three.

Case 3: Finally, if $\ra$ is even and $\rb$ is even, they can be both reduced to two. The number of leaves in $\subtreeof{\lab}$ can be then safely reduced to four or five. But actually we could still reduce from five to three: suppose the unpaired (fifth) leaf of subtree $\subtreeof{\lab}$ is covered by an out-traversal to one of the other two subtrees. At the same cost we could replace it by an in-traversal of $\subtreeof{\lab}$. Therefore in this case $\subtreeof{\lab}$ can be safely reduced to three or four.
\end{IEEEproof}

\vspace{2mm}

{\it Reduction Bound \ref{red:three-can-a-c-ab}:}
\stateThreeCanACAB
\begin{IEEEproof} Proposition~\ref{prop:red-a-and-b} guarantees that $\subtreeof{\la}$ can be safely reduced to one or two leaves. We also know that all bad links can be covered by two out-traversals (Proposition~\ref{prop:cover-of-three}) and we can always pair $\subtreeof{\lab}$ with $\subtreeof{\la}$  and $\subtreeof{\lc}$ with one of the other two subtrees. Note that $\subtreeof{\lab}$ is a more ``powerful'' version of $\subtreeof{\lb}$, therefore in the following analysis we can often generalize the statements of Reduction Bound~\ref{red:three-can-a-b-c}.

Case 1: Suppose $\ra$ is odd, being reduced to one. 
\begin{itemize}
\item If $\rc$ is odd and $\rab$ is odd, by generalizing Reduction Bound~\ref{red:three-can-a-b-c} -- Case 1, we know that $\subtreeof{\lab}$ can be safely reduced to one and $\subtreeof{\lc}$ can be safely reduced to three. But, actually, here $\subtreeof{\lc}$ can also be reduced from three to one: one leaf is enough to cover the eventual bad link separating $\subtreeof{\lc}$ from $\compsubtreeof{\lc}$.
\item If $\rc$ is odd and $\rab$ is even, by generalizing Reduction Bound~\ref{red:three-can-a-b-c} -- Case 2, we know that $\subtreeof{\lab}$ can be safely reduced to two and $\subtreeof{\lc}$ can be safely reduced to one.
\item If $\rc$ is even and $\rab$ is odd, by generalizing Reduction Bound~\ref{red:three-can-a-b-c} -- Case 2, we know that $\subtreeof{\lab}$ can be safely reduced to one and $\subtreeof{\lc}$ can be safely reduced to two.
\end{itemize}

Case 2: Suppose $\ra$ is even, being reduced to two.
\begin{itemize}
\item If $\rc$ is odd, by generalizing Reduction Bound~\ref{red:three-can-a-b-c} -- Case 2, we know that $\subtreeof{\lc}$ can be safely reduced to one and the number of leaves in $\subtreeof{\lab}$ can be safely reduced to one or two.
\item If $\rc$ is even, by generalizing Reduction Bound~\ref{red:three-can-a-b-c} -- Case 2, we know that $\subtreeof{\lc}$ can be safely reduced to two.
Then the number of leaves in $\subtreeof{\lab}$ can be safely reduced to three or four. But actually we can still reduce it from four to two: suppose we have two out-traversals between subtrees $\subtreeof{\lab}$ and $\subtreeof{\la}$ and two out-traversals between subtrees $\subtreeof{\lab}$ and $\subtreeof{\lc}$. We could then replace one out-traversal between $\subtreeof{\lab}$ and $\subtreeof{\la}$ and one out-traversal between $\subtreeof{\lab}$ and $\subtreeof{\lc}$ by one out-traversal between $\subtreeof{\la}$ and $\subtreeof{\lc}$ and one in-traversal of $\subtreeof{\lab}$. Therefore in this case $\subtreeof{\lab}$ can be safely reduced to two.
 \end{itemize}
\end{IEEEproof}

\vspace{2mm}

The following proposition will be used to prove the correctness of Reduction Bound~\ref{red:four-can}.

\vspace{2mm}

\begin{proposition}\label{prop:cover-of-four}
The bad links of a fully separated tagged component tree $\T(\ga,\gb)$ with four non-empty canonical subtrees $\subtreeof{L_1}$, $\subtreeof{L_2}$, $\subtreeof{L_3}$ and $\subtreeof{L_4}$, such that $\subtreeof{L_1 \cup L_2}$ is separated from $\subtreeof{L_3 \cup L_4}$ can be covered by two ``crossing'' paths, that either connect $\subtreeof{L_1}$ to $\subtreeof{L_3}$ and $\subtreeof{L_2}$ to $\subtreeof{L_4}$ or connect  $\subtreeof{L_1}$ to $\subtreeof{L_4}$ and $\subtreeof{L_2}$ to $\subtreeof{L_3}$.
\end{proposition}
\begin{IEEEproof}
 We simply refer to Figure~\ref{fig:topologies} (ii-c) of the paper. Note that partition subtrees $\subtreeof{1 \cup 3}$ and $\subtreeof{2 \cup 4}$ are co-rooted. It is then clear that a pair with a path between subtrees~1 and~3 and a path between subtrees~2 and~4 covers all bad links of the tagged component tree. A second choice also exists, because $\subtreeof{1 \cup 4}$ and $\subtreeof{2 \cup 3}$ are also co-rooted. Therefore a pair with a path between subtrees~1 and~4 and a path between subtrees~2 and~3 also covers all bad links of the tagged component tree.
\end{IEEEproof}

\vspace{2mm}

{\it Reduction Bound \ref{red:four-can}:}
\stateFourCan
\begin{IEEEproof} Proposition~\ref{prop:red-a-and-b} guarantees that $\subtreeof{\la}$ and $\subtreeof{\lb}$ can be safely reduced to one or two leaves. Since a tree with four canonical subtrees is a generalization of case 2 of Reduction Bound~\ref{red:three-can-a-b-c}, we also know that $\subtreeof{\lc}$ can be safely reduced to one or two leaves. 
Furthermore, all bad links can be covered by two out-traversals (Proposition~\ref{prop:cover-of-four}).
Therefore we can cover the eventual bad links by pairing $\subtreeof{\lc}$ with either $\subtreeof{\la}$ or $\subtreeof{\lb}$ and $\subtreeof{\lab}$ with either $\subtreeof{\la}$ or $\subtreeof{\lb}$.
We now need to examine the bounds for the subtree $\subtreeof{\lab}$.

Case 1: Note that, if $\ra$ is odd, $\rb$ is odd and $\rc$ is odd, they can all be reduced to one. Consequently, the number of leaves in $\subtreeof{\lab}$ can be safely reduced to one or two.

Case 2: Now let $\rb$ be odd, that is reduced to one, and let one of $\ra$ or $\rc$ be also odd, that is reduced to one, and the other one even, that is reduced to two. In order to cover the unpaired leaf in $\subtreeof{\la}$ or $\subtreeof{\lc}$, the number of leaves in $\subtreeof{\lab}$ can then be safely reduced to two or three.

Case 3: Finally, we have that $\ra$ is even and can be safely reduced to two. Furthermore, at least one value among $\rb$ and $\rc$ is also even and can be safely reduced to two. 
\begin{itemize}
\item Suppose first that either $\rb$ or $\rc$ is odd, being reduced to one. Then the number of leaves in $\subtreeof{\lab}$ can be safely reduced to three or four.
\item Now suppose that both $\rb$ and $\rc$ are even, being reduced to two. Then the number of leaves in $\subtreeof{\lab}$ can be safely reduced to four or five. But actually we can still reduce it from five to three: suppose that the unpaired (fifth) leaf of subtree $\subtreeof{\lab}$ is covered by an out-traversal to one of the other three subtrees. At most at the same cost we could replace it by an in-traversal of $\subtreeof{\lab}$. Therefore in this case $\subtreeof{\lab}$ can be safely reduced to three or four.
\end{itemize}
\vspace{-3mm}
\end{IEEEproof}

\vspace{4mm}



The following proposition helps us to complete the proof of correctness of our approach.

\vspace{3mm}

\begin{proposition} \label{prop:no-bridge}
Any tagged component tree $\T(\ga,\gb)$ has an optimal cover that
includes no \emph{bridge} (a path connecting two internal nodes of $\T(\ga,\gb)$).
\end{proposition}
\begin{IEEEproof}
Any indel-neutral bridge can be elongated to connect two leaves with at most the same cost.
We can then assume that, if an optimal cover must include a bridge, it is indel-saving.
The cases of clean leaves only or leaves that share one tag are covered by Theorems~\ref{th:one-tag} and~\ref{th:no-tag}.

Now we will prove the case of a component tree with two non-empty canonical subtrees whose leaves share no tag.
By generalizing the statements of Theorems~\ref{th:one-tag} and~\ref{th:no-tag}, we can assume that there is an optimal cover that has no bridge connecting two nodes that belong to the same canonical subtree.
Therefore, such a bridge either connects one node of each subtree to one node of the other, or if there is bad link $P$ between the subtrees, it can connect two nodes from $P$ or a node from $P$ and a node from one of the two subtrees. But, if we have such a bridge, we have no out-traversal nor semi-traversal, otherwise these would cover the same nodes covered by the bridge. This means that all the leaves of the two canonical subtrees are then covered by in-traversals or paths connecting a leaf to an internal node of the same canonical subtree.
We then have two types of cases: 

(i) $\subtreeof{\la}$ and $\subtreeof{\lb}$ -- such an indel-saving bridge would lead to an overall cost of $w=\lceil\ra/2\rceil + \lceil\rb/2\rceil + 1$. If instead we replace this bridge by an indel-saving out-semitraversal, we get the cost of $w'=\lceil(\ra-1)/2\rceil + \lceil\rb/2\rceil + 1\leq w$.

(ii) $\subtreeof{\la}$ and $\subtreeof{\lc}$:
(a) $\rc$ is odd -- such an indel-saving bridge would lead to an overall cost of $w=\lceil\ra/2\rceil + \rc (+1) + 1$ and can be replaced by an indel-neutral out-traversal, resulting in the cost of $w'=\lceil(\ra-1)/2\rceil + \rc-1 + 2=\lceil(\ra-1)/2\rceil + \rc + 1\leq w$;
(b) $\rc$ is even and $\la \geq 2$ -- such an indel-saving bridge would lead to an overall cost of $w=\lceil\ra/2\rceil + \rc + 1$ and can be replaced by two indel-neutral out-traversals, resulting in the cost of $w'=\lceil(\ra-2)/2\rceil + \rc-2 + 4=\lceil\ra/2\rceil + \rc + 1= w$;
(c) $\rc$ is even and $\la = 1$ -- such an indel-saving bridge would lead to an overall cost of $w= 1 + \rc + 1 = \rc + 2$ and can be replaced by one indel-neutral out-traversal, resulting in the cost of $w'= \rc + 2= w$. 

For trees with three or four types of leaves we can develop a similar proof, but with more cases to be analyzed.
\end{IEEEproof}



\vspace{2mm}


{\it Lemma \ref{lem:complexity}:}
\stateComplexity
\begin{IEEEproof}
Let $n=|\gg|+|\aa|+|\bb|$.
The relational graph $R(\ga,\gb)$, its number of cycles, their indel-potentials and the components can be computed in $\mathcal O(n)$ time~\cite{BER-MIX-STO-2005,BER-MIX-STO-2006b,BRA-WIL-STO-2011}. 
	The tagged component tree can also be constructed in $\mathcal O(n)$ time~\cite{BER-MIX-STO-2005}.
	For our transformation of the chained tagged component tree $\Tch(\ga,\gb)$ into the tagged component tree $\T(\ga,\gb)$, we traverse $\Tch$ in a bottom-up manner and recursively perform flower-contractions. For each removed tagged good leaf, its tag set is added to the one of its neighbor. While doing this, we obtain the sets of leaves $\la$, $\lb$, $\lc$ and $\lab$. 
	
	We then need to obtain the non-empty canonical subtrees and bad links separating them. 
	For each leaf class $L$, we again traverse the tree in a bottom-up manner. For each leaf in $L$ we proceed until we find a branching vertex, that we will call subroot of $L$. After all leaves of $L$ are visited, we repeat the procedure for all subroots of disjoint subtrees of $L$, until we get a single connected subtree.
	After repeating this procedure for all non-empty leaf classes, 
	the bad vertices that were not visited are then  \emph{bad links}. 
	All traversals of the tree can be done in $\mathcal O(k)$ time, where $k \leq n$ is the number of nodes of $\T(\ga,\gb)$. 
	
	For the reduction, we
	can obtain the set of 
	balanced in-traversals of each canonical subtree in $\mathcal O(k)$ time~\cite{ERD-SOU-STO-2011}. The reduction of each tagged canonical subtree requires another traversal of the tree and also takes $\mathcal O(k)$ time.
	
	After reducing the tagged canonical subtrees, but before performing the reduction of the clean subtree $\subtreeof{\lc}$ we still need to 
	examine whether the tree has the solo leaf.
	The candidates can be found by detecting leaves in short leaf-branches, a procedure already described in \cite{BER-MIX-STO-2005}, but we still need to check for each candidate whether it is indeed a solo leaf. 
	 The only approach we could devise so far tests all possible hypotheses (no solo leaf and each clean short leaf-branch of $\T(\ga,\gb)$ as solo leaf) and takes one that results in the smallest cover cost. 
	 In order to get the cover cost of each possible residual subtree, we need to examine its topology.
	Non-canonical partition subtrees and tag mates can be obtained by a single traversal of the residual tree, that takes $\mathcal O(k)$ time.
	Once we have obtained all required properties, we can look up the optimal cover cost at the residual leaf composition enumeration (in Appendix~2).
	 Therefore, the step of determining the solo leaf together with reducing the clean subtree and obtaining the cover of the residual tree has time complexity $\mathcal O(k\ell)$, where $\ell$ is the number of leaves in $\subtreeof{\lc}$. Since $\ell \leq k \leq n$, the time complexity of this step is $\mathcal O(n^2)$ and it dominates the complexity of the whole procedure.
	
	In summary, the time complexity of our approach is $\mathcal O(n^2)$.
\end{IEEEproof}

\clearpage

\subsection*{Algorithms}

Algorithm~\ref{alg:residual} computes the residual tree of any tagged component tree based on Reduction Bounds~\ref{red:two-can} to~\ref{red:four-can}.
We observe that the reduction of tagged subtrees do not affect the solo leaf candidates. In fact a reduction can increase the length of a leaf-branch, turning a short into a long leaf-branch, as shown in Figure~\ref{fig:destroy-solo}. However, this is a very limited behaviour that only happens when the reduction is not balanced, as in the reduction from $3$ to one leaf, and can be prevented by the choice of the leaf to be kept in this reduction (see more details below). Therefore it is safe to detect the solo leaf after reducing all tagged subtrees. 

\begin{algorithm}
\caption{$\textsc{ResidualTreeWithOptimalCoverCost}$}
\label{alg:residual}
\algsetup{indent=2em}
\linespread{1.2}
\small
\begin{algorithmic}
  \REQUIRE tagged component tree $\T$
  \ENSURE residual tree $\T^R$ and optimal cover cost $w$
  \STATE ~
  \STATE $\la := \text{set of A-leaves in }\T$; $\lb :=  \text{set of B-leaves in }\T$; $\lab := \text{set of AB-leaves in }\T$;
  \STATE $\ra := |\la|$; $\rb := |\lb|$; $\rab := |\lab|$; $\T^R :=\T$; 
  \STATE ~
  
  \STATE // Subtree $\subtreeof{\la}$ can be reduced to one or two leaves in all Reduction Bounds
  \IF {$|\la| \geq 4$}
    \STATE $(\T^R,\la) :=\textsc{PerformBalancedSimultaneousReduction}(\T^R,\la,\text{\bf true}, \text{\bf null})$; ~~~// Algorithm~\ref{alg:balanced-reduction}
  \ENDIF
  \IF {$|\la| = 3$}
     \STATE $(\T^R,\la):=\textsc{ReduceFrom3to1}(\T^R,\la)$; ~~~// Algorithm~\ref{alg:reduceFrom3to1}
  \ENDIF
  \STATE ~
  \STATE // Subtree $\subtreeof{\lb}$ can be reduced to one or two leaves in all Reduction Bounds
  \IF {$|\lb| \geq 4$}
    \STATE $(\T^R,\lb) :=\textsc{PerformBalancedSimultaneousReduction}(\T^R,\lb,\text{\bf true}, \text{\bf null})$; ~~~// Algorithm~\ref{alg:balanced-reduction} 
  \ENDIF
  \IF {$|\lb| = 3$}
     \STATE $(\T^R,\lb):=\textsc{ReduceFrom3to1}(\T^R,\lb)$; ~~~// Algorithm~\ref{alg:reduceFrom3to1}
  \ENDIF
  \STATE ~
  \STATE $w = \frac{\ra-|\la|}{2} + \frac{\rb-|\lb|}{2}$;
  \STATE ~
    \IF {($|\lb| > |\la|$)}
     \STATE ``swap'' $\asymb$-nodes with $\bsymb$-nodes in $\T^R$; 
  \ENDIF
  \STATE ~
  \IF {$|\lab| \geq 3$}
    \STATE $\rc^{R} := \text{number of clean leaves in }\T^{R}$; ~~~// $\rc^{R}$ stores the residual number of leaves of $\subtreeof{\lc}$
  \STATE {\bf if} $\rc^{R}$ is odd {\bf then} $\rc^{R} := 1$;
  \STATE {\bf else if} $\rc^{R} > 0$ {\bf then} $\rc^{R} := 2$;
  
  \STATE $\textsc{canReduceTo2}:=\text{\bf false}$;
  \IF {$|\lab|$ is even}
     \STATE // Subtree $\subtreeof{\lab}$ can usually be reduced to two leaves, 
     \STATE // except for particular cases of Reduction Bounds~\ref{red:three-can-a-b-ab} to~\ref{red:four-can}.
     \IF {($|\lb| = 0)$ \OR ($\rc^{R} = 0$ \AND $|\la| + |\lb| < 4$) \OR ($\rc^{R} > 0$ \AND $|\la|+|\lb|+\rc^{R} < 5$)} 
         \STATE $\textsc{canReduceTo2}:=\text{\bf true}$;
    \ENDIF
  \ENDIF
  
  \IF {$|\lab| \geq 5$ \OR $\textsc{canReduceTo2}$}
    \STATE $(\T^R,\lab) :=\textsc{PerformBalancedSimultaneousReduction}(\T^R,\lab,\textsc{canReduceTo2},\text{\bf null})$; ~~~// Algorithm~\ref{alg:balanced-reduction}
  \ENDIF 
  \IF {$|\lab| = 3$}
     \STATE // Subtree $\subtreeof{\lab}$ can usually be reduced to one leaf, 
     \STATE // except for particular cases of Reduction Bounds~\ref{red:three-can-a-b-ab} to~\ref{red:four-can}.
      \IF {($|\lb| = 0$  \AND $|\la| + \rc^{R} < 4$) \OR ($\rc^{R} = 0$ \AND $|\la| + |\lb| < 3$) \OR ($|\lb| > 0$  \AND $\rc^{R} > 0$ \AND $|\la|+|\lb|+\rc^{R} < 4$)} 
        \STATE $(\T^R,\lab):=\textsc{ReduceFrom3to1}(\T^R,\lab)$; ~~~// Algorithm~\ref{alg:reduceFrom3to1}
     \ENDIF
    \ENDIF 
  \ENDIF
  \STATE ~
  \STATE $w = w + \frac{\rab-|\lab|}{2}$;
   \STATE ~
  \STATE $(\T^R, w) := \textsc{ExhaustiveSearchOfResidualTree}(\T^R, w)$; ~~~// Algorithm~\ref{alg:reduce-clean} reduces $\subtreeof{\lc}$ and determines the existence of solo leaf
  \STATE ~
  \STATE Return $(\T^R,w)$;
  
\end{algorithmic}
\linespread{\lnsp}
\end{algorithm}



\clearpage

Reducing a canonical subtree from $\ell \geq 4$ to $2$ (if $\ell$ is even) or $3$ (if $\ell$ is odd) is described in Algorithm~\ref{alg:balanced-reduction}: it suffices to take any set of balanced in-traversals (keeping the solo leaf if the reduced subtree is clean). In fact, this algorithm uses a generalization of a balanced reduction.
Let $\mathcal P$ be a 
set of balanced in-traversals of some canonical subtree of $\T(\ga,\gb)$.
A \emph{simultaneous $\mathcal P$-reduction} of $\T(\ga,\gb)$ results in a smaller
tagged component tree as follows.
Let $T'$ be the tagged tree obtained from $\T(\ga,\gb)$ by replacing each bad node $b$ of each $P \in {\mathcal P}$ by a good node $g$, such that $\tagset(g)=\tagset(b)$.
Then $\Tred{-{\mathcal P}}(\ga,\gb)$ is the flower-contracted version of
$T'$. 

\begin{algorithm}
\caption{$\textsc{PerformBalancedSimultaneousReduction}$}
\label{alg:balanced-reduction}
\algsetup{indent=2em}
\linespread{1.2}
\small
\begin{algorithmic}
  \REQUIRE tagged component tree $\Tred{R}$, class of leaves $L$, boolean $\textsc{canReduceTo2}$, solo leaf candidate
  $s$ 
  \ENSURE smaller tagged component tree (in which $\subtreeof{L}$ is reduced to two leaves, if $|L|$ is even and $\textsc{canReduceTo2}$ is {\bf true}; or to three leaves, if $|L|$ is odd; or to four leaves, if $|L|$ is even and $\textsc{canReduceTo2}$ is {\bf false}) and updated set $L$
  \STATE ~
   \STATE $L':= \text{copy of }L$;
    \IF {$|L'|$ is odd}
      \STATE $u := s$;
  \IF {($u = \text{\bf null}$)} 
        \STATE $u:=$ any leaf of $\subtreeof{L'}$; 
        ~~~// if in a tagged subtree or if there is no solo leaf candidate
      \ENDIF
      \STATE $L' := L' \backslash \{u\}$; 
    \ENDIF
    \STATE ~
    \STATE // after this step $|L'|$ is even
    \STATE ~
    \STATE $\mathcal P := $ 
    set of $\frac{|L'|}{2}$ in-traversals covering $\subtreeof{L'}$ ~~~// Algorithm~\ref{alg:covering-traversals}
    \STATE ~
    \STATE // if originally $|L|$ was even
    \IF {$|L|$ is even} 
       \IF {($s \ne \text{\bf null}$)}
        \STATE remove from $\mathcal P$ the in-traversal that has $s$ as an endpoint;
        \ELSE
         \STATE remove from $\mathcal P$ any in-traversal; 
       \ENDIF
    \ENDIF
    \STATE ~
    \IF {($\ell$ is odd) \OR ({\bf not} $\textsc{canReduceTo2}$)}
      \STATE remove from $\mathcal P$ any in-traversal;
    \ENDIF
    \STATE ~
    \STATE $\T^{R} := \T^{R-{\mathcal P}}$; ~~~// apply the simultaneous $\mathcal P$-reduction in $\T^R$
    \STATE remove from $L$ all leaves except for those that remain in $\T^{R}$;
    \STATE ~
    \STATE Return $(\T^{R}, L)$;
 \end{algorithmic}
 \linespread{\lnsp}
 \end{algorithm}
 
\vspace{2mm}

Reducing a canonical subtree from $\ell = 3$ to $1$ is described in Algorithm~\ref{alg:reduceFrom3to1}: here we need to take one leaf to be kept, while the other two are used for the reduction. In order to prevent the risk of creating a new bad leaf with the reduction and/or the risk of selecting an unsafe reduction, it is necessary to be careful with the decision of the leaf to be kept, that we call \emph{essential leaf}.
Sometimes two or even the three leaves are suitable to be essential, but, since it suffices to identify a single one, the first one that is found is called \emph{the} essential leaf. 
It is identified as follows:
 \begin{itemize}
 \item It is the leaf of a leaf-branch of~$\subtreeof{L}$ that is also a leaf-branch of~$\T(\ga,\gb)$, if there is any. This case avoids the ``destruction'' of a possible solo leaf, as illustrated in Figure~\ref{fig:destroy-solo}. It also avoids other unsafe reductions, as illustrated in Figure~\ref{fig:red-from-3-ex1}, parts~(i) and~(ii).
 \item Otherwise at least one node of each leaf-branch $P_i$ of~$\subtreeof{L}$ is connected to~$\T(\ga,\gb)\backslash\subtreeof{L}$; then let $F(P_i)$ be the subgraph of~$\T(\ga,\gb)\backslash\subtreeof{L}$ that is connected to the leaf-branch $P_i$ and the set ${\mathcal S}(P_i)$ be the superset of the tag sets of the leafs of $\T(\ga,\gb)$ that are in $F(P_i)$. If, for two leaf-branches $P_1$ and $P_2$ of~$\subtreeof{L}$, we have ${\mathcal S}(P_1) \cap {\mathcal S}(P_2) \ne \emptyset$, then any of the two leaves in $P_1$ or $P_2$ is the essential leaf. This case avoids that a canonical subtree become isolated, as illustrated in Figure~\ref{fig:red-from-3-ex1}, part~(iii).
 
 \item Otherwise any leaf of~$\subtreeof{L}$ is the essential leaf.
 \end{itemize}

\begin{algorithm}
\caption{$\textsc{ReduceFrom3to1}$}
\label{alg:reduceFrom3to1}
\algsetup{indent=2em}
\linespread{1.2}
\small
\begin{algorithmic}
  \REQUIRE tagged component tree $\Tred{R}$ and class of leaves $L$ (with $|L|=3$)
  \ENSURE smaller tagged component tree (in which $\subtreeof{L}$ is reduced to one leaf) and updated set $L$
  \STATE ~
  
  \STATE $u:=$ essential leaf of $\subtreeof{L}$; 
 
  \STATE $L := L \backslash \{u\}$; 
  \STATE $P := \text{path connecting } L[1]\text{ to }L[2]$; 
  \STATE $\T^R := \T^{R-P}$; ~~~// apply the $P$-reduction in $\T^R$
    \STATE Return $(\T^R, \{u\})$;
 \end{algorithmic}
 \linespread{\lnsp}
 \end{algorithm}

 \begin{figure}[ht]
 \lfsize
	\setlength{\unitlength}{0.7pt}
 \center
  \begin{tabular}{c@{~~~~~}c}
 \begin{picture}(100,130)
        \put(50,120){\makebox(0,0){\boldmath$\subtreeof{\la}$}}
		\put(50,100){\TriTree
		  {\badvertex\rightlabel{$u$}}
		   {\UnTree{\badvertex}{\UnTree{\badvertex}{\UnTree{\badvertex}{\badAvertex\bottomlabel{$a_1$}}}}}
		   {\UnTree{\badvertex}{\UnTree{\badvertex}{\UnTree{\badvertex}{\badAvertex\bottomlabel{$a_2$}}}}}
		   {\UnTree{\badvertex\rightlabel{$v$}}{\UnTree{\badvertex}{\UnTree{\badvertex}{\badAvertex\bottomlabel{$a_3$}}}}}
		}
\end{picture}
 &
 \begin{tabular}{c@{~~~~~~}c@{~~~~~~~~~~~~}c@{~~~~~~~~~~~~}c}
& {\large \boldmath $\T$} & {\normalsize \bf Reduced by \boldmath $a_2 \travone a_3$} & {\normalsize \bf Reduced by \boldmath $a_1 \travone a_2$}\\[.5em]
 & & {\normalsize \bf Leaf \boldmath $a_1$ is essential} & \\[1.5em]
 \begin{picture}(10,125)
    \put(5,80){\makebox(0,0){\bf \large (i)}}
 	\end{picture} &
\begin{picture}(120,145)
		\put(60,135){\ParTrees
		   {\TriTree{\badvertex\toplabel{$u$}}
		       {\badvertex\bottomlabel{$c_1$}}  {\UnTree{\badvertex}{\UnTree{\badvertex}{\UnTree{\badvertex}{\badAvertex\bottomlabel{$a_1$}\leftlabel{$*$\!\!\!}}}}}
		       {\UnTree{\badvertex}{\UnTree{\badvertex}{\UnTree{\badvertex}{\badAvertex\bottomlabel{$a_2$}\rightlabel{\!\!\!$*$}}}}}
		   }
		   {\UnTree{\badvertex\toplabel{$v$}}
		    {\BiTreeNarrow{\badvertex}
		     {\UnTree{\badvertex}{\badAvertex\bottomlabel{$a_3$}}}
		     {\UnTree{\badvertex}{\badvertex\bottomlabel{$c_2$}}}
		    }
		   }
		}
  \put(60,25){\makebox(0,0){$\optcost{4}$}}
  \put(60,5){\makebox(0,0){$a_{1\!} \travone a_3$, $a_{2\!} \travtwo c_2$, $\costonepath{c_1}$ ; or}}
  \put(60,-15){\makebox(0,0){$a_{1\!} \travone a_2$, 
  $a_{1\!} \travone a_3$, $c_{1\!} \travtwo c_2$}}
\end{picture} &
\begin{picture}(120,145)
		\put(60,135){\TriTree{\goodAvertex\toplabel{$(a_2,u,v,a_3)$}}
		     {\badvertex\bottomlabel{$c_1$}}
		     {\UnTree{\badvertex}{\UnTree{\badvertex}{\UnTreeLeft{\badvertex}{\badAvertex\bottomlabel{$a_1$}}}}}
		     {\UnTree{\badvertex}{\badvertex\bottomlabel{$c_2$}}}
		   }
  \put(60,45){\makebox(0,0){(safe)}}
  \put(60,25){\makebox(0,0){$\optcost{3}$}}
  \put(60,5){\makebox(0,0){$a_{1\!} \travtwo c_2$, $\costonepath{c_1}$ ; or}}
  \put(60,-15){\makebox(0,0){$a_{1\!}  \semitravone (a_2,u,v,a_3)$, $c_{1\!} \travtwo c_2$}}
\end{picture}& 
\begin{picture}(120,145)
		\put(60,135){\TriTree{\badvertex}
		     {\UnTree{\badvertex\leftlabel{$v$}}{\UnTree{\goodAvertex\leftlabel{$(a_1,u,a_2)$~~~~~~~~~}}{\badvertex\bottomlabel{$c_1$}}}}
		     {\UnTree{\badvertex}{\badAvertex\bottomlabel{$a_3$}}}
		     {\UnTree{\badvertex}{\badvertex\bottomlabel{$c_2$}}}
		   }
  \put(60,45){\makebox(0,0){(safe)}}
  \put(60,25){\makebox(0,0){$\optcost{3}$}}
  \put(60,5){\makebox(0,0){$c_{1\!} \travtwo c_2$, $a_{3\!} \semitravone (a_1,u,a_2)$}}
\end{picture}\\[4em]
 \begin{picture}(10,125)
    \put(5,80){\makebox(0,0){\bf \large (ii)}}
 	\end{picture} &
\begin{picture}(120,145)
		\put(60,135){\ParTrees
		   {\TriTree{\badvertex\toplabel{$u$}}
		       {\badvertex\bottomlabel{$c$}}  {\UnTree{\badvertex}{\UnTree{\badvertex}{\UnTree{\badvertex}{\badAvertex\bottomlabel{$a_1$}\leftlabel{$*$\!\!\!}}}}}
		       {\UnTree{\badvertex}{\UnTree{\badvertex}{\UnTree{\badvertex}{\badAvertex\bottomlabel{$a_2$}\rightlabel{\!\!\!$*$}}}}}
		   }
		   {\UnTree{\badvertex\toplabel{$v$}}
		    {\BiTreeNarrow{\badvertex}
		     {\UnTree{\badvertex}{\badAvertex\bottomlabel{$a_3$}}}
		     {\UnTree{\badvertex}{\badABvertex\bottomlabel{$x$}}}
		    }
		   }
		}
  \put(60,25){\makebox(0,0){$\optcost{3}$}}
  \put(60,5){\makebox(0,0){$a_{2} \travone a_3$, $a_{1\!} \travone x$, $\costonepath{c}$}}
\end{picture} &
\begin{picture}(120,145)
		\put(60,135){\TriTree{\goodAvertex\toplabel{$(a_2,u,v,a_3)$}}
		     {\badvertex\bottomlabel{$c$}}
		     {\UnTree{\badvertex}{\UnTree{\badvertex}{\UnTreeLeft{\badvertex}{\badAvertex\bottomlabel{$a_1$}}}}}
		     {\UnTree{\badvertex}{\badABvertex\bottomlabel{$x$}}}
		   }
  \put(60,45){\makebox(0,0){(safe)}}
  \put(60,25){\makebox(0,0){$\optcost{2}$}}
  \put(60,5){\makebox(0,0){$a_{1\!} \travone x$, $\costonepath{c}$}}
\end{picture} & 
\begin{picture}(120,145)
		\put(60,135){\TriTree{\badvertex}
		     {\UnTree{\badvertex\leftlabel{$v$}}{\UnTree{\goodAvertex\leftlabel{$(a_1,u,a_2)$~~~~~~~~~}}{\badvertex\bottomlabel{$c$}}}}
		     {\UnTree{\badvertex}{\badAvertex\bottomlabel{$a_3$}}}
		     {\UnTree{\badvertex}{\badABvertex\bottomlabel{$x$}}}
		   }
  \put(60,45){\makebox(0,0){(unsafe)}}
  \put(60,25){\makebox(0,0){$\optcost{3}$}}
  \put(60,5){\makebox(0,0){$c_{\!} \travtwo x$, $a_{3} \travone x$}}
\end{picture}\\
\end{tabular}
\end{tabular}
\caption{Examples showing that a careless reduction on $\subtreeof{\la}$ can ``destroy'' a solo leaf. In (i) the given $\T$ has an optimal cover with a short path covering clean leaf $c_1$. This solo leaf in non-mandatory: there is also an alternative optimal cover including only traversals. After the reduction $a_{1\!}  \travone a_2$, the leaf $c_1$ is in a long leaf-branch and cannot be covered by a short path, nevertheless this reduction is safe. In (ii) we have a $\T$ with a mandatory solo leaf. Again, after the reduction $a_{1\!}  \travone a_2$, the leaf $c_1$ is in a long leaf-branch and cannot be covered by a short path, and in this case the reduction is indeed unsafe. Note that this can only happen when the reduction takes an in-traversal connecting consecutive leaves in the assumed circular order, and this is only possible when reducing from 3 to 1. Fortunately, this side effect is easily prevented by the choice of the essential leaf. }\label{fig:destroy-solo}
\end{figure}
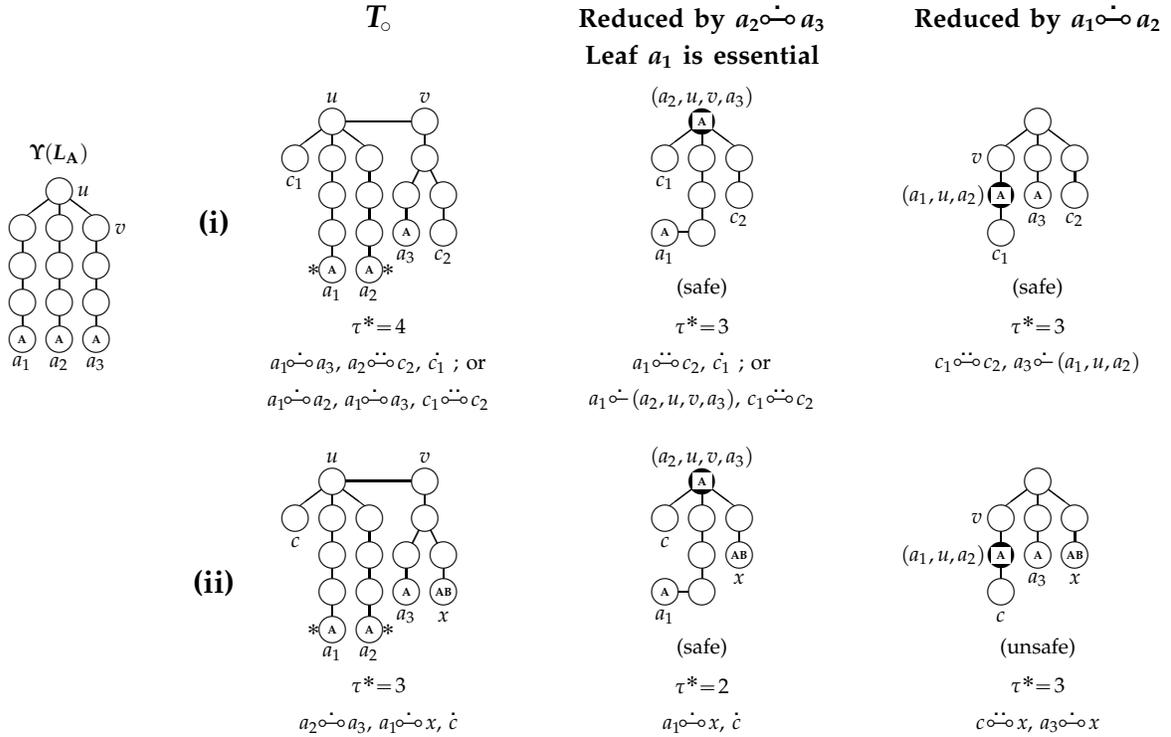

 \begin{figure}[ht]
 \def\lfsize{\scriptsize}
 \lfsize
	\setlength{\unitlength}{0.7pt}
 \center
 \begin{tabular}{c@{~~~~~}c}
 \begin{picture}(100,130)
        \put(50,120){\makebox(0,0){\boldmath$\subtreeof{\la}$}}
		\put(50,100){\TriTree
		  {\badvertex\rightlabel{$v$}}
		   {\UnTree{\badvertex\leftlabel{$u$}}
		     {\UnTree{\badvertex}{\badAvertex\bottomlabel{$a_1$}}}
		   }
		   {\UnTree{\badvertex}{\UnTree{\badvertex}{\badAvertex\bottomlabel{$a_2$}}}}
		   {\UnTree{\badvertex}{\UnTree{\badvertex}{\badAvertex\bottomlabel{$a_3$}}}}
		}
\end{picture}
 &
 \begin{tabular}{c@{~~~~~~}c@{~~~~~~~~~~~}c@{~~~~~~}c@{~~~~~~}c}
 &	{\normalsize \boldmath $\T$} & {\normalsize \bf Reduced by \boldmath $a_1 \travone a_2$} & {\normalsize \bf Reduced by \boldmath $a_1 \travone a_3$} & {\normalsize \bf Reduced by \boldmath $a_2 \travone a_3$} \\[.5em]
 & & {\normalsize \bf Leaf \boldmath $a_3$ is essential} & {\normalsize \bf Leaf \boldmath $a_2$ is essential} & \\[1.5em]
 \begin{picture}(10,120)
    \put(5,70){\makebox(0,0){\bf \large (i)}}
 	\end{picture} &
 \begin{picture}(100,120)
		\put(50,110){\ParTrees
		  {\UnTree{\badvertex\leftlabel{$u$}}
		   {\BiTreeNarrow{\badvertex}
		     {\badAvertex\bottomlabel{$a_1$}}
		     {\badBvertex\bottomlabel{$b$}}
		   }
		  }
		  {\BiTreeNarrow{\badvertex\rightlabel{$v$}}
		     {\UnTree{\badvertex}{\UnTree{\badvertex}{\badAvertex\bottomlabel{$a_2$}\leftlabel{$*$\!\!\!}}}}
		     {\UnTree{\badvertex}{\UnTree{\badvertex}{\badAvertex\bottomlabel{$a_3$}\rightlabel{\!\!\!$*$}}}}
		   }
		}
		\put(50,20){\makebox(0,0){$\optcost{3}$}}
		\put(50, 5){\makebox(0,0){$a_{1\!} \travone a_2$, $b_{\!} \travtwo a_3$}}
\end{picture} &
\begin{picture}(100,120)
		\put(50,110){\BiTreeNarrow{\goodAvertex\leftlabel{$(a_1,u,v,a_2)$~~~~~~~~~~~}}
		   {\badBvertex\bottomlabel{$b$}}
		   {\UnTree{\badvertex}{\UnTree{\badvertex}{\badAvertex\bottomlabel{$a_3$}}}}
		 }  
		\put(50,20){\makebox(0,0){$\optcost{2}$}}
		\put(50, 5){\makebox(0,0){$b_{\!} \travtwo a_3$}}
\end{picture} &
\begin{picture}(100,120)
		\put(50,110){\BiTreeNarrow{\goodAvertex\leftlabel{$(a_1,u,v,a_3)$~~~~~~~~~~~}}
		   {\badBvertex\bottomlabel{$b$}}
		   {\UnTree{\badvertex}{\UnTree{\badvertex}{\badAvertex\bottomlabel{$a_2$}}}}
		 }  
		\put(50,20){\makebox(0,0){$\optcost{2}$}}
		\put(50, 5){\makebox(0,0){$b_{\!} \travtwo a_2$}}
\end{picture}
&
\begin{picture}(100,120)
		\put(50,90){\TriTree{\badvertex}
		     {\badAvertex\bottomlabel{$a_1$}}
		     {\badBvertex\bottomlabel{$b$}}
		     {\badAvertex\bottomlabel{$u$}\toplabel{~~~~~~~~~~~$(a_2,v,a_3)$}}
		 }  
		\put(50,20){\makebox(0,0){$\optcost{2}$}}
		\put(50, 5){\makebox(0,0){$a_{1\!} \travone u$, $\costonepath{b}$}}
\end{picture}\\[.5em]
 & & (safe) & (safe) & (new leaf but safe)\\[1.5em]
\begin{picture}(10,120)
    \put(5,70){\makebox(0,0){\bf \large (ii)}}
 	\end{picture} &
 \begin{picture}(100,120)
		\put(50,110){\ParTrees
		  {\UnTree{\badvertex\leftlabel{$u$}}
		   {\BiTreeNarrow{\badvertex}
		     {\badAvertex\bottomlabel{$a_1$}}
		     {\UnTree{\badvertex}{\badBvertex\bottomlabel{$b$}}}
		   }
		  }
		  {\BiTreeNarrow{\badvertex\rightlabel{$v$}}
		     {\UnTree{\badvertex}{\UnTree{\badvertex}{\badAvertex\bottomlabel{$a_2$}\leftlabel{$*$\!\!\!}}}}
		     {\UnTree{\badvertex}{\UnTree{\badvertex}{\badAvertex\bottomlabel{$a_3$}\rightlabel{\!\!\!$*$}}}}
		   }
		}
		\put(50,20){\makebox(0,0){$\optcost{3}$}}
		\put(50, 5){\makebox(0,0){$a_{1\!} \travone a_2$, $b_{\!} \travtwo a_3$}}
\end{picture} &
\begin{picture}(100,120)
		\put(50,110){\BiTreeNarrow{\goodAvertex\leftlabel{$(a_1,u,v,a_2)$~~~~~~~~~~~}}
		   {\UnTree{\badvertex}{\badBvertex\bottomlabel{$b$}}}
		   {\UnTree{\badvertex}{\UnTree{\badvertex}{\badAvertex\bottomlabel{$a_3$}}}}
		 }  
		\put(50,20){\makebox(0,0){$\optcost{2}$}}
		\put(50, 5){\makebox(0,0){$b_{\!} \travtwo a_3$}}
\end{picture} &
\begin{picture}(100,120)
		\put(50,110){\BiTreeNarrow{\goodAvertex\leftlabel{$(a_1,u,v,a_3)$~~~~~~~~~~~}}
		   {\UnTree{\badvertex}{\badBvertex\bottomlabel{$b$}}}
		   {\UnTree{\badvertex}{\UnTree{\badvertex}{\badAvertex\bottomlabel{$a_2$}}}}
		 }  
		\put(50,20){\makebox(0,0){$\optcost{2}$}}
		\put(50, 5){\makebox(0,0){$b_{\!} \travtwo a_2$}}
\end{picture} &
\begin{picture}(100,120)
		\put(50,95){\TriTree{\badvertex}
		     {\badAvertex\bottomlabel{$a_1$}}
		     {\UnTree{\badvertex}{\badBvertex\bottomlabel{$b$}}}
		     {\badAvertex\bottomlabel{$u$}\toplabel{~~~~~~~~~~~$(a_2,v,a_3)$}}
		 }
		\put(50,20){\makebox(0,0){$\optcost{3}$}}
		\put(50, 5){\makebox(0,0){$a_{1\!} \travone u$, $a_{1\!} \travtwo b$}}
\end{picture}\\[.5em]
&  & (safe) & (safe) & (new leaf and unsafe)\\[1.5em]
\begin{picture}(10,120)
    \put(5,70){\makebox(0,0){\bf \large (iii)}}
 	\end{picture} &
 \begin{picture}(100,120)
		\put(50,110){\ParTreesWide
		  {\UnTree{\badvertex\leftlabel{$u$}}
		   {\BiTreeNarrow{\badvertex}
		     {\badAvertex\bottomlabel{$a_1$}}
		     {\UnTree{\badvertex}{\badvertex\bottomlabel{$c$}}}
		   }
		  }
		  {\BiTreeNarrow{\badvertex\rightlabel{$v$}}
		     {\BiTreeLeft{\badvertex}
		       {\UnTree{\badvertex}{\badBvertex\bottomlabel{$b_1$}}}
		       {\UnTree{\badvertex}{\badAvertex\bottomlabel{$a_2$}\cornerlabel{\!\!\!\!$*$}}}}
		     {\UnTree{\badvertex}
  		       {\BiTreeRight{\badvertex}
		        {\badAvertex\bottomlabel{$a_3$}\cornerlabel{\!\!\!\!$*$}}
		        {\badBvertex\bottomlabel{$b_2$}}
		       }
		     }
		   }
		}
		\put(50,20){\makebox(0,0){$\optcost{4}$}}
		\put(50, 5){\makebox(0,0){$a_{1\!} \travone a_2$, $a_{3\!} \travtwo c$, $b_{1\!} \travone b_2$}}
\end{picture} &
\begin{picture}(100,120)
	\put(50,110){\TriTree{\goodAvertex\leftlabel{$(a_1,u,v,a_2)$~~~~~~~~~~~~}}
		   {\UnTree{\badvertex}{\badvertex\bottomlabel{$c$}}}
		   {\UnTree{\badvertex}{\badBvertex\bottomlabel{$b_1$}}}
		   {\UnTree{\badvertex}
  		       {\BiTreeRight{\badvertex}
		        {\badAvertex\bottomlabel{$a_3$}}
		        {\badBvertex\bottomlabel{$b_2$}}
		       }
		     }
		 }
		\put(50,20){\makebox(0,0){$\optcost{3}$}}
		\put(50, 5){\makebox(0,0){$a_{3\!} \travtwo c$, $b_{1\!} \travone b_2$}}
\end{picture} &
\begin{picture}(100,120)
	\put(50,110){\TriTree{\goodAvertex\leftlabel{$(a_1,u,v,a_3)$~~~~~~~~~~~~}}
		   {\UnTree{\badvertex}{\badvertex\bottomlabel{$c$}}}
		   {\badBvertex\bottomlabel{$b_2$}}
		   {\BiTreeRight{\badvertex}
		    {\UnTree{\badvertex}{\badAvertex\bottomlabel{$a_2$}}}
		    {\UnTree{\badvertex}{\badBvertex\bottomlabel{$b_1$}}}
		   }
		 }
		\put(50,20){\makebox(0,0){$\optcost{3}$}}
		\put(50, 5){\makebox(0,0){$a_{2\!} \travtwo c$, $b_{1\!} \travone b_2$}}
\end{picture} &
\begin{picture}(100,120)
		\put(50,110){\BiTree{\badvertex\leftlabel{$u$}}
		   {\BiTreeNarrow{\badvertex}
		     {\badAvertex\bottomlabel{$a_1$}}
		     {\UnTree{\badvertex}{\badvertex\bottomlabel{$c$}}}
		   }
		   {\BiTreeNarrow{\goodAvertex\cornerlabel{~~~~~~~~$(a_2,v,a_3)$}}
		    {\UnTree{\badvertex}{\badBvertex\bottomlabel{$b_1$}}}
		    {\badBvertex\bottomlabel{$b_2$}}
		   }
		 }  
		\put(50,20){\makebox(0,0){$\optcost{4}$}}
		\put(50, 5){\makebox(0,0){$a_{1\!} \travtwo b_1$, $c_{\!} \travtwo b_2$}}
\end{picture}\\[0.5em]
 & & (safe) & (safe) & (unsafe)\\
  \end{tabular}
 \end{tabular}
 \caption{Examples of reduction of canonical subtree $\subtreeof{\la}$ from 3 leaves to one leaf. In all examples the possible essential leaves are $a_2$ and $a_3$, marked with a $*$. In (i) and (ii) only the leaf-branch ending in $a_1$ intersects $\T$. Indeed, if $a_1$ is kept by the reduction, node $u$ becomes a leaf in the reduced tree, and the reduction can be safe or unsafe. If any of the possible essential leaves is kept, the number of leaves is reduced by two and the reduction is safe. In (iii) all three leaf-branches of $\subtreeof{\la}$ intersect $\T$. Since two branches intersect a leaf-branch of $\subtreeof{\lb}$, their respective leaves are the possible essential leaves. The reduction is indeed safe if one of these two leaves is kept, otherwise it is unsafe.}\label{fig:red-from-3-ex1}
\end{figure}
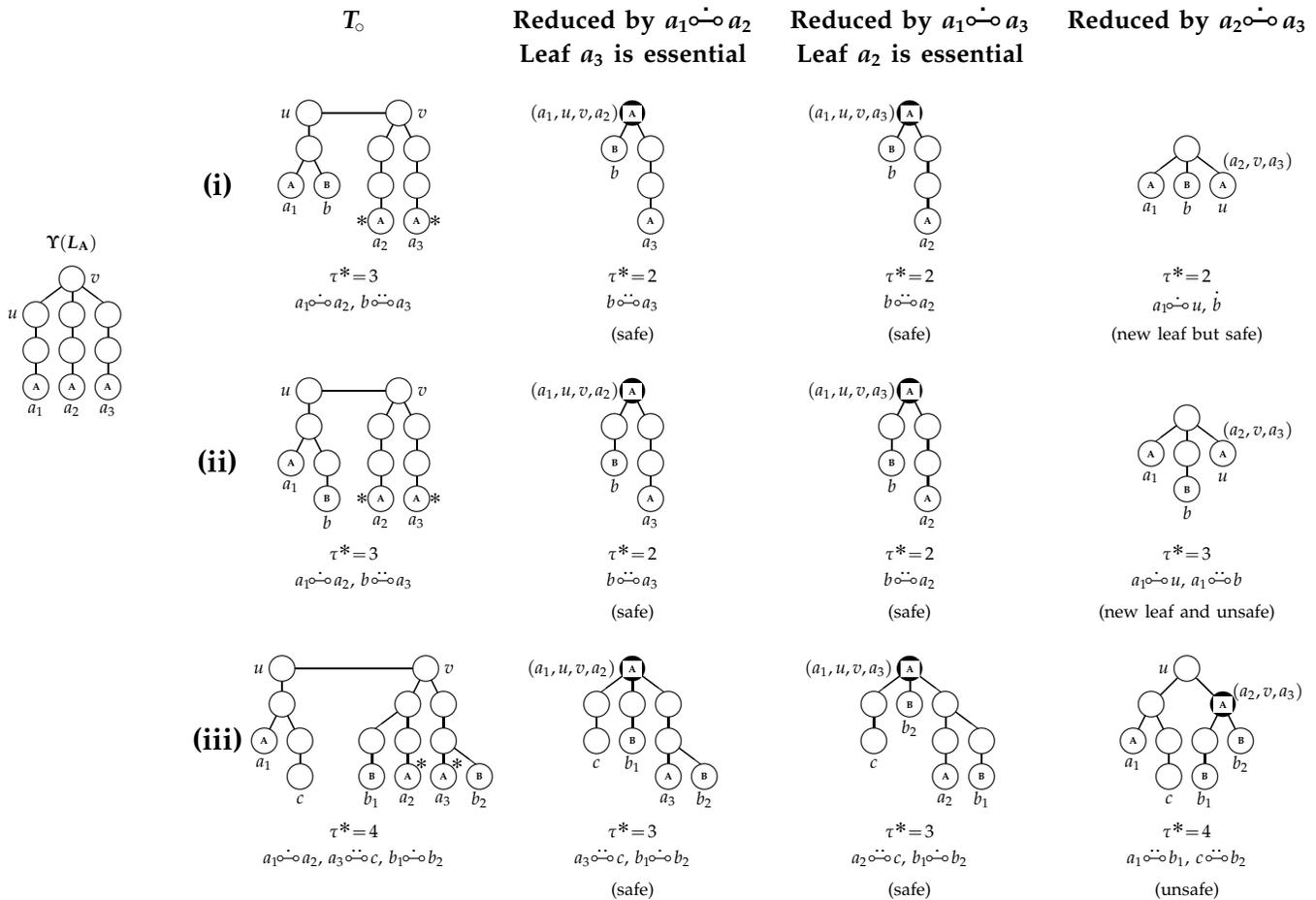


\clearpage
The only approach we could devise for determining the existence of the solo leaf tests all possible hypotheses (no solo leaf and each clean short leaf-branch of $\T(\ga,\gb)$ as solo leaf) and takes one that results in the smallest cover cost. For that reason, reducing the clean subtree is done together with searching the solo leaf, as described in Algorithm~\ref{alg:reduce-clean}.

\begin{algorithm}
\caption{$\textsc{ExhaustiveSearchOfResidualTree}$}
\label{alg:reduce-clean}
\algsetup{indent=2em}
\linespread{1.2}
\small
\begin{algorithmic}
  \REQUIRE semi-residual tree $\T^{R}$ and partial reduction cost $w_\text{red}$
  \ENSURE residual tree and optimal cover cost
  \STATE ~
  \STATE $\lc:= \text{set of clean leaves of~} \T^{R}$;
  \STATE $\rc:= |\lc|$;
  
  \STATE $\textsc{canReduceTo1} := \text{\bf false}$;
  \IF {$\rc$ is odd}
      \STATE // Subtree $\subtreeof{\lc}$ can always be reduced to one leaf, except for Reduction Bound~\ref{red:three-can-a-b-c}, first case.
      \STATE $\ra:= \text{number of~} \asymb\text{-leaves of~} \T^{R}$;
      \STATE $\rb:= \text{number of~} \bsymb\text{-leaves of~} \T^{R}$;
      \STATE $\rab:= \text{number of~} \absymb\text{-leaves of~} \T^{R}$;
      \IF {$\ra \ne 1$ \OR $\rb \ne 1$ \OR $\rab > 0$}
       \STATE $\textsc{canReduceTo1} := \text{\bf true}$;
     \ENDIF
  \ENDIF
  \STATE ~
  \STATE // First we compute the residual tree assuming that there is no \emph{a priori} solo leaf candidate
  \STATE $T_\text{min} := \text{copy of } \T^R$; 
  \STATE $L:= \text{copy of } \lc$;
  \IF {$\rc \geq 4$}
      \STATE $(T_\text{min}, L)= :=\textsc{PerformBalancedSimultaneousReduction}(T_\text{min},L,\text{\bf true}, \text{\bf null})$; ~~~// Algorithm~\ref{alg:balanced-reduction}
  \ENDIF
  \STATE $w_\text{clean\_red}:=\rc - |L|$; ~~~// the cost of the reduction of $\subtreeof{\lc}$ is the same, independently of the existence of solo leaf
  \IF {$\textsc{canReduceTo1}$}
     \STATE $(T_\text{min},L):=\textsc{ReduceFrom3to1}(T_\text{min},L)$; ~~~// Algorithm~\ref{alg:reduceFrom3to1}
     \STATE $w_\text{clean\_red} := w_\text{clean\_red} + 2$; 
  \ENDIF
  \STATE $w_\text{min} := \textsc{OptimalCoverCostOf}(T_\text{min})$; ~~~// Look up in Appendix 2
  \STATE ~
  \STATE $S :=$ set of clean leaves in short leaf-branches of $\T^{R}$;  ~~~// solo leaf candidates
  \STATE // For each element $s$ of $S$, we do the reduction of subtree $\subtreeof{\lc}$ assuming that $s$ is a potential solo leaf
  \FOR {$i=1$ \TO $|S|$}
    \STATE $T := \text{copy of } \T^R$; ~~~// Stores the residual tree keeping leaf $S[i]$
     \STATE $L:= \text{copy of } \lc$;
    \IF {$\rc \geq 4$}
      \STATE $(T,L) :=\textsc{PerformBalancedSimultaneousReduction}(T,L,\text{\bf true}, S[i])$; ~~~// Algorithm~\ref{alg:balanced-reduction}
    \ENDIF
    \IF {$\textsc{canReduceTo1}$}
      \STATE $L := L \backslash \{S[i]\}$; 
      \STATE $P := \text{path connecting } L[1]\text{ to }L[2]$; 
      \STATE $T := T^{-P}$; ~~~// apply the $P$-reduction in $T$
    \ENDIF
    \STATE $w := \textsc{OptimalCoverCostOf}(T)$; ~~~// Look up in Appendix 2
    \IF {$w < w_\text{min}$}
      \STATE $w_\text{min} := w$;
      \STATE $T_\text{min} := T$;
    \ENDIF
  \ENDFOR
  \STATE ~
  \STATE $w := w_\text{red} + w_\text{clean\_red} + w_\text{min}$;
  \STATE ~
  \STATE Return $(T_\text{min},w)$;
  
\end{algorithmic}
\linespread{\lnsp}
\end{algorithm}

\clearpage
\setcounter{page}{1}

\newcommand{\sololeaf}{
\unitlength1ex
\begin{picture}(2.1,0.95)%
\put(1.1,0.3){\makebox(0,0){-}}
\put(1.6,0.5){\circle{0.5}}
\put(0.8,0.7){\makebox(0,0){$\subtree$}}
\end{picture}
}

\newcommand{\prunedtree}{
\unitlength1ex
\begin{picture}(2.5,0.95)%
\put(1.4,0.3){\makebox(0,0){\textendash}}
\put(1.5,0.5){\makebox(0,0){\tiny \boldmath$+$}}
\put(2.1,0.5){\circle{0.5}}
\put(0.8,0.7){\makebox(0,0){$\subtree$}}
\end{picture}
}

\newcommand{\compprunedtree}{
\unitlength1ex
\begin{picture}(2.5,0.95)%
\put(1.4,0.3){\makebox(0,0){\textendash}}
\put(1.5,0.5){\makebox(0,0){\tiny \boldmath$+$}}
\put(2.1,0.5){\circle{0.5}}
\put(0.8,1.0){\makebox(0,0){$\compsubtree$}}
\end{picture}
}

\newcommand{\circled}[1]{\unitlength1ex\begin{picture}(2.5,2.5)%
\put(0.95,0.75){\circle{2.2}}
\put(0.98,0.75){\makebox(0,0){#1}}\end{picture}}

\newcommand\aleaf{\circled{\scriptsize$\asymb$}}
\newcommand\bleaf{\circled{\scriptsize$\bsymb$}}
\newcommand\ableaf{\circled{\scriptsize$\asymb\!\bsymb$}}
\newcommand\uleaf{\circled{\scriptsize$\noasymb\!\nobsymb$}}

\newcommand{\red}{\ggg}

\newcommand{\fullyco}{\subtree_{\coroot}\,\,}\
\newcommand{\fullysep}{\subtree_{\sepsymb}\,}
\newcommand{\notsep}[1]{#1\,\coroot\compsubtree}
\newcommand{\notsepinpruned}[1]{#1\,\coroot\compprunedtree}
\newcommand{\sep}[1]{#1\sepsymb\!\compsubtree}
\newcommand{\sepfrom}[2]{#1\sepsymb#2}

\newcommand\notsepfrom[2]{#1\,\coroot\,#2}
\newcommand{\connect}[2]{#1\!\!\qcoroot\!\!#2}
\newcommand\setgmate[2]{#1\mateat#2}
\newcommand\setmate[3]{\setgmate{#2}{#3}}

\newcommand{\travSW}{\travone\,\!\!}
\newcommand{\travNW}{\travtwo\,\!\!}
\newcommand{\semitravW}{\semitravone\,\!\!}

\newcommand{\otS}[2]{#1\travSW#2}
\newcommand{\otN}[2]{#1\!\travNW#2}
\newcommand{\itS}[1]{#1\travSW#1}
\newcommand{\itN}{\uset\!\travNW\uset}
\newcommand{\tSr}[2]{#1\travSW\!\left(#2\right)}
\newcommand{\tNr}[2]{#1\!\travNW\!\left(#2\right)}

\newcommand{\ostS}[2]{#1\!\semitravW #2}
\newcommand{\ostSr}[2]{\left(#1\right)\!\!\semitravW #2}
\newcommand{\twoti}{2$\times$}

\newcommand{\mycancel}{\bcancel}

\section*{Appendix 2: Enumeration of Optimal Covers for Residual Trees}
In this section we enumerate optimal covers and their respective costs for all topologies of residual trees that have at least one leaf with non-empty tag set and at least two leaves that do not share a tag.
Let $\ra$ be the number of $\asymb$-leaves, $\rb$ be the number of $\bsymb$-leaves, $\rc$ be the number of $\notag$-leaves and $\rab$ be number of $\absymb$-leaves in a residual tree. 
Recall that $\ra \leq 2$ for all residual trees. Due to the symmetry between $\asymb$-nodes and $\bsymb$-nodes, we assume without loss of generality that $\ra\geq\rb$.
The enumeration is kept as extensive as necessary but as simple as possible.

\subsection*{Residual trees are identified by their leaf composition}
 A leaf composition is represented by a tuple $\leco \ra\rb\rc\rab$ of four non negative values that always has $\ra$ in the first position, $\rb$ in the second position, $\rc$ in the third position and $\rab$ in the fourth position. 
 
Examples:

\begin{tabularx}{\linewidth}{ l X }
    
    $\leco 0021$: & 
    A leaf composition in which subtrees $\subtreeof{\la}$ and $\subtreeof{\lb}$ are empty, subtree $\subtreeof{\lc}$ has two leaves and subtree $\subtreeof{\lab}$ has one leaf.\\
    \leco 2124: & 
    A leaf composition in which subtree $\subtreeof{\la}$ has two leaves, subtree $\subtreeof{\lb}$ has one leaf, subtree $\subtreeof{\lc}$ has two leaves and subtree $\subtreeof{\la}$ has four leaves.\\

\end{tabularx}

\subsection*{Nomenclature}

\subsubsection*{General topologies of residual trees}
\begin{tabularx}{\linewidth}{ l X }
$\fullyco$: & 
		The tagged component tree $\T^{R}$ is fully co-rooted.\\
    $\fullysep$: & 
		The tagged component tree $\T^{R}$ is fully separated.\\
	$\sololeaf$: & 
		There is a short leaf-branch in $\T^{R}$ with a clean leaf.\\
\end{tabularx}

\vspace{6mm}

\noindent Relations between partition subtrees are described by the following notation, presented with the help of examples.

\subsubsection*{Partition subtrees and their specific relations in possible topologies of residual trees}

\begin{tabularx}{\linewidth}{ l X }
    $\aset$: & 
		The canonical subtree $\subtreeof{\la}$.\\
    $\connect{\uset}{\bset}$: & 
		The partition subtree $\subtreeof{\lc \cup \lb}$,
		in which the canonical subtrees
		$\subtreeof{\lc}$ and $\subtreeof{\lb}$ may be co-rooted or separated. \\
    $\notsepfrom{\aset}{\abset}$: & 
		The partition subtree $\subtreeof{\la \cup \lab}$, in which the subtrees $\subtreeof{\la}$ and $\subtreeof{\lab}$
		are co-rooted. \\
	$\notsep{\uset}$: & 
		The subtree $\subtreeof{\lc}$ is non-isolated, that is, $\subtreeof{\lc}$ and its complementary subtree $\compsubtreeof{\lc}=\subtreeof{\la \cup \lb \cup \lab}$ are co-rooted.\\
    $\sepfrom{\bset}{\uset}$: & 
		The partition subtree $\subtreeof{\lb \cup \lc}$, in which the subtree $\subtreeof{\lb}$ is separated from $\subtreeof{\lc}$.\\
    $\sep{\aset}$: & 
		The subtree $\subtreeof{\la}$ is isolated, that is, $\subtreeof{\la}$ is separated from 
		its complementary subtree $\compsubtreeof{\la}=\subtreeof{\lb \cup \lab \cup \lc}$.\\
	$\setgmate{\asymb}{\uset}$: & 
		The extended subtree $\extsubtreeof{\lc}$ has an $\asymb$-mate.\\
	$\prunedtree$: & 
		The tree obtained by pruning from $\T^{R}$ a clean short leaf-branch.\\
	$\notsepinpruned{\aset}$: & 
		The subtree $\subtreeof{\la}$ is non-isolated in a $\prunedtree$.\\
		
\end{tabularx}

\vspace{5mm}

\noindent The relations above can be applied to non-canonical subtrees: 

$\notsep{\left(\connect{\aset}{\bset}\right)}~~,~~
\notsepfrom{\uset}{\left(\sepfrom{\aset}{\bset}\right)}~~,~~
\sepfrom{\left(\notsepfrom{\abset}{\bset}\right)}{\aset}~~,~~
\sep{\left(\connect{\abset}{\uset}\right)}~~,~~
\setmate{\aset}{\asymb}{\left(\connect\bset\uset\right)}~~,~~
\setgmate{\asymb}{\left(\notsepfrom\bset\uset\right)}$

\vspace{5mm}

\subsubsection*{Logical operators}

	~And: \myand;~~Or: \myor;~~Absence of given properties: $\mycancel{\setmate{\aset}{\asymb}{\bset}}~~,~~\mycancel{\sololeaf}$

\subsubsection*{Identifiers of the optimal covers}
\begin{tabularx}{\linewidth}{ l X }
    I, II: & Basic optimal covers\\
	M: & Tag mate\\
	S: & Short path (usually covering a solo leaf)\\
	nR: & Not reducible\\
	$\red$: &
		Further reduction is possible for this type of topology\\
	W: & Worst case
\end{tabularx}

\vspace{6mm}

\noindent Optimal covers are described by the following notation, presented here with the help of examples.

\subsubsection*{Descriptions of optimal covers}
\begin{tabularx}{\linewidth}{ l X }
	$\aleaf$: & 
		A short path (whose cost is one) covering an $\asymb$-leaf. \\
	$\itS{\abset}$: & 
		A covering in-traversal using two $\absymb$-leaves that have not been covered
		before. The dot on top indicates that the cost of this traversal is one.\\
	$\otN{\uset}{\abset}$: & 
		A covering out-traversal using one $\notag$- and one $\absymb$-leaf that have not
		been covered before. The dots on top indicate that the cost of this traversal
		is two.\\
    $\tSr{\aset}{\abset}$: & 
		The parentheses around the subtree indicate that the respective leaf has
		already been covered previously.\\
	$\ostS{\abset}{\asymb}$: & 
		A semi-traversal (of cost one) using an $\absymb$-leaf with an internal node that is an $\asymb$-mate. The specific (extended) subtree that contains the $\asymb$-mate is given in the topology description.\\
   $|~2$: &
		The cost of the cover is given to the right of the vertical bar.\\ 
  $\cup$ \leco2221: &
		The optimal cover in
		this case consists of the given in-traversal plus the cost of the optimal cover of the appropriate topology of the reduced leaf composition \leco2221.\\
	$|+\!1$: &
		A ``+'' means that the cost has to be added to that of the reduced leaf composition.\\
\end{tabularx}

\vspace{5mm}

For each leaf composition and topology we describe one optimal cover, even though other optimal covers may also exist.

\subsection*{Enumeration of residual leaf compositions of Group 1}
Residual leaf compositions with exactly two non-empty canonical subtrees that do not share a tag, defined by Reduction Bound~\ref{red:two-can}. This includes~11 leaf compositions, divided in three subgroups:

\begin{center}
\footnotesize
\begin{tabular}{cc|cc|cc}
\toprule
 \multicolumn{6}{c}{Reduction Bound \ref{red:two-can}}\\
 \midrule
 \multicolumn{2}{c}{$\leco\ra\rb00$} & \multicolumn{2}{c}{$\leco\ra0\rc0$}  & \multicolumn{2}{c}{$\leco00\rc\rab$}\\
 \midrule
 $\leco 1100$ & $\leco 2100$ & $\leco 1010$ & $\leco 2010$ & $\leco 0011$  & $\leco 0021$ \\
& $\leco 2200$ & $\leco 1020$ & $\leco 2020$ & $\leco 0012$ & $\leco 0022$ \\
\bottomrule
\end{tabular}
\end{center}

\vspace{6mm} 

\noindent{\bf Residual leaf compositions of subgroup $\leco \ra\rb00$}
\vspace{3mm}

        \begin{tabular}{l@{~~~~~~~~}lll|l}
	     \multicolumn{2}{l}{$\leco 1100$}~\\
		 & any topology & I & $\otN\aset\bset$ & 2
	    \end{tabular}
        \vspace{6mm}
          
        \begin{tabular}{l@{~~~~~~~~}llll|l}
	            \multicolumn{2}{l}{$\leco 2100$}~\\
		          & $\fullyco$ & & S & $\bleaf$, $\itS\aset$ & 2 \\
		          & else & $\setmate{\bset}{\bsymb}{\aset}$ & M & $\itS\aset$~, $\ostS\bset\bsymb$ & 2 \\
		          & & else & W & $\itS\aset$~, $\tNr\bset\aset$ & 3
        \end{tabular}
                    
        \vspace{6mm}
                    
        \begin{tabular}{l@{~~~~~~~~}llll|l}
	            \multicolumn{2}{l}{$\leco 2200$}~\\
		         & $\fullyco$ & & I & $\itS\aset$~, $\itS\bset$ & 2 \\
		         & short bad link & & S & $\itS\aset$~, $\itS\bset$~, bad link cut & 3 \\
		         & else & $\setmate{\aset}{\asymb}{\bset}$
                                        & Ma & $\itS\aset$~, $\ostSr\aset\asymb$~,
                                               $\itS\bset$  & 3 \\
		         & & $\setmate{\bset}{\bsymb}{\aset}$ & Mb & $\itS\aset$~, $\ostSr\bset\bsymb$~, $\itS\bset$
		          & 3 \\
		         & & else & W & \twoti$\otN\aset\bset$ & 4
        \end{tabular}
        
 \vspace{6mm}        
\noindent{\bf Residual leaf compositions of subgroup $\leco \ra0\rc0$}
\vspace{3mm}

        \begin{tabular}{l@{~~~~~~~~}lll|l}
		     \multicolumn{2}{l}{$\leco 1010$}~\\
				& any topology & I & $\otN\aset\uset$ & 2
		     \end{tabular}
		     
		     \vspace{6mm}
		     
        \begin{tabular}{l@{~~~~~~~~}llll|l}
	             \multicolumn{2}{l}{$\leco 1020$}~\\
	              & $\sololeaf$ && S & $\uleaf$, $\otN\uset\aset$ & 3 \\
		          & else & $\fullyco$ & Sa & $\aleaf$, $\itN$ & 3 \\
		          & & $\setmate{\aset}{\asymb}{\uset}$ & M & $\itN$~, $\ostS\aset\asymb$ & 3 \\
		          & & else & W & $\otN\uset\aset$~, $\tNr\uset\aset$ & 4
        \end{tabular}
        \vspace{6mm}
                    
                    \begin{tabular}{l@{~~~~~~~~}lll|l}
	            \multicolumn{2}{l}{$\leco 2010$}~\\
   		         & $\fullyco$ & S & $\uleaf$, $\itS\aset$ & 2 \\
		         & else & W & $\otN\uset\aset$~, $\tSr\aset\aset$ & 3
		    \end{tabular}
                    \vspace{6mm}
                    
                    \begin{tabular}{l@{~~~~~~~~}lll|l}    
	            \multicolumn{2}{l}{$\leco 2020$}~\\
		          & $\fullyco$ & I & $\itN$~, $\itS\aset$ & 3 \\
		          & else & W & \twoti$\otN\uset\aset$ & 4
		    \end{tabular}

 \vspace{6mm}
\noindent{\bf Residual leaf compositions of subgroup $\leco 00\rc\rab$}
\vspace{3mm}

                    \begin{tabular}{l@{~~~~~~~~}lll|l}
	   	    \multicolumn{2}{l}{$\leco 0011$}~\\	
				& any topology & I & $\otN\uset\abset$ & 2
		    \end{tabular}
                    \vspace{6mm}

                    \begin{tabular}{l@{~~~~~~~~}lll|l}
	            \multicolumn{2}{l}{$\leco 0012$}~\\
		         & $\fullyco$ & S & $\uleaf$, $\itS\abset$ & 2 \\
		         & else & W & $\otN\uset\abset$~, $\tSr\abset\abset$ & 3
		    \end{tabular}
                    \vspace{6mm}
                    
                    \begin{tabular}{l@{~~~~~~~~}llll|l}
	            \multicolumn{2}{l}{$\leco 0021$}~\\
	            & $\sololeaf$ && S & $\uleaf$, $\otN\uset\abset$ & 3 \\
		         & else & $\fullyco$ & Sa & $\ableaf$, $\itN$ & 3 \\ 
		        & & $\setmate{\abset}{\asymb}{\uset}$ & Ma & $\itN$~, $\ostS\abset\asymb$ & 3 \\
			 & & $\setmate{\abset}{\bsymb}{\uset}$ & Mb & $\itN$~, $\ostS\abset\bsymb$ & 3 \\
			 
			 & & else & W & $\otN\uset\abset$~, $\tNr\uset\abset$ & 4
                    \end{tabular}
                    \vspace{6mm}

           \begin{tabular}{l@{~~~~~~~~}lll|l}
	            \multicolumn{2}{l}{$\leco 0022$}~\\
		          & $\fullyco$ & I & $\itN$~, $\itS\abset$ & 3 \\
		          & else & W & \twoti$\otN\uset\abset$ & 4
		    \end{tabular}


\subsection*{Enumeration of residual leaf compositions of Group 2}
Residual leaf compositions with exactly three non-empty canonical subtrees. This includes~25 leaf compositions, divided in three subgroups:

\begin{center}
\footnotesize
\begin{tabular}{c@{~}|@{~}c@{~}|@{~}c}
\begin{tabular}{c|cc}
\toprule
 \multicolumn{3}{c}{Reduction Bound~\ref{red:three-can-a-b-c}}\\
 \midrule
  \multicolumn{1}{c}{Case 1} & \multicolumn{2}{c}{Case 2}\\ 
 $\leco11\rc0$ & $\leco21\rc0$ & $\leco22\rc0$\\
 \midrule
$\leco 1110$ & $\leco 2110$ & $\leco 2210$\\
$\leco 1120$ & $\leco 2120$ & $\leco 2220$\\
$\leco 1130$ & &\\
\myemptycell
\bottomrule
\end{tabular} &
\begin{tabular}{c|c|c}
\toprule
\multicolumn{3}{c}{Reduction Bound~\ref{red:three-can-a-b-ab}}\\
\midrule
\multicolumn{1}{c}{Case 1} & \multicolumn{1}{c}{Case 2} & \multicolumn{1}{c}{Case 3}\\
$\leco110\rab$ & $\leco210\rab$ & $\leco220\rab$\\
\midrule
$\leco 1101$ & $\leco 2101$ & $\leco 2201$\\
$\leco 1102$ & $\leco 2102$ & $\leco 2202$\\
& $\leco 2103$ & $\leco 2203$\\
& & $\leco 2204$\\
\bottomrule
\end{tabular} & 
\begin{tabular}{ccc|c}
\toprule
\multicolumn{4}{c}{Reduction Bound~\ref{red:three-can-a-c-ab}}\\
\midrule
\multicolumn{3}{c}{Case 1} & \multicolumn{1}{c}{Case 2}\\
 $\leco101\rab$ & $\leco102\rab$ & $\leco201\rab$ & $\leco202\rab$\\
 \midrule
  $\leco 1011$ & $\leco 1021$ & $\leco 2011$ & $\leco 2021$\\
  $\leco 1012$ & $\leco 1022$ & $\leco 2012$ & $\leco 2022$\\
  & & & $\leco 2023$ \\
  \myemptycell
\bottomrule
\end{tabular}
\end{tabular}
\end{center}

 \vspace{6mm}

\noindent{\bf Reduction Bound~\ref{red:three-can-a-b-c}: residual leaf compositions of subgroup $\leco 11\rc0$ + $\leco 21\rc0$ + $\leco 22\rc0$}
\vspace{3mm}

	\begin{tabular}{l@{~~~~~~~~}lllll|l}
		\multicolumn{2}{l}{$\leco 1110$}~\\
			& $\notsep{\uset}$ &&& S & $\uleaf$, $\otN\aset\bset$ & 3 \\
			& else & $\notsep{\aset}$ && Sa & $\aleaf$, $\otN\bset\uset$ & 3 \\
			& 	& $\notsep{\bset}$ && Sb & $\bleaf$, $\otN\aset\uset$ & 3 \\
			& 	& else & $\setmate{\aset}{\asymb}{\left(\sepfrom{\bset}{\uset}\right)}$ & Ma &
				$\otN\bset\uset$~, $\ostS\aset\asymb$ & 3\\
			& 		& & $\setmate{\bset}{\bsymb}{\left(\sepfrom{\aset}{\uset}\right)}$ & Mb & 
					$\otN\aset\uset$~, $\ostS\bset\bsymb$ & 3\\
			& 		& & else & W & $\otN\aset\uset$~, $\tNr\bset\uset$ & 4
		\end{tabular}
\vspace{6mm}

	\begin{tabular}{l@{~~~~~~~~}lll|l}
		\multicolumn{2}{l}{$\leco 1120$}~\\
			& any topology & I & $\otN\aset\uset$~, $\otN\uset\bset$ & 4
		\end{tabular}
\vspace{6mm}

\begin{tabular}{l@{~~~~~~~~}lll|l}
		\multicolumn{2}{l}{$\leco 1130$}~\\
		& 
		$\sololeaf$ \myand $\sep{\aset}$ \myand $\sep{\bset}$
		& \multirow{2}{*}{S} &
		\multirow{2}{*}{$\uleaf$, $\otN\aset\uset$~, $\otN\bset\uset$} &
		\multirow{2}{*}{5} \\
		&  
		\myand \mycancel{$\setmate{\aset}{\asymb}{\left(\sepfrom{\bset}{\uset}\right)}$}  \myand 
		\mycancel{$\setmate{\bset}{\bsymb}{\left(\sepfrom{\aset}{\uset}\right)}$} & & \\
		& else & $\red$ & $\itN ~\cup\ \leco 1110$ & +2
		\end{tabular}
\vspace{6mm}

    \begin{tabular}{l@{~~~~~~~~}llll|l}
             \multicolumn{2}{l}{$\leco 2110$}~\\
		& $\notsep{\aset}$ && I & $\itS\aset$~, $\otN\bset\uset$ & 3 \\
		& else & $\notsep{\uset}$ \myand
		$\setmate{\bset}{\bsymb}{\aset}$ & SM & $\uleaf$, $\itS\aset$~,
		$\ostS\bset\bsymb$ & 3 \\
		& & else & W & $\otN\bset\aset$~, $\otN\aset\uset$ & 4
		\end{tabular}
\vspace{6mm}
        
	\begin{tabular}{l@{~~~~~~~~}lllll|l}
		\multicolumn{2}{l}{$\leco 2120$}~\\
			& $\fullyco$ &&& Sb & $\bleaf$, $\itN$~, $\itS\aset$ & 4 \\
			& $\exists \prunedtree$ & $\notsepinpruned{\aset}$ && S & $\uleaf$, $\otN\uset\bset$~, $\itS\aset$ &
			4 \\
			& $\notsep{\aset}$	& $\sep{\uset}$ & $\setmate{\bset}{\bsymb}{\uset}$ & & & \\
			&	& $\sepfrom{\left(\notsepfrom{\aset}{\uset}\right)}{\bset}$ &
				$\setmate{\bset}{\bsymb}{\left(\notsepfrom{\aset}{\uset}\right)}$ & M & $\itS\aset$~, $\itN$~,
				$\ostS\bset\bsymb$ & 4 \\
			& $\sep{\aset}$ & $\notsep{\uset}$ &
			$\setmate{\bset}{\bsymb}{\aset}$ & & &\\
			& &&&~&~\\
			& else &&& W & $\itS\aset$~, $\otN\bset\uset$~, $\tNr\uset\aset$ & 5
		\end{tabular}
\vspace{6mm}


	\begin{tabular}{l@{~~~~~~~~}llll|l}
		\multicolumn{2}{l}{$\leco 2210$}~\\
			& $\fullyco$ && S & $\uleaf$, $\itS\aset$~, $\itS\bset$ & 3 \\
			& $\notsep{\aset}$ && Ia & $\itS\aset$~, $\itS\bset$~,
			$\tNr\uset\bset$ & 4 \\
			& $\notsep{\bset}$ && Ib & $\itS\aset$~, $\itS\bset$~,
			$\tNr\uset\aset$ & 4 \\
			& else & $\setmate{\aset}{\asymb}{\bset}$ & Ma &
			$\itS\bset$~, $\otN\aset\uset$~, $\ostS\aset\asymb$ & 4 \\
			& & $\setmate{\bset}{\bsymb}{\aset}$ & Mb & $\itS\aset$~,
			$\otN\bset\uset$~, $\ostS\bset\bsymb$ & 4 \\
			& & else & W & $\itS\aset$~, $\otN\bset\uset$~, $\tNr\bset\aset$ & 5
        \end{tabular}
        \vspace{6mm}
        
         \begin{tabular}{l@{~~~~~~~~}lllll|l}
            \multicolumn{2}{l}{$\leco 2220$}~\\
			& $\fullyco$ && & I & $\itS\aset$~, $\itS\bset$~, $\itN$ & 4 \\
			& $\notsep{\aset}$ &&& IIa & $\itS\aset$~, \twoti$\otN\bset\uset$ & 5 \\
			& $\notsep{\bset}$ &&& IIb & $\itS\bset$~, \twoti$\otN\aset\uset$ & 5
			\\
			& $\notsep\uset$ & $\setmate{\aset}{\asymb}{\bset}$ && Ma & $\itN$~,
 			$\itS\bset$~, $\itS\aset$~, $\ostSr\aset\asymb$ & 5 \\
 			& & $\setmate{\bset}{\bsymb}{\aset}$ && Mb & $\itN$~, $\itS\aset$~, $\itS\bset$~, $\ostSr\bset\bsymb$ & 5 \\
			& $\fullysep$
				& $\setmate{\aset}{\asymb}{\bset}$ & $\sololeaf$ & SMa & $\uleaf$,
				$\otN\uset\aset$~, $\itS\bset$~, $\ostS\aset\asymb$ & 5 \\
			& 	& & $\setmate{\aset}{\asymb}{\uset}$ & MMa & $\itN$~, $\itS\bset$~,
				\twoti$\ostS\aset\asymb$ & 5\\
			& 	& $\setmate{\bset}{\bsymb}{\aset}$ & $\sololeaf$ & SMb & $\uleaf$,
				$\otN\uset\bset$~, $\itS\aset$~, $\ostS\bset\bsymb$ & 5 \\
			&	& & $\setmate{\bset}{\bsymb}{\uset}$ & MMb & $\itN$~, $\itS\aset$~,
				\twoti$\ostS\bset\bsymb$ & 5 \\
			& & else && W & $\otN\aset\bset$~, $\otN\bset\uset$~, $\otN\uset\aset$ & 6
		\end{tabular}
\vspace{6mm}


 \vspace{6mm}

\noindent{\bf Reduction Bound~\ref{red:three-can-a-b-ab}: residual leaf compositions of subgroup  $\leco110\rab$ + $\leco210\rab$ + $\leco220\rab$}
\vspace{3mm}

   \begin{tabular}{l@{~~~~~~~~}lll|l}
		\multicolumn{2}{l}{$\leco 1101$}~\\
			& any topology & I & $\otS\aset\abset$~, $\tSr\bset\abset$ & 2
		\end{tabular}
\vspace{6mm}

	\begin{tabular}{l@{~~~~~~~~}lll|l}
		\multicolumn{2}{l}{$\leco 1102$}~\\
			& any topology & I & $\otS\aset\abset$~, $\otS\abset\bset$ & 2
		\end{tabular}
\vspace{6mm}

	\begin{tabular}{l@{~~~~~~~~}lll|l}
		\multicolumn{2}{l}{$\leco 2101$}~\\
		& $\notsep{\aset}$ & I & $\itS\aset$~, $\otS\abset\bset$ & 2 \\
		& else & W & $\otS\abset\aset$~, $\otN\aset\bset$ & 3
		\end{tabular}
\vspace{6mm}

	\begin{tabular}{l@{~~~~~~~~}lll|l}
		\multicolumn{2}{l}{$\leco 2102$}~\\
			& any topology & I & \twoti $\otS\aset\abset$~, $\tSr\bset\abset$ & 3
		\end{tabular}
\vspace{6mm}

	\begin{tabular}{l@{~~~~~~~~}lll|l}
		\multicolumn{2}{l}{$\leco 2103$}~\\
		& $\sep{\aset}$ & nR & $\otS\bset\abset$~, \twoti$\otS\aset\abset$ & 3 \\
		& else & $\red$ & $\itS\abset ~\cup\ \leco 2101$ & +1
		\end{tabular}
\vspace{6mm}

        \begin{tabular}{l@{~~~~~~~~}llll|l}
	       \multicolumn{2}{l}{$\leco 2201$}~\\	
			& $\notsep{\aset}$ && Ia & $\itS\aset$~, $\otS\bset\abset$~,
			$\tSr\bset\abset$ & 3 \\
			& $\notsep{\bset}$ && Ib & $\itS\bset$~, $\otS\aset\abset$~,
			$\tSr\aset\abset$ & 3 \\
			& else & $\setmate{\aset}{\asymb}{\bset}$ & Ma & $\otS\aset\abset$~,
			$\itS\bset$~, $\ostS\aset\asymb$ & 3 \\
			& & $\setmate{\bset}{\bsymb}{\aset}$ & Mb & $\otS\bset\abset$~, $\itS\aset$~,
			$\ostS\bset\bsymb$ & 3 \\
			& &else & W & $\otS\aset\abset$~, $\otN\aset\bset$~, $\tSr\bset\abset$ & 4
		\end{tabular}
\vspace{6mm}

	\begin{tabular}{l@{~~~~~~~~}lll|l}
		\multicolumn{2}{l}{$\leco 2202$}~\\
			& $\notsep{\aset}$ & Ia & $\itS\aset$~, \twoti$\otS\bset\abset$ & 3 \\
			& $\notsep{\bset}$ & Ib & $\itS\bset$~, \twoti$\otS\aset\abset$ & 3 \\
			& else & W & $\otS\aset\abset$~, $\otS\abset\bset$~,
                                $\otN\bset\aset$ & 4
        \end{tabular}
\vspace{6mm}

	\begin{tabular}{l@{~~~~~~~~}l@{~~~}l@{~~}l|l}
		\multicolumn{2}{l}{$\leco 2203$}~\\
		& $\sep{\aset}$ \myand
		$\sep{\bset}$ \myand & \multirow{2}{*}{nR}
		& \multirow{2}{*}{\twoti$\otS\aset\abset$~, $\otS\bset\abset$~, $\tSr\bset\abset$} &
		\multirow{2}{*}{4} \\
		& \mycancel{$\setmate{\aset}{\asymb}{\bset}$} 
		\myand
		\mycancel{$\setmate{\bset}{\bsymb}{\aset}$} & &
		&\\[-.4em]
		& else & $\red$ & $\itS\abset ~\cup\ \leco 2201$ & +1
		\end{tabular}
\vspace{6mm}

\begin{tabular}{l@{~~~~~~~~}lll|l}
		\multicolumn{2}{l}{$\leco 2204$}~\\
		& $\sep{\aset}$ \myand
		$\sep{\bset}$ 
		 & nR &
		\twoti$\otS\aset\abset$~, \twoti$\otS\bset\abset$ & 4 \\
		& else & $\red$ & $\itS\abset ~\cup\ \leco 2202$ & +1
		\end{tabular}

 \vspace{6mm}
\noindent{\bf Reduction Bound \ref{red:three-can-a-c-ab}: residual leaf compositions of subgroup $\leco101\rab$ + $\leco102\rab$ + $\leco201\rab$ + $\leco202\rab$}
\vspace{3mm}

	\begin{tabular}{l@{~~~~~~~~}lll|l}
		\multicolumn{2}{l}{$\leco 1011$}~\\
			& $\notsep{\uset}$ & S & $\uleaf$, $\otS\aset\abset$ & 2\\
			& else & W & $\otS\abset\aset$~, $\tNr\uset\aset$ & 3
		\end{tabular}
\vspace{6mm}

	\begin{tabular}{l@{~~~~~~~~}lll|l}
		\multicolumn{2}{l}{$\leco 1012$}~\\
			& any topology & I & $\otS\aset\abset$~, $\otN\abset\uset$ & 3
		\end{tabular}
\vspace{6mm}

	\begin{tabular}{l@{~~~~~~~~}lll|l}
		\multicolumn{2}{l}{$\leco 1021$}~\\
			& $\notsep{\uset}$ & I & $\otS\aset\abset$~, $\itN$ & 3 \\
			& else & W & $\otN\aset\uset$~, $\otN\uset\abset$ & 4
		\end{tabular}
\vspace{6mm}

	\begin{tabular}{l@{~~~~~~~~}llll|l}
		\multicolumn{2}{l}{$\leco 1022$}~\\
			& $\notsep{\uset}$ && I & $\itN$~, $\otS\aset\abset$~,
			$\tSr\abset\aset$ & 4 \\
			& else & $\sololeaf$ & S & $\uleaf$,
                                                   $\otN\uset\abset$~,
                                                   $\otS\abset\aset$ &
                                                                       4 \\
			& & $\setmate{\abset}{\asymb}{\uset}$ & Ma & $\itN$~, $\otS\aset\abset$~,
			$\ostS\abset\asymb$ & 4 \\
			& & $\setmate{\abset}{\bsymb}{\uset}$ & Mb & $\itN$~, $\otS\aset\abset$~,
				$\ostS\abset\bsymb$ & 4 \\
				 
		        & & else & W & $\otN\aset\uset$~, $\otN\uset\abset$~,
				 $\tSr\abset\aset$ & 5
		\end{tabular}
\vspace{6mm}

    \begin{tabular}{l@{~~~~~~~~}lll|l}
		\multicolumn{2}{l}{$\leco 2011$}~\\
			& any topology & I & $\otN\uset\aset$~, $\otS\aset\abset$ & 3
		\end{tabular}
\vspace{6mm}

	\begin{tabular}{l@{~~~~~~~~}lll|l}
		\multicolumn{2}{l}{$\leco 2012$}~\\
			& $\notsep{\uset}$ & S & $\uleaf$, \twoti$\otS\aset\abset$ & 3\\
			& else & W & \twoti$\otS\aset\abset$~, $\tNr\uset\abset$ & 4
		\end{tabular}
\vspace{6mm}

	\begin{tabular}{l@{~~~~~~~~}llll|l}
		\multicolumn{2}{l}{$\leco 2021$}~\\
			& $\notsep{\uset}$ && I & $\itN$~, $\otS\aset\abset$~,
			$\tSr\aset\abset$ & 4\\
			& else & $\sololeaf$ & S & $\uleaf$, $\otN\uset\aset$~, $\otS\aset\abset$ & 4 \\
			& & $\setmate{\aset}{\asymb}{\uset}$ & Ma & $\itN$~, $\otS\abset\aset$~, $\ostS\aset\asymb$ & 4 \\
			& & $\setmate{\abset}{\bsymb}{\uset}$ \myand $\notsep{\aset}$ & Mb
			& $\itN$~, $\itS\aset$~, $\ostS\abset\bsymb$ & 4 \\
			& &else & W & $\otN\abset\uset$~,
                                      $\otN\uset\aset$~,
                                      $\tSr\aset\abset$ & 5
		\end{tabular}
\vspace{6mm}

	\begin{tabular}{l@{~~~~~~~~}lll|l}
		\multicolumn{2}{l}{$\leco 2022$}~\\
		    & $\notsep{\uset}$ & I & $\itN$~, \twoti$\otS\aset\abset$ & 4\\
		    & else & W & $\otS\aset\abset$~, $\otN\abset\uset$~, $\otN\uset\aset$ & 5
		\end{tabular}
\vspace{6mm}

	\begin{tabular}{l@{~~~~~~~~}ll@{\;}l|l}
		\multicolumn{2}{l}{$\leco 2023$}~\\
		& $\sep\aset$ \myand $\sep\uset$ \myand
		$\setmate{\abset}{\bsymb}{\uset}$ & \multirow{2}{*}{M} &
		\multirow{2}{*}{$\itN$~, \twoti$\otS\aset\abset$~, $\ostS\abset\bsymb$} & \multirow{2}{*}{5} \\[-.4em] 
		& \myand \mycancel{$\setmate{\aset}{\asymb}{\uset}$} \myand \mycancel{$\sololeaf$} & & & \\[.3em]
		& else & $\red$ & $\itS\abset ~\cup\ \leco 2021$ & +1
		\end{tabular}
\vspace{6mm}

\subsection*{Enumeration of residual leaf compositions of Group 3}
Residual leaf compositions with four non-empty canonical subtrees. This includes~20 leaf compositions:

\begin{center}
\footnotesize

\begin{tabular}{c|cc|ccc}
\toprule
\multicolumn{6}{c}{Reduction Bound \ref{red:four-can}}\\
\midrule
\multicolumn{1}{c}{Case 1} & \multicolumn{2}{c}{Case 2} &\multicolumn{3}{c}{Case 3}\\
$\leco111\rab$ & $\leco112\rab$ & $\leco211\rab$ & $\leco212\rab$ & $\leco221\rab$ & $\leco222\rab$ \\
\midrule
\leco 1111 &\leco 1121 &\leco 2111 & \leco 2121 &\leco 2211 &\leco 2221\\
\leco 1112 &\leco 1122 &\leco 2112 & \leco 2122 &\leco 2212 &\leco 2222\\
 &\leco 1123 &\leco 2113 & \leco 2123 &\leco 2213 &\leco 2223\\
& & & \leco 2124 &\leco 2214 &\leco 2224\\
\bottomrule
\end{tabular}
\end{center}

\vspace{6mm}

    \begin{tabular}{l@{~~~~~~~~}lll|l}
		\multicolumn{2}{l}{\leco 1111}~\\
		& $\notsepfrom{\left(\connect{\aset}{\abset}\right)}{\left(\connect{\bset}{\uset}\right)}$ & Ia &
		$\otS\aset\abset$~, $\otN\bset\uset$ & 3 \\
		& $\notsepfrom{\left(\connect{\bset}{\abset}\right)}{\left(\connect{\aset}{\uset}\right)}$ (else) & Ib &
		$\otS\bset\abset$~, $\otN\aset\uset$ & 3
		\end{tabular}
\vspace{6mm}

	\begin{tabular}{l@{~~~~~~~~}lll|l}
		\multicolumn{2}{l}{\leco 1112}~\\
			& $\notsep{\uset}$ & S & $\uleaf$, $\otS\aset\abset$~,
			$\otS\abset\bset$ & 3 \\
			& else & W & $\otS\aset\abset$~, $\otS\abset\bset$~, $\tNr\uset\abset$ & 4
		\end{tabular}
\vspace{6mm}

\begin{tabular}{l@{~~~~~~~~}lllll|l}
		\multicolumn{2}{l}{\leco 1121}~\\
		& $\notsep{\uset}$ & & & I & $\itN$~, $\otS\aset\abset$~,
		$\tSr\bset\abset$ & 4 \\
		& else & $\notsepfrom{\left(\sepfrom{\aset}{\uset}\right)}{\left(\connect{\bset}{\abset}\right)}$ 
		& $\sololeaf$ & S1 & $\uleaf$, $\otN\uset\aset$~, $\otS\bset\abset$ & 4 \\
			& & & $\setmate{\aset}{\asymb}{\uset}$ & Ma & $\itN$~,
			$\otS\bset\abset$~, $\ostS\aset\asymb$ & 4 \\ 
			& & $\notsepfrom{\left(\sepfrom{\bset}{\uset}\right)}{\left(\connect{\aset}{\abset}\right)}$
		    & $\sololeaf$ & S2 & $\uleaf$, $\otN\uset\bset$~, $\otS\aset\abset$ & 4\\
			& & & $\setmate{\bset}{\bsymb}{\uset}$ & Mb & $\itN$~,
			$\otS\aset\abset$~, $\ostS\bset\bsymb$ & 4 \\ 
			& & else & & W & $\otN\aset\uset$~, $\otN\uset\bset$~, $\tSr\abset\aset$ & 5
		\end{tabular}
\vspace{6mm}

	\begin{tabular}{l@{~~~~~~~~}lll|l}
		\multicolumn{2}{l}{\leco 1122}~\\
			& $\notsep{\uset}$ & I & $\itN$~, $\otS\aset\abset$~, $\otS\bset\abset$ & 4 \\
			& else & W & $\otN\aset\uset$~, $\otN\uset\abset$~, $\otS\abset\bset$ & 5
		\end{tabular}
\vspace{6mm}

	\begin{tabular}{l@{~~~~~~~~}l@{~~~}ll|l}
		\multicolumn{2}{l}{\leco 1123}~\\
	& $\sep{\left(\sepfrom{\uset}{\aset}\right)}$ \myand $\setmate{\abset}{\asymb}{\uset}$ \myand $\mycancel{\setmate{\bset}{\bsymb}{\uset}}$& Ma & $\itN$~,$\otS\aset\abset$~,
			$\otS\abset\bset$~, $\ostS\abset\asymb$ & 5 \\
	& $\sep{\left(\sepfrom{\uset}{\bset}\right)}$ \myand $\setmate{\abset}{\bsymb}{\uset}$ \myand $\mycancel{\setmate{\aset}{\asymb}{\uset}}$ & Mb & $\itN$~,$\otS\aset\abset$~,
			$\otS\abset\bset$~, $\ostS\abset\bsymb$ & 5 \\
	& else  & $\red$ & $\itS\abset ~\cup\ \leco1121$ & +1
		\end{tabular}
	\vspace{6mm}

\begin{tabular}{l@{~~~~~~~~}lll|l}
		\multicolumn{2}{l}{\leco 2111}~\\
			& $\notsep{\uset}$ \myand $\notsep{\aset}$ \myand
			$\notsep{\left(\connect{\aset}{\uset}\right)}$ & S & $\uleaf$, $\itS\aset$~,
			$\otS\abset\bset$ & 3 \\
			& else & W & $\otN\uset\aset$~, $\otS\aset\abset$~,
			$\tSr\bset\abset$ & 4
		\end{tabular}
		
\vspace{6mm}

	\begin{tabular}{l@{~~~~~~~~}lll|l}
		\multicolumn{2}{l}{\leco 2112}~\\
			& any topology & I & $\otS\aset\abset$~, $\otS\bset\abset$~, $\otN\aset\uset$ & 4
		\end{tabular}
\vspace{6mm}

	\begin{tabular}{l@{~~~~~~~~}lll|l}
		\multicolumn{2}{l}{\leco 2113}~\\
			& $\sep\aset$ \myand $\notsep{\uset}$ & \multirow{2}{*}{S} &  \multirow{2}{*}{$\uleaf$, \twoti$\otS\aset\abset$~, $\otS\abset\bset$} &  \multirow{2}{*}{4} \\
			& $\sep{\left(\notsepfrom{\aset}{\uset}\right)}$  & &  &  \\
			& else & $\red$ & $\itS\abset ~\cup\ \leco2111$ & +1
		\end{tabular}
\vspace{6mm}

	\begin{tabular}{l@{~~~~~~~~}lll|l}
		\multicolumn{2}{l}{\leco 2121}~\\
			& $\notsep{\uset}$ \myand $\notsep{\aset}$ \myand $\notsep{\left(\connect{\aset}{\uset}\right)}$ & I
			& $\itN$~, $\itS\aset$~, $\otS\abset\bset$ & 4 \\
			& else & W & $\otS\abset\aset$~, $\otN\aset\uset$~, $\otN\uset\bset$ & 5
		\end{tabular}
\vspace{6mm}

	\begin{tabular}{l@{~~~~~~~~}llll|l}
		\multicolumn{2}{l}{\leco 2122}~\\
			& $\notsep{\uset}$ && I & $\itN$~, \twoti$\otS\aset\abset$~,
			$\tSr\bset\abset$ & 5 \\
			& else & $\sololeaf$ & S & $\uleaf$, $\otN\uset\aset$~, $\otS\aset\abset$~,
			$\otS\bset\abset$ & 5 \\
			&	& $\setmate{\aset}{\asymb}{\uset}$ & Ma & $\itN$~, $\otS\aset\abset$~,
				$\otS\bset\abset$~, $\ostS\aset\asymb$ & 5 \\
			&	& $\setmate{\abset}{\bsymb}{\uset}$ \myand $\notsep{\aset}$ & Mb1~~ & $\itN$~,
				$\itS\aset$~, $\otS\bset\abset$~, $\ostS\abset\bsymb$ & 5 \\
			&	& $\setmate{\bset}{\bsymb}{\uset}$ \myand
				$\notsep{\left(\connect{\bset}{\uset}\right)}$ & Mb2 &
				$\itN$~, \twoti$\otS\aset\abset$~, $\ostS\bset\bsymb$ & 5 \\
			&	& else & W & \twoti$\otS\aset\abset$~, $\otN\bset\uset$~, $\tNr\uset\aset$ & 6
		\end{tabular}
\vspace{6mm}

	\begin{tabular}{l@{~~~~~~~~}l@{~~~~~~}ll|l}
		\multicolumn{2}{l}{\leco 2123}~\\
			& $\sep\aset$ \myand $\notsep{\uset}$ & \multirow{2}{*}{I} & \multirow{2}{*}{ $\itN$~, \twoti$\otS\aset\abset$~, $\otS\abset\bset$}  & \multirow{2}{*}{5} \\
			& $\sep{\left(\notsepfrom{\aset}{\uset}\right)}$ &  & \\
			
			& else  & $\red$ & $\itS\abset ~\cup\ \leco2121$ & +1
		\end{tabular}
\vspace{6mm}

	\begin{tabular}{l@{~~~~~~~~}lll|l@{}}
		\multicolumn{2}{l}{\leco 2124}~\\
			& $\sep\uset$ \myand $\sep\aset$ & & & \\[-.4em]
			& \myand $\sep{\left(\connect{\bset}{\uset}\right)}$ \myand \mycancel{$\sololeaf$} & M & $\itN$~, \twoti$\otS\aset\abset$~, $\otS\abset\bset$~, $\ostS\abset\bsymb$ & 6 \\ [-.4em]
			& \myand \mycancel{$\setmate{\aset}{\asymb}{\uset}$} \myand $\setmate{\abset}{\bsymb}{\uset}$ & & &\\
			& $~~~$&~&~&~\\
			& else & $\red$ & $\itS\abset ~\cup\ \leco2122$ & +1
		\end{tabular}
\vspace{6mm}

\begin{tabular}{l@{~~~~~~~~}llll|l}
		\multicolumn{2}{l}{\leco 2211}~\\
			& $\notsep{\aset}$ & & Ia & $\itS\aset$~, $\otN\uset\bset$~, $\otS\bset\abset$ & 4 \\
			& $\notsep{\bset}$ & & Ib & $\itS\bset$~, $\otN\uset\aset$~, $\otS\aset\abset$ & 4 \\
			& $\notsep{\uset}$ &
			$\setmate{\aset}{\asymb}{\bset}$ & SMa & $\uleaf$, $\otS\abset\aset$~,
			$\itS\bset$~, $\ostS\aset\asymb$ & 4 \\
			& 	& $\setmate{\bset}{\bsymb}{\aset}$ & SMb & $\uleaf$, $\otS\abset\bset$~,
				$\itS\aset$~, $\ostS\bset\bsymb$ & 4 \\
			& else & & W & $\otS\abset\aset$~, $\otN\aset\bset$~, $\otN\bset\uset$ & 5
		\end{tabular}
\vspace{6mm}

\begin{tabular}{l@{~~~~~~~~}llll|l}
		\multicolumn{2}{l}{\leco 2212}~\\
			& $\notsep{\uset}$ & $\notsep{\aset}$ \myand
			$\notsep{\left(\connect{\aset}{\uset}\right)}$ & S1 & $\uleaf$, $\itS\aset$~,
			\twoti$\otS\bset\abset$ & 4 \\
			& & $\notsep{\bset}$ \myand $\notsep{\left(\connect{\bset}{\uset}\right)}$ & S2 &
			$\uleaf$, $\itS\bset$~, \twoti$\otS\aset\abset$ & 4 \\
			& else & & W & $\otN\uset\aset$~, \twoti$\otS\bset\abset$~, $\tSr\aset\abset$ & 5
		\end{tabular}
\vspace{6mm}

	\begin{tabular}{l@{~~~~~~~~}llll|l}
		\multicolumn{2}{l}{\leco 2213}~\\
			& \multicolumn{2}{l}{$\sep\aset$ \myand $\sep\bset$}\\
			& & $\sep\uset$ & \multirow{2}{*}{nR} &
			\multirow{2}{*}{\twoti$\otS\aset\abset$~, $\otS\abset\bset$~, $\otN\bset\uset$} & \multirow{2}{*}{5} \\[-.4em] 
			& & $\notsep\uset$ \myand \mycancel{$\setmate{\aset}{\asymb}{\bset}$} \myand \mycancel{$\setmate{\bset}{\bsymb}{\aset}$} & & & \\[.3em]
			& else & & $\red$ & $\itS\abset ~\cup\ \leco2211$ & +1
		\end{tabular}
\vspace{6mm}

	\begin{tabular}{l@{~~~~~~~~}ll@{~~~~~}ll|l}
		\multicolumn{2}{l}{\leco 2214}~\\
		& \multicolumn{2}{l}{$\notsep{\uset}$ \myand} \\
		&    & \{~$\sep{\aset}$ \myor $\sep{\left(\notsepfrom{\aset}{\uset}\right)}$~\} \myand&
		    \multirow{2}{*}{S} & \multirow{2}{*}{$\uleaf$, \twoti$\otS\aset\abset$~,
		    \twoti$\otS\bset\abset$} & \multirow{2}{*}{5} \\
		  &  & \{~$\sep{\bset}$ \myor
                       $\sep{\left(\notsepfrom{\bset}{\uset}\right)}$~\} & & &\\ 
			& else & & $\red$ & $\itS\abset ~\cup\ \leco 2212$ & +1
		\end{tabular}
\vspace{6mm}

        \begin{tabular}{l@{~~~~~~~~}lllll|l}
		\multicolumn{2}{l}{\leco 2221}~\\
			&  $\exists \prunedtree$ & $\notsepinpruned{\aset}$ & & S1 & $\uleaf$, $\itS\aset$~, $\otN\bset\uset$~, $\otS\bset\abset$& 5 \\
			& & $\notsepinpruned{\bset}$ & & S2 & $\uleaf$, $\itS\bset$~, $\otN\aset\uset$~, $\otS\aset\abset$& 5 \\
		 	& $\notsep{\aset}$	& $\notsep{\uset}$~~~\myand & $\notsep{\left(\notsepfrom{\aset}{\uset}\right)}$ & Ia &
		 		$\itS\aset$~, $\itN$~, $\otS\bset\abset$~, $\tSr\bset\abset$ & 5 \\
		 	& 	& $\setmate{\bset}{\bsymb}{\uset}$ & & \multirow{1}{*}{Mb} & \multirow{2}{*}{$\itS\aset$~,  $\itN$~,
				$\otS\bset\abset$~, $\ostS\bset\bsymb$} & \multirow{2}{*}{5} \\
		 	 & $\notsep{\uset}$ & $\setmate{\bset}{\bsymb}{\aset}$ & &
		 	 \multirow{1}{*}{Mb} & &  \\
		 	&	& $\setmate{\aset}{\asymb}{\bset}$ & & \multirow{1}{*}{Ma} & \multirow{2}{*}{$\itS\bset$~, $\itN$~,
				$\otS\aset\abset$~, $\ostS\aset\asymb$} & \multirow{2}{*}{5} \\
			& $\notsep{\bset}$ &
                                             $\setmate{\aset}{\asymb}{\uset}$ & & \multirow{1}{*}{Ma} & &  \\
			& 	& $\notsep{\uset}$~~~\myand & $\notsep{\left(\notsepfrom{\bset}{\uset}\right)}$ & Ib & $\itS\bset$~, $\itN$~,
				$\otS\aset\abset$~, $\tSr\aset\abset$ & 5 \\
		 & else & & & W & $\otN\aset\uset$~, $\otS\aset\abset$~, $\otN\bset\uset$~,
		 $\tSr\bset\abset$ & 6
		\end{tabular}
\vspace{6mm}

	\begin{tabular}{l@{~~~~~~~~}llll|l}
		\multicolumn{2}{l}{\leco 2222}~\\
			& $\notsep{\uset}$ ~~~~~~& $\notsep{\aset}$ \myand
			$\notsep{\left(\connect{\aset}{\uset}\right)}$ & Ia & $\itN$~, $\itS\aset$~, \twoti$\otS\bset\abset$ & 5 \\
			 & & $\notsep{\bset}$ \myand $\notsep{\left(\connect{\bset}{\uset}\right)}$ & Ib & $\itN$~, $\itS\bset$~, \twoti$\otS\aset\abset$ & 5 \\
			& else & & W & $\otN\uset\aset$~, $\otS\aset\abset$~, $\otS\abset\bset$~,
			$\otN\bset\uset$ & 6
		\end{tabular}
\vspace{6mm}

   \begin{tabular}{l@{~~~~~~~~}l@{~~~~~~}ll|l@{}}
		\multicolumn{2}{l}{\leco 2223}~\\
		& \{~$\sep{\aset}$ \myor $\sep{\left(\connect{\aset}{\uset}\right)}$~\} \myand &~&~\\
		& \{~$\sep{\bset}$ \myor $\sep{\left(\connect{\bset}{\uset}\right)}$~\} &~&~\\
		& ~~~~~~~~~~~~~~$\notsep{\uset}$ &I & $\itN$~, \twoti$\otS\aset\abset$~, $\otS\bset\abset$~, $\tSr\bset\abset$ & 6 \\
		& ~~~~~~~~~~~~~~$\sololeaf$ &S & $\uleaf$, $\otN\uset\aset$~, $\otS\aset\abset$~, \twoti$\otS\abset\bset$ & 6 \\
               & $~~~$&~&~&~\\
               & $\sep\aset$ \myand $\sep\bset$ \myand $\sep\uset$ &~&~\\
	       & ~~~~~~~~~~~~~~$\setmate{\aset}{\asymb}{\uset}$ & Ma & $\itN$~,
			\twoti$\otS\bset\abset$~, $\otS\aset\abset$~, $\ostS\aset\asymb$ & 6 \\
	       & ~~~~~~~~~~~~~~$\setmate{\bset}{\bsymb}{\uset}$ & Mb & $\itN$~,
			\twoti$\otS\aset\abset$~, $\otS\bset\abset$~, $\ostS\bset\bsymb$ & 6 \\
	       & $~~~$&~&~&~\\
	       & else & $\red$ & $\itS\abset ~\cup\ \leco2221$ & +1
		\end{tabular}
\vspace{6mm}

	\begin{tabular}{l@{~~~~~~~~}ll@{~~~~~}l@{~~}l|l}
		\multicolumn{2}{l}{\leco 2224}~\\
		& \multicolumn{2}{l}{$\notsep{\uset}$ \myand} \\
			& & \{~$\sep{\aset}$ \myor $\sep{\left(\notsepfrom{\aset}{\uset}\right)}$~\} \myand&
			\multirow{2}{*}{I} & \multirow{2}{*}{$\itN$~, \twoti$\otS\aset\abset$~,
			\twoti$\otS\bset\abset$} & \multirow{2}{*}{6}\\
			& & \{~$\sep{\bset}$ \myor $\sep{\left(\notsepfrom{\bset}{\uset}\right)}$~\} & & &\\
			& else & & $\red$ & $\itS\abset ~\cup\ \leco 2222$ & +1
		\end{tabular}
		
\vspace{6mm}

\vspace{20mm}

\begin{figure}[ht]
	\setlength{\unitlength}{0.75pt}
	\lfsize
	\centering
	{\bf Non-reducible topologies of leaf composition \boldmath$\leco 2223$}\\
	\vspace{2mm}
    \begin{tabular}{c@{~~~~~}c@{~~~~~}c}
		\begin{tabular}{c}
       {\bf (i) \boldmath$\subtreeof{\lc}$ non-isolated}\\
    	\begin{picture}(200,190)
		\put(100,160){\PenTree{\badvertex}
		     {\UnTree{\badvertex}
		       {\BiTreeLeft{\goodvertex}
		        {\Leaf{\badAvertex\bottomlabel{$a_1$}}}
		        {\Leaf{\badAvertex\bottomlabel{$a_2$}}}
		       }
		     }
		     {\UnTree{\badvertex}{\badvertex\bottomlabel{$c_1$}}}
		     {\UnTree{\badvertex}{\badvertex\cornerlabel{$c_2$}}}
		     {\UnTree{\badvertex}
		        {\BiTreeLeft{\goodvertex}
		          {\Leaf{\badBvertex\bottomlabel{$b_1$}}}
		          {\Leaf{\badBvertex\bottomlabel{$b_2$}}}
		        }
		     }
		        {\TriTreeRight{\goodvertex}
		          {\Leaf{\badABvertex\bottomlabel{$x_1$}}}
		          {\Leaf{\badABvertex\bottomlabel{$x_2$}}}
		          {\Leaf{\badABvertex\bottomlabel{$x_3$}}}
		        }
		    }
		\put(100,45){\makebox(0,0){$\optcost{6}$}}
		\put(100,25){\makebox(0,0){$c_{1\!} \travtwo c_2$, $a_1 \travone x_1$, $a_2 \travone x_2$,}}
		\put(100, 5){\makebox(0,0){$b_1 \travone x_3$, $b_2 \travone x_3$}}
		\end{picture}
		\end{tabular} & 
		\begin{tabular}{c}
         {\bf (ii) \boldmath$\subtreeof{\la \cup \lc}$ isolated and solo leaf}\\
		\begin{picture}(180,190)
		\put(90,180){\ParTreesWide
		       {\UnTree{\badvertex}
		        {\TriTree{\goodvertex}
		          {\Leaf{\badAvertex\bottomlabel{$a_1$}}}
		          {\Leaf{\badAvertex\bottomlabel{$a_2$}}}
		          {\UnTree{\badvertex}
		            {\BiTreeNarrow{\badvertex}
		             {\UnTree{\badvertex}{\badvertex\bottomlabel{$c$}}}
		             {\Leaf{\badvertex\bottomlabel{$s$}}}
		            }
		          }
		       }
		      }
		      {\BiTreeWide{\badvertex} 
		       {\UnTree{\badvertex}
		        {\BiTreeRight{\goodvertex}
		          {\Leaf{\badBvertex\bottomlabel{$b_1$}}}
		          {\Leaf{\badBvertex\bottomlabel{$b_2$}}}
		        }
		       } 
		        {\TriTree{\goodvertex}
		          {\Leaf{\badABvertex\bottomlabel{~$x_1$}}}
		          {\Leaf{\badABvertex\bottomlabel{$x_2$}}}
		          {\Leaf{\badABvertex\bottomlabel{$x_3$}}}
		        }
		      } 
		    }
	    \put(90,45){\makebox(0,0){$\optcost{6}$}}
		\put(90,25){\makebox(0,0){$\costonepath{s}$, $c \travtwo a_1$, $a_2 \travone x_3$,}}
		\put(90, 5){\makebox(0,0){$b_1 \travone x_1$, $b_{2} \travone x_2$}}
		\end{picture}
		\end{tabular}  & 
		\begin{tabular}{c}
        {\bf (iii) tag mate at \boldmath$\extsubtreeof{\lc}$} \\
		\begin{picture}(180,190)
		\put(90,160)
		  {\QuaTreeWide{\badvertex}
		       {\UnTree{\badvertex}
		        {\BiTreeLeft{\goodvertex}
		          {\Leaf{\badAvertex\bottomlabel{$a_1$}}}
		          {\Leaf{\badAvertex\bottomlabel{$a_2$}}}
		        }
		       } 
		       {\UnTree{\badBvertex\rightlabel{$m$}}
		        {\BiTreeLeft{\goodvertex}
		          {\UnTree{\badvertex}{\badvertex\bottomlabel{$c_1$}}}
		          {\UnTree{\badvertex}{\badvertex\bottomlabel{$c_2$}}}
		        }
		       } 
		       {\UnTree{\badvertex}
		        {\BiTreeLeft{\goodvertex}
		          {\Leaf{\badBvertex\bottomlabel{$b_1$}}}
		          {\Leaf{\badBvertex\bottomlabel{$b_2$}}}
		        }
		       } 
		        {\TriTree{\goodvertex}
		          {\Leaf{\badABvertex\bottomlabel{$x_1$}}}
		          {\Leaf{\badABvertex\bottomlabel{$x_2$}}}
		          {\Leaf{\badABvertex\bottomlabel{$x_3$}}}
		        }
		    }
		\put(90,45){\makebox(0,0){$\optcost{6}$}}
		\put(90,25){\makebox(0,0){$c_{1\!} \travtwo c_2$, $a_1 \travone x_1$, $a_{2} \travone x_2$,}}
		\put(90, 5){\makebox(0,0){$b_1 \travone x_3$, $b_{2\!} \semitravone m$}}
		\end{picture}
		\end{tabular}
	   \endtab \\
	\caption{All non-reducible topologies of leaf composition \leco 2223 have a bad link between $\subtreeof{\lab}$ and $\subtreeof{\la}$ and a bad link between $\subtreeof{\lab}$ and $\subtreeof{\lb}$. A symmetric case of (iii) 
	is omitted.}
	\label{fig:leco2223nR}
	\end{figure}

\begin{figure}[ht]
    \setlength{\unitlength}{0.75pt}
	\centering
	\lfsize
	{\bf Topologies of leaf composition \boldmath$\leco 2223$ reducible to \boldmath$\leco 2221$}\\
	\vspace{3mm}
		
	\begin{tabular}{c@{~~}c@{~~}c}  
    \multicolumn{3}{c}{\bf (i) \boldmath$\subtreeof{\la}$ non-isolated and solo leaf}\\
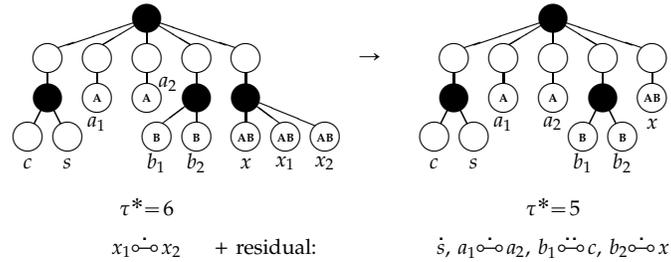
    \begin{picture}(180,130)
		\put(90,120){\PenTree{\goodvertex}
		     {\UnTree{\badvertex}
		       {\BiTreeNarrow{\goodvertex}
		        {\Leaf{\badvertex\bottomlabel{$c$}}}
		        {\Leaf{\badvertex\bottomlabel{$s$}}}
		       }
		     }
		     {\UnTree{\badvertex}{\badAvertex\bottomlabel{$a_1$}}}
		     {\UnTree{\badvertex}{\badvertex\cornerlabel{$a_2$}\tagA}}
		     {\UnTree{\badvertex}
		        {\BiTreeLeft{\goodvertex}
		          {\Leaf{\badBvertex\bottomlabel{$b_1$}}}
		          {\Leaf{\badBvertex\bottomlabel{$b_2$}}}
		        }
		     }
		     {\UnTree{\badvertex}
		        {\TriTreeRight{\goodvertex}
		          {\Leaf{\badABvertex\bottomlabel{$x$}}}
		          {\Leaf{\badABvertex\bottomlabel{$x_1$}}}
		          {\Leaf{\badABvertex\bottomlabel{$x_2$}}}
		        }
		     }
		    }
		\put(90,25){\makebox(0,0){$\optcost{6}$}}
		\put(90,5){\makebox(0,0){$x_1 \travone x_2$}}
		\put(150, 5){\makebox(0,0){+ residual:}}
		\end{picture}&
		\begin{picture}(20,130)
		\put(15,100){\makebox(0,0){$\rightarrow$}}
		\end{picture} &
		\begin{picture}(160,130)
		\put(80,120){\PenTree{\goodvertex}
		     {\UnTree{\badvertex}
		       {\BiTreeNarrow{\goodvertex}
		        {\Leaf{\badvertex\bottomlabel{$c$}}}
		        {\Leaf{\badvertex\bottomlabel{$s$}}}
		       }
		     }
		     {\UnTree{\badvertex}{\badAvertex\bottomlabel{$a_1$}}}
		     {\UnTree{\badvertex}{\badAvertex\bottomlabel{$a_2$}}}
		     {\UnTree{\badvertex}
		        {\BiTreeNarrow{\goodvertex}
		          {\Leaf{\badBvertex\bottomlabel{$b_1$}}}
		          {\Leaf{\badBvertex\bottomlabel{$b_2$}}}
		        }
		     }
		     {\UnTree{\badvertex}{\badABvertex\bottomlabel{$x$}}}
		    }
		\put(80,25){\makebox(0,0){$\optcost{5}$}}
		\put(80,5){\makebox(0,0){$\costonepath{s}$, $a_1 \travone a_2$, $b_{1\!} \travtwo c$, $b_2 \travone x$}}
		\end{picture}
		\end{tabular}
		
		\vspace{3mm}
		
		\begin{tabular}{c@{~~}c@{~~}c}
		\multicolumn{3}{c}{\bf (ii) \boldmath$\subtreeof{\la}$ and \boldmath$\subtreeof{\lc}$ non-isolated}\\
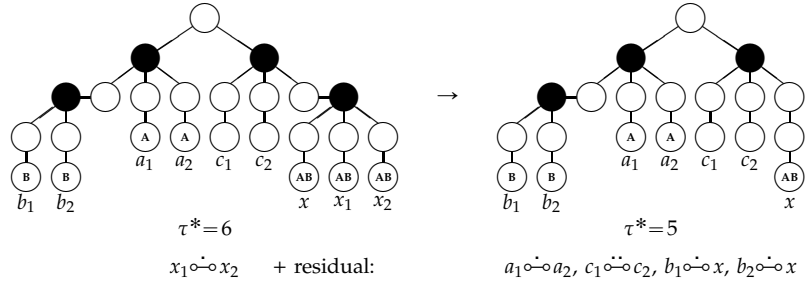
		\begin{picture}(200,140)
		\put(100,130){\BiTreeWide{\badvertex}
		     {\TriTree{\goodvertex}
		      {\UnTreeLeft{\badvertex}
		       {\BiTreeLeft{\goodvertex}
		        {\UnTree{\badvertex}{\badBvertex\bottomlabel{$b_1$}}}
		        {\UnTree{\badvertex}{\badBvertex\bottomlabel{$b_2$}}}
		       }
		      }
		      {\UnTree{\badvertex}{\badAvertex\bottomlabel{$a_1$}}}
		      {\UnTree{\badvertex}{\badAvertex\bottomlabel{$a_2$}}}
		     }
		     {\TriTree{\goodvertex}
		      {\UnTree{\badvertex}{\badvertex\bottomlabel{$c_1$}}}
		      {\UnTree{\badvertex}{\badvertex\bottomlabel{$c_2$}}}
		      {\UnTreeRight{\badvertex}
		       {\TriTree{\goodvertex}
		        {\UnTree{\badvertex}{\badABvertex\bottomlabel{$x$}}}
		        {\UnTree{\badvertex}{\badABvertex\bottomlabel{$x_1$}}}
		        {\UnTree{\badvertex}{\badABvertex\bottomlabel{$x_2$}}}
		       }
		      }
		     }
		    }
		\put(100,25){\makebox(0,0){$\optcost{6}$}}
		\put(100,5){\makebox(0,0){$x_1 \travone x_2$}}
		\put(160, 5){\makebox(0,0){+ residual:}}
		\end{picture}&
		\begin{picture}(20,140)
		\put(15,90){\makebox(0,0){$\rightarrow$}}
		\end{picture} &
		\begin{picture}(180,140)
		\put(110,130){\BiTreeWide{\badvertex}
		     {\TriTree{\goodvertex}
		      {\UnTreeLeft{\badvertex}
		       {\BiTreeLeft{\goodvertex}
		        {\UnTree{\badvertex}{\badBvertex\bottomlabel{$b_1$}}}
		        {\UnTree{\badvertex}{\badBvertex\bottomlabel{$b_2$}}}
		       }
		      }
		      {\UnTree{\badvertex}{\badAvertex\bottomlabel{$a_1$}}}
		      {\UnTree{\badvertex}{\badAvertex\bottomlabel{$a_2$}}}
		     }
		     {\TriTree{\goodvertex}
		      {\UnTree{\badvertex}{\badvertex\bottomlabel{$c_1$}}}
		      {\UnTree{\badvertex}{\badvertex\bottomlabel{$c_2$}}}
		      {\UnTree{\badvertex}
		       {\UnTree{\badvertex}{\badABvertex\bottomlabel{$x$}}}
		      }
		     }
		    }
		\put(90,25){\makebox(0,0){$\optcost{5}$}}
		\put(90,5){\makebox(0,0){$a_1 \travone a_2$, $c_{1\!} \travtwo c_2$, $b_1 \travone x$, $b_2 \travone x$}}
		\end{picture}
		\end{tabular}
		
		\vspace{3mm}
		
		\begin{tabular}{c@{~~}c@{~~}c} 
	\multicolumn{3}{c}{\bf (iii) \boldmath$\subtreeof{\la}$ non-isolated and \boldmath$\bsymb$-mate at \boldmath$\extsubtreeof{\lc}$}\\
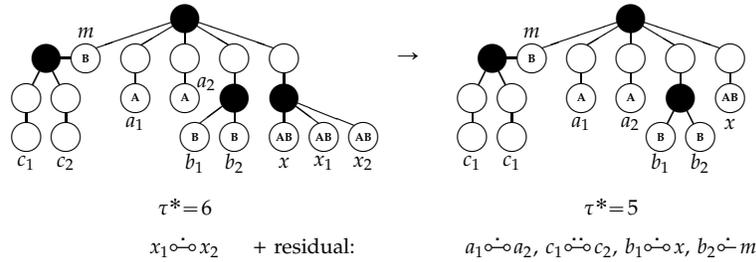
		\begin{picture}(180,130)
		\put(90,120){\PenTree{\goodvertex}
		     {\UnTreeLeft{\badBvertex\toplabel{$m$}}
		       {\BiTreeNarrow{\goodvertex}
		        {\UnTree{\badvertex}{\badvertex\bottomlabel{$c_1$}}}
		        {\UnTree{\badvertex}{\badvertex\bottomlabel{$c_2$}}}
		       }
		     }
		     {\UnTree{\badvertex}{\badAvertex\bottomlabel{$a_1$}}}
		     {\UnTree{\badvertex}{\badvertex\cornerlabel{$a_2$}\tagA}}
		     {\UnTree{\badvertex}
		        {\BiTreeLeft{\goodvertex}
		          {\Leaf{\badBvertex\bottomlabel{$b_1$}}}
		          {\Leaf{\badBvertex\bottomlabel{$b_2$}}}
		        }
		     }
		     {\UnTree{\badvertex}
		        {\TriTreeRight{\goodvertex}
		          {\Leaf{\badABvertex\bottomlabel{$x$}}}
		          {\Leaf{\badABvertex\bottomlabel{$x_1$}}}
		          {\Leaf{\badABvertex\bottomlabel{$x_2$}}}
		        }
		     }
		    }
		\put(90,25){\makebox(0,0){$\optcost{6}$}}
		\put(90,5){\makebox(0,0){$x_1 \travone x_2$}}
		\put(150, 5){\makebox(0,0){+ residual:}}
		\end{picture}&
		\begin{picture}(20,130)
		\put(15,100){\makebox(0,0){$\rightarrow$}}
		\end{picture} &
		\begin{picture}(180,130)
		\put(100,120){\PenTree{\goodvertex}
		     {\UnTreeLeft{\badBvertex\toplabel{$m$}}
		       {\BiTreeNarrow{\goodvertex}
		        {\UnTree{\badvertex}{\badvertex\bottomlabel{$c_1$}}}
		        {\UnTree{\badvertex}{\badvertex\bottomlabel{$c_1$}}}
		       }
		     }
		     {\UnTree{\badvertex}{\badAvertex\bottomlabel{$a_1$}}}
		     {\UnTree{\badvertex}{\badAvertex\bottomlabel{$a_2$}}}
		     {\UnTree{\badvertex}
		        {\BiTreeNarrow{\goodvertex}
		          {\Leaf{\badBvertex\bottomlabel{$b_1$}}}
		          {\Leaf{\badBvertex\bottomlabel{$b_2$}}}
		        }
		     }
		     {\UnTree{\badvertex}{\badABvertex\bottomlabel{$x$}}}
		    }
		\put(90,25){\makebox(0,0){$\optcost{5}$}}
		\put(90,5){\makebox(0,0){$a_1 \travone a_2$, $c_{1\!} \travtwo c_2$, $b_1 \travone x$, $b_{2\!} \semitravone m$}}
		\end{picture}
		\end{tabular}
		
		\vspace{3mm}
		
		\begin{tabular}{c@{~~}c@{~~}c}
		\multicolumn{3}{c}{\bf (iv) \boldmath$\subtreeof{\lc}$ non-isolated and \boldmath$\bsymb$-mate at \boldmath$\extsubtreeof{\la}$}\\
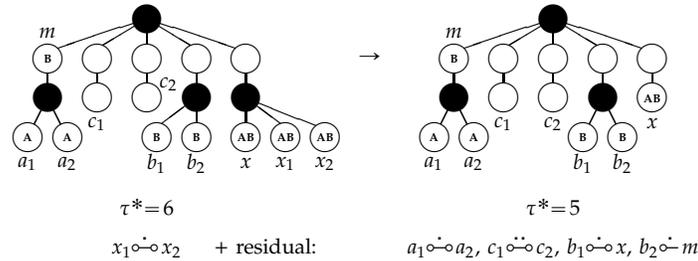
    \begin{picture}(180,130)
		\put(90,120){\PenTree{\goodvertex}
		     {\UnTree{\badBvertex\toplabel{$m$}}
		       {\BiTreeNarrow{\goodvertex}
		        {\Leaf{\badAvertex\bottomlabel{$a_1$}}}
		        {\Leaf{\badAvertex\bottomlabel{$a_2$}}}
		       }
		     }
		     {\UnTree{\badvertex}{\badvertex\bottomlabel{$c_1$}}}
		     {\UnTree{\badvertex}{\badvertex\cornerlabel{$c_2$}}}
		     {\UnTree{\badvertex}
		        {\BiTreeLeft{\goodvertex}
		          {\Leaf{\badBvertex\bottomlabel{$b_1$}}}
		          {\Leaf{\badBvertex\bottomlabel{$b_2$}}}
		        }
		     }
		     {\UnTree{\badvertex}
		        {\TriTreeRight{\goodvertex}
		          {\Leaf{\badABvertex\bottomlabel{$x$}}}
		          {\Leaf{\badABvertex\bottomlabel{$x_1$}}}
		          {\Leaf{\badABvertex\bottomlabel{$x_2$}}}
		        }
		     }
		    }
		\put(90,25){\makebox(0,0){$\optcost{6}$}}
		\put(90,5){\makebox(0,0){$x_1 \travone x_2$}}
		\put(150, 5){\makebox(0,0){+ residual:}}
		\end{picture}&
		\begin{picture}(20,130)
		\put(15,100){\makebox(0,0){$\rightarrow$}}
		\end{picture} &
		\begin{picture}(160,130)
		\put(80,120){\PenTree{\goodvertex}
		     {\UnTree{\badBvertex\toplabel{$m$}}
		       {\BiTreeNarrow{\goodvertex}
		        {\Leaf{\badAvertex\bottomlabel{$a_1$}}}
		        {\Leaf{\badAvertex\bottomlabel{$a_2$}}}
		       }
		     }
		     {\UnTree{\badvertex}{\badvertex\bottomlabel{$c_1$}}}
		     {\UnTree{\badvertex}{\badvertex\bottomlabel{$c_2$}}}
		     {\UnTree{\badvertex}
		        {\BiTreeNarrow{\goodvertex}
		          {\Leaf{\badBvertex\bottomlabel{$b_1$}}}
		          {\Leaf{\badBvertex\bottomlabel{$b_2$}}}
		        }
		     }
		     {\UnTree{\badvertex}{\badABvertex\bottomlabel{$x$}}}
		    }
		\put(80,25){\makebox(0,0){$\optcost{5}$}}
		\put(80,5){\makebox(0,0){$a_1 \travone a_2$, $c_{1\!} \travtwo c_2$, $b_1 \travone x$, $b_{2\!} \semitravone m$}}
		\end{picture}
		\end{tabular} 
		
		\vspace{3mm}
		
		\begin{tabular}{c@{~~}c@{~~}c}
		\multicolumn{3}{c}{\bf (v) all canonical subtrees isolated, no solo leaf, no tag mate at \boldmath$\extsubtreeof{\lc}$}\\
		\begin{picture}(180,145)
		\put(90,135){\QuaTreeWide{\goodvertex}
		       {\UnTree{\badvertex}
		        {\BiTreeLeft{\goodvertex}
		          {\Leaf{\badAvertex\bottomlabel{$a_1$}}}
		          {\Leaf{\badAvertex\bottomlabel{$a_2$}}}
		        }
		       } 
		       {\UnTree{\badvertex}
		        {\BiTreeLeft{\goodvertex}
		          {\UnTree{\badvertex}{\badvertex\bottomlabel{$c_1$}}}
		          {\UnTree{\badvertex}{\badvertex\bottomlabel{$c_2$}}}
		        }
		       } 
		       {\UnTree{\badvertex}
		        {\BiTreeLeft{\goodvertex}
		          {\Leaf{\badBvertex\bottomlabel{$b_1$}}}
		          {\Leaf{\badBvertex\bottomlabel{$b_2$}}}
		        }
		       } 
		       {\UnTree{\badvertex}
		        {\TriTree{\goodvertex}
		          {\Leaf{\badABvertex\bottomlabel{$x$}}}
		          {\Leaf{\badABvertex\bottomlabel{$x_1$}}}
		          {\Leaf{\badABvertex\bottomlabel{$x_2$}}}
		        }
		       }
		    }
		\put(90,25){\makebox(0,0){$\optcost{7}$}}
		\put(90,5){\makebox(0,0){$x_1 \travone x_2$}}
		\put(150, 5){\makebox(0,0){+ residual:}}
		\end{picture} &
		\begin{picture}(20,145)
		\put(15,115){\makebox(0,0){$\rightarrow$}}
		\end{picture} &
        \begin{picture}(170,145)
		\put(85,135){\QuaTreeWide{\goodvertex}
		       {\UnTree{\badvertex}
		        {\BiTreeLeft{\goodvertex}
		          {\Leaf{\badAvertex\bottomlabel{$a_1$}}}
		          {\Leaf{\badAvertex\bottomlabel{$a_2$}}}
		        }
		       } 
		       {\UnTree{\badvertex}
		        {\BiTreeLeft{\goodvertex}
		          {\UnTree{\badvertex}{\badvertex\bottomlabel{$c_1$}}}
		          {\UnTree{\badvertex}{\badvertex\bottomlabel{$c_2$}}}
		        }
		       } 
		       {\UnTree{\badvertex}
		        {\BiTreeLeft{\goodvertex}
		          {\Leaf{\badBvertex\bottomlabel{$b_1$}}}
		          {\Leaf{\badBvertex\bottomlabel{$b_2$}}}
		        }
		       } 
		       {\UnTree{\badvertex}{\badABvertex\bottomlabel{$x$}}}
		    }
		\put(85,25){\makebox(0,0){$\optcost{6}$}}
		\put(85,5){\makebox(0,0){$a_{1\!} \travtwo c_1$, $a_2 \travone x$, $b_{1\!} \travtwo c_2$, $b_2 \travone x$}}
		\end{picture}
    \end{tabular}\\
	\caption{There are five reducible topologies of leaf composition \leco 2223. Symmetric cases of (i), (ii), (iii) and (iv) are omitted.}
	\label{fig:leco2223R}
	\end{figure}
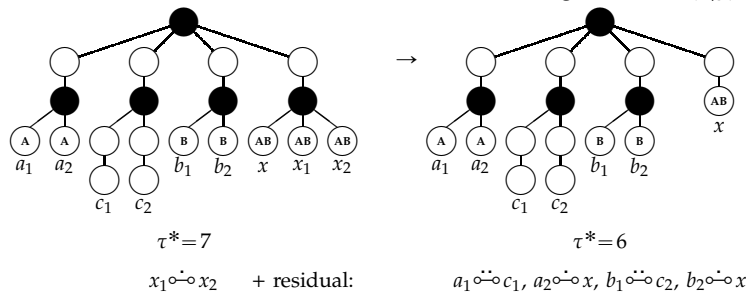


\end{document}